\documentclass[12pt,a4paper]{article}
\usepackage[margin=1in]{geometry}
\usepackage{setspace}
\usepackage[font=normalsize,labelfont=bf]{caption}
\onecolumn
\usepackage[round,authoryear]{natbib}
\usepackage{nicefrac}
\usepackage{booktabs}
\usepackage{multirow}
\usepackage{float}
\usepackage{algorithm}
\usepackage[toc,page]{appendix}
\usepackage{subcaption}
\usepackage[colorlinks,citecolor=blue,urlcolor=blue]{hyperref}
\usepackage{placeins}
\usepackage{authblk}

\usepackage{graphicx}
\graphicspath{{Fig/}}

\usepackage{xcolor}
\usepackage{authblk}
\usepackage{amsmath,amssymb,amsthm}

\newcommand{\ORCID}[1]{\href{https://orcid.org/#1}{\textsuperscript{ORCID}}}

\newtheoremstyle{thmstyleone}
{6pt}{6pt}{\itshape}{}{\bfseries}{.}{.5em}{}
\newtheoremstyle{thmstyletwo}
{6pt}{6pt}{\itshape}{}{\bfseries}{.}{.5em}{}
\newtheoremstyle{thmstylethree}
{6pt}{6pt}{\normalfont}{}{\bfseries}{.}{.5em}{}

\theoremstyle{thmstyleone}
\newtheorem{theorem}{Theorem}
\newtheorem{proposition}{Proposition}
\newtheorem{lemma}{Lemma}

\theoremstyle{thmstyletwo}
\newtheorem{example}{Example}
\newtheorem{remark}{Remark}

\theoremstyle{thmstylethree}
\newtheorem{definition}{Definition}

\usepackage[font=normalsize,labelfont=bf]{caption}

\begin{document}
	\title{\bf Cox Regression on the Plane}
	
	\author{
		Yael Travis-Lumer$^{1,*}$,
		Micha Mandel$^2$,
		Ido Didi Fabian$^{3,4}$,\\
		Rebecca A. Betensky$^5$,
		Malka Gorfine$^6$
	}
	
	\date{}
	
	\maketitle
	
	\begin{center}
		\small
		$^1$Technion -- Israel Institute of Technology, Israel\\
		$^2$Hebrew University of Jerusalem, Israel\\
		$^3$Sheba Medical Center, Tel Aviv University, Israel\\
		$^4$London School of Hygiene \& Tropical Medicine, UK\\
		$^5$New York University, USA\\
		$^6$Tel Aviv University, Israel
		
		\vspace{0.5em}
		
		*Address for correspondence: Yael Travis-Lumer, Faculty of Data and Decision Sciences, Technion -- Israel Institute of Technology, Technion City, 3200003 Haifa, Israel.
		Email: \href{mailto:travis-lumer@technion.ac.il}{travis-lumer@technion.ac.il}
	\end{center}
	
%	\title{Cox Regression on the Plane}
%	
%	\author[1,*]{Yael Travis-Lumer\ORCID{0000-0002-2897-4451}}
%	\author[2]{Micha Mandel\ORCID{0000-0002-4270-9777}}
%	\author[3,4]{Ido Didi Fabian\ORCID{0000-0002-0846-5131}}
%	\author[5]{Rebecca A. Betensky\ORCID{0000-0002-3793-1437}}
%	\author[6]{Malka Gorfine\ORCID{0000-0002-1577-6624}}
%	
%	\affil[1]{Faculty of Data and Decision Sciences, Technion -- Israel Institute of Technology, Technion City, 3200003 Haifa, Israel}
%	\affil[2]{Department of Statistics and Data Science, The Hebrew University of Jerusalem, Mount Scopus, 9190501 Jerusalem, Israel}
%	\affil[3]{Goldschleger Eye Institute, Sheba Medical Center, Tel Aviv University, Derech Sheba 2, 5262000 Ramat Gan, Israel}
%	\affil[4]{International Centre for Eye Health, London School of Hygiene \& Tropical Medicine, Keppel Street, WC1E 7HT London, UK}
%	\affil[5]{Department of Biostatistics, New York University, 708 Broadway, 10003 New York, NY, USA}
%	\affil[6]{Department of Statistics and Operations Research, Tel Aviv University, Ramat Aviv, 6997801 Tel Aviv, Israel}
%	
%	\affil[*]{Address for correspondence: Yael Travis-Lumer, Faculty of Data and Decision Sciences, Technion -- Israel Institute of Technology, Haifa 3200003, Israel. \href{mailto:travis-lumer@technion.ac.il}{travis-lumer@technion.ac.il}}
%	
%	\date{}
%	
%	\maketitle

\abstract{The Cox proportional hazards model is the most widely used regression model in univariate survival analysis, yet extensions to bivariate survival data remain scarce. We propose two novel extensions based on a Lehmann-type representation of the survival function. The first, the simple Lehmann model, is a direct extension that retains a straightforward structure. The second, the generalized Lehmann model, allows greater flexibility by incorporating three distinct regression parameters and includes the simple Lehmann model as a special case. The models admit a direct interpretation in terms of survival probabilities, providing a transparent, fully semiparametric framework for assessing covariate effects on both marginal survival probabilities and their dependence, without requiring specification of a copula or frailty distribution. To estimate the regression parameters, we build on a pseudo-observation–based approach for bivariate survival data and extend it to the generalized model via a two-step procedure. We establish consistency and asymptotic normality of the resulting estimators. The proposed approach is illustrated through simulation studies and an application to data from the Global Retinoblastoma Outcome Study.}

\noindent\textbf{Keywords:}
Bivariate survival function, Pseudo-observations, Generalized estimating equations, Hazard rates, Multivariate failure times

% \boxedtext{
% \begin{itemize}
% \item Key boxed text here.
% \item Key boxed text here.
% \item Key boxed text here.
% \end{itemize}}

\maketitle

\section{Introduction}
The Cox proportional hazards model \citep{cox_regression_1972} is a cornerstone of survival analysis, widely used to assess covariate effects on failure times. Its success stems from its semiparametric structure, where covariate effects enter multiplicatively and the baseline hazard remains unspecified. Extending this framework to bivariate survival data with correlated failure times is challenging, as it requires jointly modeling the marginals and their dependence while preserving interpretability.

Our motivating example concerns children diagnosed with retinoblastoma, the most common pediatric eye cancer. In bilateral cases, times to enucleation of the two eyes are naturally correlated. The Global Retinoblastoma Outcome Study (GROS) \citep{global_retinoblastoma_study_group_global_2020,fabian_global_2022} is a large, multicenter international study of over 4,000 newly diagnosed patients recruited from treatment centers worldwide. For these children, clinically relevant questions involve joint and conditional probabilities, such as the probability that both eyes require enucleation within a given time period, or that one eye requires enucleation within a given time frame given that the other has not yet been enucleated. Such quantities cannot be inferred from marginal models alone, highlighting the need for flexible and interpretable methods for bivariate survival analysis.

Several extensions of the Cox model to bivariate survival data have been proposed. Shared frailty models \citep{clayton_model_1978, clayton_multivariate_1985, nielsen1992counting,gorfine_prospective_2006,zeng_maximum_2007} specify marginal Cox models conditional on a common frailty term, while copula-based approaches \citep{oakes1982association, shih_inferences_1995} combine marginal Cox models with a parametric dependence structure. More recently, \citet{marra_copula_2020} incorporated regression models for both marginal hazards and the association parameter within a parametric copula framework. While flexible, these approaches rely on parametric assumptions on the dependence structure and often yield covariate effects that are not directly interpretable at the joint survival level.

An alternative extension of the Cox model specifies three proportional hazards models: two for the single-failure hazards and one for the double-failure hazard. \citet[Section 4]{pons_nonparametric_1992} briefly proposed this approach, combining two marginal Cox models with an additional model for the joint failure rate. %\citet{tien_proportional_2002} proposed a related framework for bivariate interval-censored data, restricting estimation to time points where the two event times coincide.
More recently, \citet{prentice_regression_2021} extended such multiplicative hazard models to multivariate failure times. A key limitation of these approaches is that the resulting covariate-adjusted hazards do not, in general, correspond to a valid bivariate survival distribution, as we show in Section~\ref{sec:hazard_functions}.

Despite the central role of the Cox model in univariate survival analysis, existing extensions do not simultaneously preserve its semiparametric structure, yield a valid bivariate survival distribution, and retain transparent covariate interpretation. Our approach addresses this gap by extending a structural property of the Cox model rather than introducing additional parametric components.

In this paper, we propose two semiparametric extensions of the Cox model for bivariate survival data based on a generalized Lehmann representation. The models directly parameterize the covariate-adjusted bivariate survival function, yielding regression coefficients with a direct interpretation on the survival scale, and can be viewed as proportional derivatives of the log-survival function. The models impose parametric assumptions only on covariate effects, leaving the baseline dependence structure unspecified, thereby reducing the risk of misspecification associated with parametric copula models. Although developed for general bivariate distributions, the framework can accommodate exchangeable event times through symmetry constraints, as in our motivating retinoblastoma application. As illustrated in the GROS analysis (Section~\ref{sec:data}), even copulas favored by standard model selection criteria may imply an incorrect dependence structure and fail to preserve structural features such as exchangeability, whereas the proposed approach yields a flexible and clinically interpretable characterization of joint and conditional survival.

We also develop a two-step estimation method extending the pseudo-observations approach. To our knowledge, this is the first two-step pseudo-observations–based method for bivariate survival regression. We establish consistency and asymptotic normality. An appealing feature of the approach is that model assumptions need only hold at a finite set of time points, aligning naturally with inference on clinically meaningful survival probabilities, as in the GROS application.

%\color{blue}Conceptually, the proposed framework differs from copula-based regression approaches in the level at which dependence is modeled. Copula models construct the joint survival distribution through a chosen copula family, and model dependence through copula parameters. By contrast, our approach models covariate effects directly on the bivariate survival function itself via a generalized Lehmann representation, permitting flexible baseline dependence without requiring specification of a parametric copula family. \color{black}

\color{black}
%Conceptually, the proposed framework differs from copula-based regression approaches in the level at which dependence is modeled. Copula models construct the joint survival distribution through a chosen copula family and characterize dependence through copula parameters. By contrast, our approach models covariate effects directly on the bivariate survival function itself via a generalized Lehmann representation, allowing flexible baseline dependence without requiring specification of a parametric copula family.
Conceptually, the proposed framework differs from copula-based regression approaches in the level at which dependence is modeled. Copula methods construct the joint survival function through a specified copula family and characterize dependence via copula parameters. In contrast, our approach models covariate effects directly on the bivariate survival function through a generalized Lehmann representation, allowing flexible baseline dependence without specifying a copula family.
\color{black}

\section{Preliminaries}

Let $(T_1, T_2)$ denote continuous bivariate survival times, and let $Z\in \mathbb{R}^p$ denote a time-independent covariate vector.
Let $S(t_1,t_2)=\Pr(T_1>t_1,T_2>t_2)$ denote the joint survival function, with marginals $S_{T_1}(t_1)$ and $S_{T_2}(t_2)$. For infinitesimal increments $ds,dt$, write $\Pr(T_1\in ds,T_2\in dt)$ for the joint probability over $[s,s+ds]\times[t,t+dt]$.
The joint failure hazard is defined as
$$\Lambda_{11}(ds,dt)
= \Pr(T_1\in ds,T_2\in dt\mid T_1\ge s,T_2\ge t)
= \frac{\Pr(T_1\in ds,T_2\in dt)}{S(s^-,t^-)}.$$
The single-failure hazards are
$\Lambda_{10}(ds,t)
= \Pr(T_1\in ds\mid T_1\ge s,T_2>t)$,  and $\Lambda_{01}(s,dt)
= \Pr(T_2\in dt\mid T_1>s,T_2\ge t).$
Thus, $\Lambda_{10}$ and $\Lambda_{01}$ are conditional univariate hazards given survival of the other component. Define the cumulative hazard vector as $\Lambda(s,t)=\big(\Lambda_{10}(s,t),\Lambda_{01}(s,t),\Lambda_{11}(s,t)\big)^T$. If $S$ admits a density $f$,
\[
\Lambda_{11}(ds,dt)=\lambda_{11}(s,t)\,ds\,dt,\quad
\Lambda_{10}(ds,t)=\lambda_{10}(s,t)\,ds,\quad
\Lambda_{01}(s,dt)=\lambda_{01}(s,t)\,dt,
\]
where $\lambda_{11}(s,t)=\frac{f(s,t)}{S(s^-,t^-)},\quad
\lambda_{10}(s,t)=\frac{\int_t^\infty f(s,v)\,dv}{S(s^-,t)},\quad
\lambda_{01}(s,t)=\frac{\int_s^\infty f(u,t)\,du}{S(s,t^-)}$.
The marginal hazards satisfy $\lambda_{T_1}(t_1)=\lambda_{10}(t_1,0)$ and $\lambda_{T_2}(t_2)=\lambda_{01}(0,t_2)$.

Define $A(t_1,t_2)=\log S(t_1,t_2)$. By \citet{dabrowska_kaplan-meier_1988}, for $(t_1,t_2)\in [0,\tau_1]\times [0,\tau_2]$ with $S(\tau_1,\tau_2)>0$, the bivariate survival function admits the decomposition
\begin{equation}\label{eq1}
	S(t_1,t_2)
	= S_{T_1}(t_1)\,S_{T_2}(t_2)\,
	\exp\left\{\int_0^{t_1}\int_0^{t_2}A(du,dv)\right\}.
\end{equation}
Thus, $S(t_1,t_2)$ can be expressed as the product of its marginals and a multiplicative term capturing dependence, given by
$\nicefrac{S(t_1,t_2)}{\left(S_{T_1}(t_1)S_{T_2}(t_2)\right)}.$
Further details and alternative representations in terms of the cumulative hazard vector are provided in Appendices~\ref{Appendix A} and \ref{app:B} of the supplementary material.

\section{The two models}\label{sec:Lehmann}
\subsection{The simple Lehmann model}\label{sec:Lehmann_simple}
In the univariate setting, the Cox model $\lambda_T(t \mid Z) = \lambda_T^0(t)\exp(\beta^T Z)$ admits the equivalent survival representation $S_T(t \mid Z) = \left[ S_T^0(t) \right]^{\exp(\beta^T Z)}$, known as the Lehmann alternative form. 
A bivariate extension of the Lehmann representation is % of the Cox model is
\begin{equation}\label{eq:Lehmann}
	S(t_1,t_2 \mid Z) = \left[ S^0(t_1,t_2) \right]^{\exp(b^T Z)},
\end{equation}

where $S^0(t_1,t_2)$ is a baseline bivariate survival function and $b \in \mathbb{R}^p$ is a vector of regression coefficients. We refer to \eqref{eq:Lehmann} as the \emph{simple Lehmann model}. It induces a common Cox proportional hazards effect on both marginals, so that covariates have identical effects on the log-survival of $T_1$ and $T_2$. At the joint level, covariates act through a single multiplicative factor on the log-survival function, preserving the baseline dependence structure and introducing no separate regression effect on dependence; in particular, a positive coefficient corresponds to a faster decline of the joint survival surface. The model is therefore interpretable on the survival scale, although it does not, in general, correspond to a proportional hazards model for the joint (double-failure) hazard; see Section~\ref{sec:hazard_functions}.

\subsection{The generalized Lehmann model}
In some applications, covariates may affect the two marginals and their dependence differently. Using the decomposition in (1), this motivates a natural generalization in which separate regression coefficients are assigned to the marginal components and to the dependence component, each admitting a Cox-type representation; see Appendix~\ref{app:Lehmann_details} of the supplementary material for details. Specifically,
\begin{equation}\label{eq:Cox model_general}
	\begin{aligned}
		S_{T_1}(t_1\mid Z)&= \left[S_{T_1}^0(t_1)\right]^{\exp(\alpha^T Z)},
		\quad
		S_{T_2}(t_2\mid Z)= \left[S_{T_2}^0(t_2)\right]^{\exp(\beta^T Z)}, \\
		\exp\Big\{\int_0^{t_1} \int_0^{t_2} A(du,dv; Z)\Big\}
		&=
		\left[\exp\Big\{\int_0^{t_1} \int_0^{t_2} A^0(du,dv)\Big\}\right]^{\exp(\gamma^T Z)}.
	\end{aligned}
\end{equation}
%\begin{equation}\label{eq:Cox model_general}
%	\begin{aligned}
%		S_{T_1}(t_1\mid Z)&= \left[S_{T_1}^0(t_1)\right]^{\exp(\alpha^T Z)}, \\
%		S_{T_2}(t_2\mid Z)&= \left[S_{T_2}^0(t_2)\right]^{\exp(\beta^T Z)}, \\
%		\exp\Big\{\int_0^{t_1} \int_0^{t_2} A(du,dv; Z)\Big\}
%		&=
%		\left[\exp\Big\{\int_0^{t_1} \int_0^{t_2} A^0(du,dv)\Big\}\right]^{\exp(\gamma^T Z)}.
%	\end{aligned}
%\end{equation}
Equivalently,
\begin{equation}\label{eq:overall_survival}
	S(t_1,t_2 \mid Z)
	=
	\left[S_{T_1}^0(t_1)\right]^{\exp(\alpha^T Z)}
	\left[S_{T_2}^0(t_2)\right]^{\exp(\beta^T Z)}
	\left[\exp\Big\{\int_0^{t_1} \int_0^{t_2} A^0(du,dv)\Big\}\right]^{\exp(\gamma^T Z)}.
\end{equation}
 We refer to \eqref{eq:overall_survival} as the \emph{generalized Lehmann model}. When $\alpha=\beta=\gamma$, this reduces to the simple Lehmann model.
The model can also be expressed in terms of derivatives of the log survival function:
$\lambda_{T_1}(t_1 \mid Z)=\lambda_{T_1}^0(t_1)e^{\alpha^T Z},\quad
\lambda_{T_2}(t_2 \mid Z)=\lambda_{T_2}^0(t_2)e^{\beta^T Z},\quad
a(t_1,t_2; Z)=a^0(t_1,t_2)e^{\gamma^T Z},$
where
$\lambda_{T_1}(t_1 \mid Z) = -\nicefrac{\partial \log S(t_1,0 \mid Z)}{\partial t_1}$, $\lambda_{T_2}(t_2 \mid Z) = -\nicefrac{\partial \log S(0,t_2 \mid Z)}{\partial t_2}$, and
$a(t_1,t_2; Z)=\nicefrac{\partial^2 \log S(t_1,t_2\mid Z)}{\partial t_1\partial t_2}.$
Thus, the generalized Lehmann model specifies three multiplicative models for derivatives of the log survival function. The first two are standard proportional hazards models, while the third is a proportional model for the mixed second derivative.

The marginal parameters $\alpha$ and $\beta$ retain the usual hazard ratio interpretation. The parameter $\gamma$ governs how covariates affect dependence, which can be understood through the ratio
$\nicefrac{S^0(t_1,t_2)}{\left(S_{T_1}^0(t_1)S_{T_2}^0(t_2)\right)}$.
When this ratio equals one, the two survival times are independent, and the dependence component is absent (see Section~\ref{sec:estimation} for implications).
We now define these concepts for a generic bivariate survival function $S(t_1,t_2)$, suppressing any covariate dependence.

\begin{definition}\label{defn:PQD}
	The pair $(T_1,T_2)$ is \emph{positively quadrant dependent} (PQD), as defined by \citet{lehmann_concepts_1966}, if $S(t_1,t_2)\geq S_{T_1}(t_1)S_{T_2}(t_2)
	\quad \text{for all } (t_1,t_2)$,
	and \emph{negatively quadrant dependent} (NQD) if the inequality is reversed.
\end{definition}

By Equation~\eqref{eq1},
$\exp\left\{\int_0^{t_1}\int_0^{t_2}A(du,dv)\right\}
=
\nicefrac{S(t_1,t_2)}{\left(S_{T_1}(t_1)S_{T_2}(t_2)\right)},$
which is at least one under PQD and at most one under NQD. Accordingly, the dependence component in \eqref{eq:Cox model_general} can be written as
$\left[
\nicefrac{S^0(t_1,t_2)}{\left(S_{T_1}^0(t_1)S_{T_2}^0(t_2)\right)}
\right]^{\exp(\gamma^T Z)}$.
Thus, $\gamma$ modulates the strength of the baseline dependence structure, scaling the strength of the baseline dependence while preserving its structure. Importantly, the parameter $\gamma$ does not correspond to a conventional scalar association parameter such as Kendall’s tau or a copula parameter. Rather, it governs how covariates amplify or attenuate the baseline dependence component $\nicefrac{S^0(t_1,t_2)}{\{S_{T_1}^0(t_1)S_{T_2}^0(t_2)\}}$ while preserving its qualitative structure, including the direction of dependence. \color{black} %while the type of dependence (e.g., whether it is PQD or NQD) is determined by the baseline model.

For a scalar covariate $Z$, increasing $z$ by one unit yields
\[
S(t_1,t_2 \mid z+1)
=
[S_{T_1}(t_1 \mid z)]^{e^\alpha}
[S_{T_2}(t_2 \mid z)]^{e^\beta}
\left[
\frac{S(t_1,t_2 \mid z)}{S_{T_1}(t_1 \mid z)S_{T_2}(t_2 \mid z)}
\right]^{e^\gamma}.
\]
Hence, the overall effect of a covariate on the joint survival reflects the combined action of the marginal and dependence components: $\alpha$ and $\beta$ govern marginal survival, while $\gamma$ controls how dependence is amplified or attenuated.

Finally, the model readily accommodates different covariate vectors for the three components (e.g., $Z_1$, $Z_2$, and $Z_3$), with irrelevant coefficients set to zero.

\subsection{The corresponding hazard functions}\label{sec:hazard_functions}

Assuming continuity, the covariate-adjusted bivariate survival function admits the representation
%\[
%\begin{aligned}
	$S(t_1,t_2\mid Z)
	=\exp\Bigg\{-\Lambda_{10}(t_1,0\mid Z)-\Lambda_{01}(0,t_2\mid Z)
	 + \int_{0}^{t_1}\int_{0}^{t_2} 
	a(u,v;Z) dudv\Bigg\}$,
%\end{aligned}
%\]
where $a(u,v;Z)=\lambda_{11}(u,v\mid Z)-\lambda_{10}(u,v\mid Z)\lambda_{01}(u,v\mid Z)$.
The corresponding hazard functions satisfy
\[ \begin{aligned} \lambda_{10}(t_1,t_2\mid Z) &= \lambda_{10}(t_1,0\mid Z)-\int_{0}^{t_2}a(t_1,v; Z)\,dv,\\ \lambda_{01}(t_1,t_2\mid Z) &= \lambda_{01}(0,t_2\mid Z)-\int_{0}^{t_1}a(u,t_2; Z)\,du,\\ \lambda_{11}(t_1,t_2\mid Z) &= \lambda_{10}(t_1,t_2\mid Z)\lambda_{01}(t_1,t_2\mid Z)+a(t_1,t_2; Z). \end{aligned} \]
 Under continuity, the hazard functions satisfy the constraint
\[
-\frac{\partial\lambda_{01}(t_1,t_2\mid Z)}{\partial t_1}
=
a(t_1,t_2; Z)
=
-\frac{\partial\lambda_{10}(t_1,t_2\mid Z)}{\partial t_2},
\]
which follows from equality of mixed partial derivatives; see Appendix~\ref{app:C} of the supplementary material for details.
Consequently, specifying three proportional hazards models independently (for the two single-failure hazards and the double-failure hazard), as in \citealt{pons_nonparametric_1992,prentice_regression_2021}, does not, in general, yield a valid bivariate survival distribution (see Appendix~\ref{app:C}).

\allowdisplaybreaks
\begin{proposition}\label{prop:hazards}
	The hazard functions corresponding to the simple Lehmann model are:
	\begin{equation}\label{eq:cond_hazard_simple}
		\begin{aligned}
			\lambda_{10}(u,v\mid Z)&=\lambda_{10}^0(u,v)e^{b^TZ},	\quad
			\lambda_{01}(u,v\mid Z)=\lambda_{01}^0(u,v)e^{b^TZ},\\
			\lambda_{11}(u,v\mid Z)
			&=\lambda_{11}^0(u,v)e^{b^TZ}
			-\lambda_{10}^0(u,v)\lambda_{01}^0(u,v)\left[e^{b^TZ}-e^{2b^TZ}\right].
		\end{aligned}
	\end{equation}
	
The hazard functions corresponding to the generalized Lehmann model are:
	\begin{equation}\label{eq:cond_hazard_general}
		\begin{aligned}
			\lambda_{10}(t_1,t_2\mid Z)
			&= e^{\alpha^TZ}\lambda_{10}^0(t_1,0)-e^{\gamma^TZ}\int_{0}^{t_2}a^0(t_1,v)\,dv,\\
			\lambda_{01}(t_1,t_2\mid Z)
			&= e^{\beta^TZ}\lambda_{01}^0(0,t_2)-e^{\gamma^TZ}\int_{0}^{t_1}a^0(u,t_2)\,du,\\
			\lambda_{11}(t_1,t_2\mid Z)
			&= \lambda_{10}(t_1,t_2\mid Z)\lambda_{01}(t_1,t_2\mid Z)
			+e^{\gamma^TZ}a^0(t_1,t_2),
		\end{aligned}
	\end{equation}
	where $a^0(u,v)=\lambda_{11}^0(u,v)-\lambda_{10}^0(u,v)\lambda_{01}^0(u,v)$. 
\end{proposition}
%\vspace{-0.25\baselineskip}
\noindent
For a proof see Appendix~\ref{app:C} of the supplementary material.

\subsection{The validity of the proposed models}\label{sec:validity}

For the proposed Lehmann models to define valid bivariate survival functions, the corresponding hazard functions must be non-negative:
$\lambda_{10}(t_1,t_2\mid Z) \geq 0,\quad
\lambda_{01}(t_1,t_2\mid Z) \geq 0,\quad
\lambda_{11}(t_1,t_2\mid Z) \geq 0,$
for all $(t_1,t_2)$ in the model domain; see Appendix~\ref{app:E} of the supplementary material for details. We next provide sufficient conditions ensuring validity.

\begin{proposition}[Conditions for the simple Lehmann model]\label{prop:suff_cond_simple}
	Let $a^0(t_1,t_2)=\frac{\partial^2}{\partial t_1\partial t_2}\log S^0(t_1,t_2)$. If either:
	\begin{enumerate}
		\item $a^0(u,v)\geq 0$ for all $u,v\geq 0$; or
		\item $b^T Z \geq 0$ for all covariate vectors $Z$,
	\end{enumerate}
	then the simple Lehmann model \eqref{eq:Lehmann} defines a proper bivariate survival function.
\end{proposition}

The proof is given in Appendix~\ref{app:E_proofs} of the supplementary material. Condition (1) corresponds to log-supermodular baseline survival functions and thus implies PQD, meaning that survival in one component is associated with higher survival probability in the other. Although imposed on the baseline distribution, the model preserves the sign of dependence, allowing empirical assessment by comparing the joint survival with the product of the marginals. Condition (2) instead imposes $b^T Z\ge 0$, allowing arbitrary baseline dependence. In practice, this can be enforced by shifting covariates and constraining coefficients, or more generally by requiring $b^T z \ge 0$ over the covariate region of interest. Such shifts affect only the baseline survival function, not the interpretation of the regression coefficients. Similar non-negativity constraints arise in additive hazard models; see \citet{lu_maximum_2023}.

%The proof is given in Appendix~\ref{app:E_proofs} of the supplementary material. Condition (1) corresponds to log-supermodular baseline survival functions and hence implies PQD, meaning that survival in one component is associated with a higher probability of survival in the other. Although this condition is imposed on the baseline distribution, the model preserves the sign of dependence, so it can be assessed empirically by comparing the joint survival with the product of the marginals. Condition (2) instead imposes $b^T Z\ge 0$, allowing arbitrary baseline dependence. In practice, this can be enforced by shifting covariates and constraining the coefficients, or more generally by requiring $b^T z \ge 0$ over the covariate region of interest. Such shifts affect only the baseline survival function and not the interpretation of the regression coefficients. Similar non-negativity constraints arise in additive hazard models; see, for example, \citet{lu_maximum_2023}. 

\begin{proposition}[Conditions for the generalized Lehmann model]\label{prop:suff_cond_generalized}
	Let $a^0(t_1,t_2)=\frac{\partial^2}{\partial t_1\partial t_2}\log S^0(t_1,t_2)$. If either:
	\begin{enumerate}
		\item $\gamma^T Z \leq \min\{\alpha^T Z,\beta^T Z\}$ for all $Z$, and $a^0(u,v)\geq 0$ for all $(u,v)$; or
		\item $0 \leq \gamma^T Z \leq \min\{\alpha^T Z,\beta^T Z\}$ for all $Z$,
	\end{enumerate}
	then \eqref{eq:overall_survival} defines a proper bivariate survival function.
\end{proposition}

The proof is given in Appendix~\ref{app:E_proofs} of the supplementary material. 
Under the conditions in Proposition~\ref{prop:suff_cond_generalized}, the covariate effect on dependence is smaller than its effect on each margin. For example, if a covariate increases the risk of both outcomes, this condition requires that its effect on each outcome individually be at least as strong as its effect on their dependence. In the context of retinoblastoma, this corresponds to assuming that factors such as tumor stage or age at presentation have a stronger impact on the risk of enucleation in each eye than on the association between the two times. This restriction aligns with many practical settings, where covariates primarily act on marginal risks while inducing more limited changes in dependence. As in the simple model, such constraints can often be accommodated through appropriate reparameterization or shifting of the covariates, which affects only the baseline functions and not the interpretation of the regression coefficients.

The above conditions are sufficient but not necessary. For example, the Gumbel--Barnett copula, which induces NQD, together with $\gamma=\alpha+\beta$, yields a valid model that does not satisfy Proposition~\ref{prop:suff_cond_generalized} (see Appendix~\ref{sec:exp2D_valid} of the supplementary material). A full characterization of necessary conditions remains open.

%\color{blue}The conditions in Propositions~\ref{prop:suff_cond_simple}-\ref{prop:suff_cond_generalized} are intended primarily as theoretical guarantees ensuring that the resulting hazard functions define a valid bivariate survival distribution. In practice, estimation is performed without explicit parameter constraints, and in our simulations and data application we did not encounter invalid fitted survival functions over the observed covariate range.\color{black}

%The proposed models are semiparametric at both marginal and joint levels, avoiding specification of a copula family or frailty distribution. They act directly on the joint survival function, yielding transparent covariate interpretations without latent variables. In the generalized Lehmann model, $\gamma$ captures a separate regression effect on dependence, expressed directly on the survival scale and independent of parametric family choice.

\subsection{Dependency measures}\label{sec:dependency_measures}

Dependence between two failure times can be quantified by the cross-ratio function, introduced by \citet{clayton_model_1978}, \citet{oakes_bivariate_1989}, and \citet{hsu_assessing_1996}. For continuous times, it is defined in terms of the joint and marginal hazard functions:
$c(t_1,t_2)=\frac{\lambda_{11}(t_1,t_2)}{\lambda_{10}(t_1,t_2)\lambda_{01}(t_1,t_2)}$.
The value of $c(t_1,t_2)$ indicates positive ($>1$), no ($=1$), or negative ($<1$) local association between the two failure times at $(t_1,t_2)$. 

This notion extends to covariate-adjusted dependence via
\[
\begin{aligned}
	c(t_1,t_2\mid Z)&=\frac{\lambda_{11}(t_1,t_2\mid Z)}{\lambda_{10}(t_1,t_2\mid Z)\lambda_{01}(t_1,t_2\mid Z)},\\
	C(t_1,t_2\mid Z)&=\frac{\int_{0}^{t_1}\int_{0}^{t_2} \lambda_{11}(s_1,s_2\mid Z)\, ds_1 ds_2}{\int_{0}^{t_1}\int_{0}^{t_2} \lambda_{10}(s_1,s_2\mid Z)\lambda_{01}(s_1,s_2\mid Z)\, ds_1 ds_2},
\end{aligned}
\]
where the latter is a special case of a broader class of weighted dependence measures; see Appendix~\ref{app:D} of the supplementary material. 

\begin{proposition}\label{prop:cross-ratio}
	Under either Lehmann model, $c(t_1,t_2\mid Z)$ is greater than, equal to, or less than one depending on whether $a^0(t_1,t_2)$ is positive, zero, or negative.
\end{proposition}
\begin{proposition}\label{prop:weighted_cross-ratio}
	Under either Lehmann model, $C(t_1,t_2\mid Z)$ is greater than, equal to, or less than one depending on whether $\int_0^{t_1}\int_0^{t_2} a^0(s_1,s_2) ds_1 ds_2=\log \left(\frac{S^0(t_1,t_2)}{S_1^0(t_1)S_2^0(t_2)}\right)$ is positive, zero, or negative.
\end{proposition}
Proofs and explicit formulations of Propositions~\ref{prop:cross-ratio}–\ref{prop:weighted_cross-ratio} are given in Appendix~\ref{app:D} of the supplementary material. These propositions show that, under both Lehmann models, the direction of dependence is determined by the baseline and preserved under covariate adjustment, directly linking to the PQD/NQD classification. 

\section{Estimation of model parameters}\label{sec:estimation}

A natural starting point is a likelihood-based approach via partial likelihood \citep{cox_regression_1972}. In the univariate Cox model, the partial likelihood depends only on regression parameters, as the baseline hazard cancels under proportional hazards. In the bivariate setting, an analogous partial likelihood can be based on the double-failure rate $\lambda_{11}(t_1,t_2\mid Z)$. However, under both Lehmann models, $\lambda_{11}(t_1,t_2\mid Z)$ is not multiplicative, so the baseline hazards do not cancel and the partial likelihood involves additional nuisance parameters.

To avoid this difficulty, we adopt a generalization of the pseudo-observations approach \citep{andersen_generalised_2003}. For the simple Lehmann model, estimation can be carried out using the existing bivariate pseudo-observations approach of \citet{travis2024pseudo}. For the generalized Lehmann model, we develop a two-step estimation procedure extending this approach and establish its asymptotic properties.

\subsection{Bivariate pseudo-observations}\label{sec:PO2D}
Let $(T_1, T_2)$ denote continuous bivariate survival times and $(C_1, C_2)$ the corresponding right censoring times, which may coincide (i.e., $C_1=C_2$). For $j=1,2$, define $\tilde T_j=\min(T_j,C_j)$ and $\Delta_j=I(T_j\le C_j)$. The observed data consist of $n$ i.i.d.\ copies $\{(\tilde T_{1i},\Delta_{1i},\tilde T_{2i},\Delta_{2i},Z_i)\}_{i=1}^n$, where $Z_i\in\mathbb{R}^p$ is a time-independent covariate vector.

Let $h(\cdot,\cdot)$ be a function of $(T_1,T_2)$, and define
$\theta = E[h(T_1,T_2)].
$
We focus on indicator functions of the form
$
h(T_1,T_2)=I(T_1>t_1^0,\,T_2>t_2^0),
$
for which $\theta=S(t_1^0,t_2^0)$, the bivariate survival probability at $(t_1^0,t_2^0)$.

 Given an estimator $\hat{\theta}$ of $\theta$, the pseudo-observation \citep{andersen_generalised_2003} for subject $i$ is
 $\hat{\theta}_i = n\hat{\theta} - (n-1)\hat{\theta}^{(-i)}$, where $\hat{\theta}^{(-i)}$ is the leave-one-out estimator.
Following \citet{andersen_generalised_2003} and \citet{travis2024pseudo}, the pseudo-observations $\hat{\theta}_i$ are treated as responses in a generalized linear model (GLM). With link function $g$, the regression coefficients are estimated from
%\begin{equation}\label{eq:GLM}
	$g\big(E[h(T_1,T_2)\mid Z]\big) = \beta_0 + \beta^T Z$.
%\end{equation}

This approach directly models covariate effects on bivariate survival functionals, avoiding complications of likelihood-based estimation in the bivariate Cox setting.
%This approach enables direct modeling of covariate effects on bivariate survival functionals, avoiding the complications of likelihood-based estimation in the bivariate Cox-type setting.

\subsection{Fitting the simple Lehmann model}\label{sec:fit_Lehmann}

The simple Lehmann model can be written as a semiparametric model using the complementary log-log (cloglog) link:
$\log\big(-\log S(t_1,t_2\mid Z)\big)
=
b_0(t_1,t_2)+b^TZ$,
where $b=(b_1,\ldots,b_p)^T$ are the regression coefficients (slopes) and $b_0(t_1,t_2)=\log\big(-\log S^0(t_1,t_2)\big)$ is the link-transformed value of the baseline survival function.
For functionals of the form $h(T_1,T_2)=I(T_1>t_1^0, T_2>t_2^0)$, the simple Lehmann model reduces to a GLM with inverse cloglog link:
\begin{equation}\label{eq:GLM_simple}
	E[h(T_1,T_2)\mid Z]
	=
	S(t_1^0,t_2^0 \mid Z)
	=
	\exp\!\left\{-\exp\left(b_0(t_1^0,t_2^0) + b^TZ\right)\right\},
\end{equation}
where $b_0(t_1^0,t_2^0)$ acts as a time-specific intercept, so that the model involves $p+1$ parameters at a given time point.

When evaluating $K$ time points $\{(t_1^j,t_2^j)\}_{j=1}^K$, each time point has its own intercept, while the slope vector is shared, yielding $p+K$ parameters. Estimation proceeds via generalized estimating equations (GEE) \citep{liang_longitudinal_1986}:
\begin{equation}\label{eq:GEE}
	U(\tilde b)=\sum_{i=1}^{n}U_i(\tilde b)=\sum_{i=1}^{n}
	\left(\frac{\partial}{\partial \tilde b} g^{-1}(\tilde b^T Z_i^*)\right)^T
	V_i^{-1}
	\left(\hat{\theta}_i - g^{-1}(\tilde b^T Z_i^*)\right)
	=
	0,
\end{equation}
where $\tilde b=(b_{01},\ldots,b_{0K},b_1,\ldots,b_p)^T$ stacks the $K$ time-specific intercepts and the shared slope coefficients, $\hat{\theta}_i\in\mathbb{R}^K$ are pseudo-observations, $g^{-1}(y)=\exp(-\exp(y))$, $V_i$ is a working covariance matrix (often the identity), and $Z_i^*$ augments $Z_i$ with time point indicators. See \citet[Section 2.4]{travis2024pseudo} for details.
%\subsubsection{Asymptotic theory}

\citet{travis2024pseudo} established consistency and asymptotic normality of the estimator $\hat{\tilde b}$, defined as the solution to the GEE \eqref{eq:GEE}, for general link functions $g$. For completeness, we summarize the result at a single bivariate time point ($K=1$), assuming the GLM in \eqref{eq:GLM_simple} with true parameter $\tilde b = (b_0,b_1,\ldots,b_p)^T$. Assume also that $(C_1,C_2)$ is independent of $(T_1,T_2)$ and $Z$, the bivariate analogue of the completely independent censoring assumption standard in the pseudo-observations literature \citep{andersen_generalised_2003,graw_pseudo-values_2009,overgaard_asymptotic_2017}.

\begin{theorem}\label{theorem}
	Let $\hat{\tilde b}$ solve the generalized estimating equations \eqref{eq:GEE}, where the pseudo-observations $\hat{\theta}_i$ are based on the \cite{dabrowska_kaplan-meier_1988} estimator. Assume independent censoring.
	Then, under suitable regularity conditions (Appendix~\ref{app:F} of the supplementary material),
	$\hat{\tilde b} \xrightarrow{p} \tilde b^*$ and
	$\sqrt{n}(\hat{\tilde b}-\tilde b^*) \xrightarrow{d} N(0, M^{-1}\Sigma_{\tilde b} M^{-1})$,
	where $M = N(\tilde b^*)$, $N(\tilde b) = E\!\left[\left(\frac{\partial}{\partial \tilde b} g^{-1}(\tilde b^T Z_i^*)\right)^T V_i^{-1}
	\frac{\partial}{\partial \tilde b} g^{-1}(\tilde b^T Z_i^*)\right]$, and $\Sigma_{\tilde b}$ is defined in Appendix~\ref{app:F} of the supplementary material.
\end{theorem}

\subsection{Fitting the generalized Lehmann-type model}\label{sec:fitting_gen_Lehmann}

We extend the pseudo-observations approach to estimate the regression parameters $\alpha$, $\beta$, and $\gamma$ by expressing the generalized Lehmann model as a trivariate GLM at a fixed time point $(t_1^0,t_2^0)$.
The three components in \eqref{eq:Cox model_general} can be written as
\begin{equation}\label{eq:Cox_gen_GLM}
	\begin{aligned}
		\log(-\log S_{T_1}(t_1^0\mid Z)) &= \log(\Lambda_{T_1}^0(t_1^0)) + \alpha^T Z, \\
		\log(-\log S_{T_2}(t_2^0\mid Z)) &= \log(\Lambda_{T_2}^0(t_2^0)) + \beta^T Z, \\
		\exp\left\{\int_0^{t_1^0}\!\int_0^{t_2^0} A(du,dv;Z)\right\}
		&=
		\left[\exp\left\{\int_0^{t_1^0}\!\int_0^{t_2^0} A^0(du,dv)\right\}\right]^{\exp(\gamma^T Z)}.
	\end{aligned}
\end{equation}

The first two equations correspond to GLMs with cloglog links. The third requires a link function that depends on the sign of $\int_0^{t_1^0}\int_0^{t_2^0} A^0(du,dv)$.
Let
$D^0(t_1^0,t_2^0)
=
\frac{S^0(t_1^0,t_2^0)}{S_{T_1}^0(t_1^0)S_{T_2}^0(t_2^0)}
=
\exp\left\{\int_0^{t_1^0}\int_0^{t_2^0} A^0(du,dv)\right\}$,
so that the third equality in \eqref{eq:Cox_gen_GLM} can be written as $
\left[\exp\left\{\int_0^{t_1^0}\!\int_0^{t_2^0} A^0(du,dv)\right\}\right]^{\exp(\gamma^T Z)}=\left[D^0(t_1^0,t_2^0)\right]^{\exp(\gamma^T Z)}$.

Thus, the third component can be expressed as a regression model with an appropriate link function, depending on whether $D^0(t_1^0,t_2^0)$ is greater or less than one. In practice, the link function is selected empirically; see Remark~\ref{rem:g3}. Since $D^0(t_1,t_2)$ is unknown, selection is based on the observed pseudo-values and performed separately at each time point to recover the transformation implied by $D^0(t_1,t_2)$.
The asymptotic results are derived under a fixed link function; accounting for data-driven selection is beyond the scope of this work. 

To determine the link function for the third equation in \eqref{eq:Cox_gen_GLM}, we consider the value of $D^0(t_1^0,t_2^0)$.
If $D^0(t_1^0,t_2^0)=1$, then
$S^0(t_1^0,t_2^0)=S_{T_1}^0(t_1^0)S_{T_2}^0(t_2^0)$
so the baseline distribution is independent. Therefore,
$S(t_1^0,t_2^0\mid Z)=S_{T_1}(t_1^0\mid Z)\,S_{T_2}(t_2^0\mid Z)$
for all $Z$, and hence no dependence parameter $\gamma$ is required. The model therefore reduces to two marginal regressions for $\alpha$ and $\beta$.

If $D^0(t_1^0,t_2^0)<1$, define $g(u,v,w)=\big(\log(-\log u),\, \log(-\log v),\, \log(-\log(w/(uv)))\big)^T$,
and if $D^0(t_1^0,t_2^0)>1$, replace the third component by $\log(\log(w/(uv)))$. Accordingly, define the scalar link $g_3(d)=\log(-\log d)$ for $D^0(t_1^0,t_2^0)<1$ and $g_3(d)=\log(\log d)$ for $D^0(t_1^0,t_2^0)>1$, where $d=w/(uv)$ is the dependence ratio.

To link this representation with pseudo-observations, for a fixed time point $(t_1^0,t_2^0)$, we define the mapping $h:\mathbb{R}^2\mapsto\mathbb{R}^3$ by
$h(T_1,T_2)=
\big(
I(T_1>t_1^0),\,
I(T_2>t_2^0),\,
I(T_1>t_1^0,T_2>t_2^0)
\big)^T$.
Then
$E[h(T_1,T_2)] =
\big(S_{T_1}(t_1^0),\, S_{T_2}(t_2^0),\, S(t_1^0,t_2^0)\big)^T\equiv \theta=(\theta_1,\theta_2,\theta_3)^T$,
and similarly for the conditional expectation given $Z$.

When $D^0(t_1^0,t_2^0)\neq 1$, the generalized Lehmann model can be written as
\begin{equation*}\label{eq:g_tilde_model}
	g\big(S_{T_1}(t_1^0\mid Z),\, S_{T_2}(t_2^0\mid Z),\, S(t_1^0,t_2^0\mid Z)\big)
	=
	\big(\alpha_0+\alpha^T Z,\ \beta_0+\beta^T Z,\ \gamma_0+\gamma^T Z\big)^T,
\end{equation*}
where $\alpha_0=\log(-\log S_{T_1}^0(t_1^0)), \
\beta_0=\log(-\log S_{T_2}^0(t_2^0))$,
and $\gamma_0 = \log\!\bigl(-\log D^0(t_1^0,t_2^0)\bigr)$ for $D^0(t_1^0,t_2^0)<1$, and $\gamma_0 = \log\!\bigl(\log D^0(t_1^0,t_2^0)\bigr)$ for $D^0(t_1^0,t_2^0)>1$.

Let $\hat{\theta}=(\hat{S}_{T_1}(t_1^0), \hat{S}_{T_2}(t_2^0), \hat{S}(t_1^0,t_2^0))^T$
denote the plug-in estimator of $\theta$ based on the Dabrowska estimator of the joint survival function, whose marginals coincide with the Kaplan--Meier estimators \citep{kaplan_nonparametric_1958}.
Define the vector pseudo-observations %\citep{overgaard_asymptotic_2017,travis2024pseudo}
$\hat{\theta}_i
=
n\hat{\theta}-(n-1)\hat{\theta}^{(-i)}= (\hat{\theta}_{1i},\hat{\theta}_{2i},\hat{\theta}_{3i})^T,
$
corresponding to the three components of $\theta$.

We estimate the parameters via a two-step GEE procedure, as described in Algorithm~\ref{alg:2step} below. Let
$\xi_1=(\alpha_0,\alpha^T,\beta_0,\beta^T)^T$ and
$\xi_2=(\gamma_0,\gamma^T)^T$. For each subject $i$, let $V_i$ and $W_i$
denote working covariance matrices (in the usual GEE sense) for the pseudo-observation vectors entering the first- and second-step estimating equations, respectively.
Define $X_i=(\hat{\theta}_{1i},\hat{\theta}_{2i})^T$.
Then $$U_1(\xi_1)
=
\sum_{i=1}^n
\left(\frac{\partial m_i(\xi_1)}{\partial \xi_1}\right)^T
V_i^{-1}
\bigl(X_i - m_i(\xi_1)\bigr),$$
where
$m_i(\xi_1)=\tilde g^{-1}(\alpha_0+\alpha^T Z_i,\beta_0+\beta^T Z_i)^T$
and $\tilde g^{-1}(u,v)=(e^{-e^u},e^{-e^v})^T$.
Also,
$$U_2(\xi_1,\xi_2)
=
\sum_{i=1}^n
\left(\frac{\partial \mu_i(\xi_2)}{\partial \xi_2}\right)^T
W_i^{-1}
\bigl(Y_i - \mu_i(\xi_2)\bigr),$$
where
$\mu_i(\xi_2)=g_3^{-1}(\gamma_0+\gamma^T Z_i)$,
and
where $Y_i=\hat{\theta}_{3i}/\{\hat S_{T_1}(t_1^0\mid Z_i)\hat S_{T_2}(t_2^0\mid Z_i)\}$ denotes a modified response, with $\hat S_{T_1}(t_1^0\mid Z_i)$ and $\hat S_{T_2}(t_2^0\mid Z_i)$ obtained by plugging $\hat{\xi_1}$
into the marginal models.

\begin{algorithm}[H]
	\caption{Two-step estimation for the generalized Lehmann model at $(t_1^0,t_2^0)$}
	\label{alg:2step}
	\begin{enumerate}
		\item Solve $U_1(\xi_1)=0$ to obtain the estimated marginal parameters $\hat{\xi}_1$.
		
		\item Compute the modified response $Y_i$ and select the scalar link $g_3$ for the dependence ratio $d$ based on $\bar Y=n^{-1}\sum_{i=1}^n Y_i$:
	$\quad g_3(d)=
		\begin{cases}
			\log(\log d), & \bar Y>1,\\
			\log(-\log d), & \bar Y<1.
		\end{cases}$

		\item Solve $U_2(\hat{\xi}_1,\xi_2)=0$ to obtain the estimated dependence parameters $\hat{\xi}_2$.
		
	%	\item Return $\hat{\xi}=(\hat{\xi}_1^T,\hat{\xi}_2^T)^T$.
	\end{enumerate}
\end{algorithm}

%\color{blue}
%%The link-selection rule in Step~2 of Algorithm~\ref{alg:2step} is based on the mean value of the modified response. Lemma~\ref{lem:link_selection} shows that this rule consistently identifies whether the dependence ratio is above or below one on average under the model.
%The link-selection rule in Step~2 of Algorithm~\ref{alg:2step} is based on the mean value of the modified response. Lemma~\ref{lem:link_selection} shows that this rule consistently identifies whether the dependence ratio is above or below one on average under the model. The procedure therefore selects the appropriate branch of the transformation, and implicitly assumes that, at a given time point, the dependence ratio remains predominantly either above or below one over the relevant covariate range.
%\color{black}

The first-step estimation of $\xi_1=(\alpha_0,\alpha^T,\beta_0,\beta^T)^T$ can be viewed as two separate marginal pseudo-observations problems, based on the Kaplan--Meier estimators for $T_1$ and $T_2$, respectively. The corresponding estimating equations are decoupled, and the marginal asymptotic properties follow from standard pseudo-observations arguments. We present them jointly mainly for notational convenience and because their joint covariance structure is needed in the second step.  

\begin{remark}
	The second-step estimating equation $U_2(\xi_1,\xi_2)$ depends on $\xi_1$ through $Y_i$, which involves the fitted marginal survival functions $\hat S_{T_1}(t_1^0\mid Z_i)$ and $\hat S_{T_2}(t_2^0\mid Z_i)$. Consequently, the estimation of $\xi_2$ must be carried out using the first-step estimator $\hat{\xi}_1$, and the variability of $\hat{\xi}_1$ must be accounted for in the asymptotic analysis. 
\end{remark}

\begin{remark}\label{rem:g3}
For multiple time points $(t_1^j,t_2^j)$, $j=1,\ldots, m$, the same rule applies to each
	$\bar Y_j=\frac{1}{n}\sum_{i=1}^n
	\frac{\hat{\theta}_{3ij}}
	{\hat{S}_{T_1}(t_1^j\mid Z_i)\hat{S}_{T_2}(t_2^j\mid Z_i)}$.
If all $\bar Y_j-1$ have the same sign, a common link may be used; otherwise, time-specific links may be employed. In practice, this choice may be guided by subject-matter considerations regarding the dependence structure. 

\end{remark}

In our motivating example of retinoblastoma, all empirical averages satisfied $\bar Y_j<1$, indicating negative dependence between the times to enucleation. Accordingly, we used a common link function across all time points. This choice is also consistent with subject-matter considerations, as times to enucleation in the two eyes are expected to exhibit negative dependence (see Section~\ref{sec:data}).

\begin{theorem}\label{thm:2-step}
	Let $\xi_1^*$ and $\xi_2^*$ solve $E\{U_1(\xi_1)\}=0$ and $E\{U_2(\xi_1^*,\xi_2)\}=0$, respectively. Assume independent censoring, the regularity conditions in Appendix~\ref{app:F}, and that $\inf_z S_{T_j}(t_j^0\mid z)>\varepsilon$ for $j=1,2$ and some $\varepsilon>0$.
	
	Then $\hat{\xi}_1 \xrightarrow{p} \xi_1^*$, $\hat{\xi}_2 \xrightarrow{p} \xi_2^*$, and
$\bigl(n^{-1/2}U_1(\xi_1^*),\, n^{-1/2}U_2(\xi_1^*,\xi_2^*)\bigr)
	\xrightarrow{d}
	N(0,\Sigma)$,
	where
$\Sigma =
\left(\begin{smallmatrix}
	\displaystyle\Sigma_1 & \displaystyle\Sigma_{12}\\
	\displaystyle\Sigma_{21} & \displaystyle\Sigma_2
\end{smallmatrix}\right)$. %= \lim_{n\to\infty} n^{-1}\mathrm{Var}(U_1,U_2)$.
 Moreover,
	$\sqrt{n}(\hat{\xi}_2 - \xi_2^*) \xrightarrow{d} N(0,\Omega)$,
	where
	\begin{equation}\label{eq:omega}
	\Omega
	=
	B_2^{-1}\!\left(
	\Sigma_2
	+ B_1 A_1^{-1} \Sigma_1 A_1^{-T} B_1^T
	- \Sigma_{21} A_1^{-T} B_1^T
	- B_1 A_1^{-1} \Sigma_{12}
	\right)\!B_2^{-T},
	\end{equation}
	and $A_1$, $B_1$, and $B_2$ are defined in Appendix~\ref{app:F}  of the supplementary material.
\end{theorem}

The proof is given in Appendix~\ref{app:F} of the supplementary material.
%The following lemma provides asymptotic justification for the empirical link-selection rule used in Algorithm~1.

%\color{blue}
%\color{blue}
%\begin{lemma}\label{lem:link_selection}
%	Assume the conditions of Theorem~\ref{thm:2-step}. Let
%	$\bar Y=\frac1n\sum_{i=1}^n
%	\frac{\hat\theta_{3i}}
%	{\hat S_{T_1}(t_1^0\mid Z_i)\hat S_{T_2}(t_2^0\mid Z_i)}$.
%	
%	Then
%	$\bar Y \xrightarrow{p} E\{\mu_i(\xi_2^*)\}$, where $\mu_i(\xi_2^*)
%	=
%	\frac{S(t_1^0,t_2^0\mid Z_i)}
%	{S_{T_1}(t_1^0\mid Z_i)S_{T_2}(t_2^0\mid Z_i)}$.
%	Consequently, if \(E\{\mu_i(\xi_2^*)\}\neq 1\), the sign of \(\bar Y-1\) consistently identifies whether the dependence ratio is above or below one on average.
%\end{lemma}
%\color{black}
%The proofs of Theorem~\ref{thm:2-step} and Lemma~\ref{lem:link_selection} are given in Appendix~\ref{app:F} of the supplementary material. %Theorem~\ref{thm:2-step} proves consistency and asymptotic normality of the regression estimates in the two-step estimation procedure, and Lemma~\ref{lem:link_selection} provides asymptotic justification for the empirical link-selection rule used in Algorithm~\ref{alg:2step}.
%\color{black}

\subsection{Variance estimates}\label{sec:var}

In Theorem~\ref{theorem}, the covariance matrix of $\hat{\tilde b}$ can be estimated using a sandwich estimator, as is standard in the pseudo-observation (GEE) framework \citep{andersen_generalised_2003}. Let  
$I(\tilde b)=\sum_{i=1}^{n}\left(\frac{\partial}{\partial \tilde b}g^{-1}(\tilde b^T Z_i^*)\right)^T V_i^{-1}\left(\frac{\partial}{\partial \tilde b}g^{-1}(\tilde b^T Z_i^*)\right)$,
and $\hat{\Sigma}_{\tilde b}=\sum_{i=1}^n U_i(\hat{\tilde b}) U_i(\hat{\tilde b})^T$.
The ordinary sandwich estimator of $\mathrm{var}(\hat{\tilde b})$ is $\hat{\mathrm{var}}(\hat{\tilde b}) = I(\hat{\tilde b})^{-1} \hat{\Sigma}_{\tilde b} I(\hat{\tilde b})^{-1}$.
Although this ordinary estimator can be slightly biased in some cases, improved estimators require second-order influence functions and are computationally cumbersome. In practice, the difference is typically negligible \citep{jacobsen_note_2016,overgaard_asymptotic_2017,furberg_bivariate_2023}, and in bivariate survival contexts, it generally provides correct or conservative coverage \citep{travis2024pseudo}.

Analogously, %to \eqref{eq:sigma_hat},
the covariance matrix $\Sigma$ in Theorem~\ref{thm:2-step} can be estimated by
\[
\begin{pmatrix}
	\hat{\Sigma}_1 & \hat{\Sigma}_{12} \\
	\hat{\Sigma}_{21} & \hat{\Sigma}_2
\end{pmatrix}
=
\frac{1}{n}
\begin{pmatrix}
	\sum_{i=1}^n U_{1i}(\hat{\xi}_1) U_{1i}(\hat{\xi}_1)^T &
	\sum_{i=1}^n U_{1i}(\hat{\xi}_1) U_{2i}(\hat{\xi}_1,\hat{\xi}_2)^T \\
	\sum_{i=1}^n U_{2i}(\hat{\xi}_1,\hat{\xi}_2) U_{1i}(\hat{\xi}_1)^T &
	\sum_{i=1}^n U_{2i}(\hat{\xi}_1,\hat{\xi}_2) U_{2i}(\hat{\xi}_1,\hat{\xi}_2)^T
\end{pmatrix}.
\]
%\[
%\bigl(\begin{smallmatrix}
%	\hat{\Sigma}_1 & \hat{\Sigma}_{12} \\
%	\hat{\Sigma}_{21} & \hat{\Sigma}_2
%\end{smallmatrix}\bigr)
%=
%\frac{1}{n}
%\bigl(\begin{smallmatrix}
%	\sum_{i=1}^n U_{1i}(\hat{\xi}_1) U_{1i}(\hat{\xi}_1)^T &
%	\sum_{i=1}^n U_{1i}(\hat{\xi}_1) U_{2i}(\hat{\xi}_1,\hat{\xi}_2)^T \\
%	\sum_{i=1}^n U_{2i}(\hat{\xi}_1,\hat{\xi}_2) U_{1i}(\hat{\xi}_1)^T &
%	\sum_{i=1}^n U_{2i}(\hat{\xi}_1,\hat{\xi}_2) U_{2i}(\hat{\xi}_1,\hat{\xi}_2)^T
%\end{smallmatrix}\bigr).
%\]

%\[
%{\small
%	\begin{pmatrix}
%		\hat{\Sigma}_1 & \hat{\Sigma}_{12} \\
%		\hat{\Sigma}_{21} & \hat{\Sigma}_2
%	\end{pmatrix}
%	=
%	\frac{1}{n}
%	\begin{pmatrix}
%		\sum_{i=1}^n U_{1i}(\hat{\xi}_1) U_{1i}(\hat{\xi}_1)^T &
%		\sum_{i=1}^n U_{1i}(\hat{\xi}_1) U_{2i}(\hat{\xi}_1,\hat{\xi}_2)^T \\
%		\sum_{i=1}^n U_{2i}(\hat{\xi}_1,\hat{\xi}_2) U_{1i}(\hat{\xi}_1)^T &
%		\sum_{i=1}^n U_{2i}(\hat{\xi}_1,\hat{\xi}_2) U_{2i}(\hat{\xi}_1,\hat{\xi}_2)^T
%	\end{pmatrix}
%}
%\]

Similarly, the matrices $A_1$, $B_1$, and $B_2$ in Equation~\eqref{eq:omega} can be estimated directly; details are provided at the end of supplementary material Appendix~\ref{app:F}.

\subsection{Additional considerations}\label{sec:exchange}
In some applications, the two failure times are exchangeable, as in the retinoblastoma example in Section~\ref{sec:data}. Exchangeability means that $S(t_1,t_2\mid Z)=S(t_2,t_1\mid Z)$. In the simple Lehmann model, this property holds if the baseline distribution is symmetric, that is, if $S^0(t_1,t_2)=S^0(t_2,t_1)$. In the generalized Lehmann model, exchangeability further requires $\alpha=\beta$. The estimation procedure can incorporate these constraints; details are given in Appendix~\ref{sec:sup_exchange} of the supplementary material. Section~\ref{sec:simulations} includes an exchangeable simulation setting analyzed both with and without the constraint, and Section~\ref{sec:data} presents an analysis of the retinoblastoma data under exchangeability.

Finally, we assessed model fit using pseudo-residual–based graphical diagnostics, extending existing univariate approaches to the bivariate setting; see Appendix~\ref{app:GOF}. %of the supplementary material.

\section{Simulation study}\label{sec:simulations}

We evaluate our approach using simulated data, with the goals of estimating the regression parameters and the covariate-adjusted bivariate survival probability. Our approach includes both the simple and generalized Lehmann models, which we compare to the flexible copula-based models of \cite{marra_copula_2020}.
We consider four data-generating mechanisms (DGMs) under the generalized Lehmann model, with baseline bivariate survival specified by Frank (NQD), Frank (PQD), Clayton (PQD), and Gumbel–Barnett (NQD; bivariate exponential) copulas. These DGMs mainly differ in baseline dependence structure, covering negative and positive dependence. Appendix~\ref{app:G} of the supplementary material gives the full details, including parameter choices, censoring, and time-point selection. Across all DGMs, subject-specific covariates $Z_i\sim\mathrm{Uniform}(0,1)$ were generated independently, and the slope parameters in the generalized Lehmann model were set to $\alpha_1=1$, $\beta_1=0.7$, and $\gamma_1=0.3$.

For \(n=800\), Table~\ref{tab:sim_main_n800} reports results for the second-step dependence parameter \(\gamma\) in the main Frank and Clayton benchmark scenarios, while Table~\ref{tab:sim_exp2D_n800} presents the Gumbel--Barnett setting and related extensions, including simple Lehmann analyses with and without exchangeability constraints; see also Table~\ref{tab:sim_exp2D_simple} in Appendix~\ref{app:G}. All benchmarks use six pre-specified bivariate time points. Accordingly, Table~\ref{tab:sim_main_n800} contains seven rows per DGM: six intercept-related parameters \(\tilde\gamma^0_1,\ldots,\tilde\gamma^0_6\), corresponding to link-transformed baseline quantities at these time points (see Appendix~\ref{sec:app_intercept_params} of the supplementary material), and the slope parameter \(\gamma_1\). Results for the marginal parameters \(\alpha\) and \(\beta\) are deferred to Table~\ref{tab:sim_marginal}, as they rely on the standard univariate pseudo-observations approach. Results for \(n=400\) are given in Tables~\ref{tab:sim_main_n400}--\ref{tab:sim_exp2D_n400} of the supplementary material and show, as expected, higher bias and lower coverage.

%Results for the second-step (dependence) parameter $\gamma$ for the main benchmark scenarios (Frank and Clayton copulas) with sample size n=800 are presented in Table~\ref{tab:sim_main_n800}, while Table~\ref{tab:sim_exp2D_n800} of the supplementary material reports the corresponding results for the Gumbel--Barnett setting and its extensions. The latter also includes supplementary analyses based on the simple Lehmann model, with and without exchangeability constraints (see also Table~\ref{tab:sim_exp2D_simple} in Appendix~\ref{app:G} of the supplementary material). All benchmark scenarios use six pre-specified bivariate time points (see Appendix~\ref{app:G} for details). Table~\ref{tab:sim_main_n800} includes seven rows per DGM: six corresponding to the intercept-related parameters $\tilde{\gamma}_0^1, \ldots, \tilde{\gamma}_0^6$, representing link-transformed baseline quantities evaluated at these time points, and one corresponding to the slope parameter $\gamma_1$. Results for the marginal parameters $\alpha$ and $\beta$ are deferred to Appendix~\ref{app:G} (Table~\ref{tab:sim_marginal}), as their estimation follows the standard pseudo-observations approach, which is well known to perform reliably in univariate settings. Results for the smaller sample size $n=400$ are reported in Tables~\ref{tab:sim_main_n400} and \ref{tab:sim_exp2D_n400} of the supplementary material and, as expected, exhibit higher bias and reduced coverage. 

For $n=800$, estimation performance is generally good across all DGMs. The intercept-related parameters $\tilde \gamma_0^j$, $j=1,\ldots,6$, in Table~\ref{tab:sim_main_n800} show low bias, moderate variability, and coverage close to the nominal level. This behavior is particularly evident in the NQD setting (Frank NQD), where both intercepts and the slope parameter $\gamma_1$ are estimated accurately, with small bias and relatively low variability.

In the PQD settings (Frank PQD and Clayton PQD), a mild systematic pattern is observed for the first intercept $\tilde \gamma_0^1$ and the slope $\gamma_1$: $\tilde \gamma_0^1$ tends to be underestimated and $\gamma_1$ slightly overestimated. This occurs when the baseline bivariate quantity is close to one, causing instability in the double logarithmic transformation and a compensatory increase in the variability of $\gamma_1$.
%In the PQD settings (Frank PQD and Clayton PQD), a mild systematic pattern is observed for the first intercept $\tilde \gamma_0^1$ and the slope parameter $\gamma_1$, where $\tilde \gamma_0^1$ tends to be underestimated and $\gamma_1$ slightly overestimated. This behavior arises when the baseline bivariate quantity is close to one, leading to instability under the double logarithmic transformation. The resulting distortion induces a compensatory shift in $\gamma_1$, increasing its variability.

Prediction accuracy was assessed using the mean absolute error (MAE) between estimated and true joint survival probabilities (see supplementary material Appendix~\ref{app:G} for details).
Figure~\ref{fig:MAEs} presents boxplots of the MAEs for DGM~1 (Frank NQD, top panel) and DGM~2 (Frank PQD, bottom panel), with sample size 
$n=800$. Results are shown for the simple and generalized Lehmann models, as well as for the copula-based approach of \cite{marra_copula_2020}.

The copula-based comparison uses the flexible framework of \cite{marra_copula_2020}, which allows
(i) flexible marginal survival models,
(ii) covariate-dependent association parameters, and
(iii) spline-based baseline estimation. Thus, the comparison is not against a simplistic parametric benchmark, but against a modern flexible copula regression framework. The key distinction is not implementation flexibility, but the level at which dependence is modeled: the copula approach parameterizes dependence through a chosen copula family, whereas the proposed framework models covariate effects directly on the bivariate survival function.

To provide the copula-based approach with substantial flexibility, we considered four copulas—Frank, Clayton, Gaussian, and Gumbel—and selected the optimal model using BIC. The selected copulas aligned with the data-generating mechanisms: Frank in DGMs 1 and 2, Clayton in DGM 3, and Frank again in DGM~4.

%Overall, all three approaches yield comparably low MAEs. None of the methods uniformly dominates the others across all time points and settings. In some cases, the generalized Lehmann model shows a slight advantage, while in others the copula-based approach performs marginally better.
Overall, predictive performance was comparable across methods, with no single approach dominating across all settings and time points. The proposed models nevertheless offer important advantages in interpretability and robustness to dependence misspecification, while maintaining competitive predictive accuracy. Notably, although the simple Lehmann model is misspecified in these settings (as $\alpha \neq \beta \neq \gamma$), it nevertheless provides accurate predictions of the bivariate survival probability.

All simulations were conducted in R (version 4.5.2) using parallel computing on a high-performance server. Our pseudo-observations implementation relies primarily on the \texttt{mhazard}, \texttt{geepack}, and \texttt{geeM} packages: \texttt{mhazard} estimates the baseline (Dabrowska) estimator, while \texttt{geepack} and \texttt{geeM} fit the GEE models for the NQD and PQD settings, respectively. Copula-based methods were implemented using \texttt{GJRM}. Code to reproduce the results is available at \url{https://github.com/Yael-Travis-Lumer/Cox-Regression-on-the-Plane}
%All simulations were conducted in R (version 4.5.2) using parallel computing on a high-performance server. Our implementation, based on pseudo-observations, relies primarily on the \texttt{mhazard}, \texttt{geepack}, and \texttt{geeM} packages, where \texttt{mhazard} is used to estimate the baseline (Dabrowska) estimator and \texttt{geepack} and \texttt{geeM} are used to fit the GEE models for the NQD and PQD settings, respectively. The copula-based methods were implemented using the \texttt{GJRM} package. Code to reproduce the results is available on GitHub at \url{https://github.com/Yael-Travis-Lumer/Cox-Regression-on-the-Plane}.

\begin{table}[!htbp]
	\centering
	\caption{Simulation results for the main benchmark scenarios with sample size $n=800$ and $500$ simulation replications. For each parameter, the table reports the true value, Monte Carlo mean, median, empirical standard deviation (SD), average estimated standard error for the two-step approach (SE), and empirical coverage of the two-step 95\% confidence interval.}
	\label{tab:sim_main_n800}
	\begin{tabular}{llrrrrrr}
		\hline
		DGM & Parameter & True & Mean & Median & SD & SE & Cov. \\
		\hline
		\multirow{7}{*}{Frank NQD}
		& $\tilde \gamma_0^1$ & -0.84 & -0.87 & -0.86 & 0.29 & 0.30 & 0.958 \\
		& $\tilde \gamma_0^2$ &  0.36 &  0.37 &  0.37 & 0.15 & 0.16 & 0.970 \\
		& $\tilde \gamma_0^3$ &  0.17 &  0.17 &  0.17 & 0.10 & 0.11 & 0.962 \\
		& $\tilde \gamma_0^4$ &  0.52 &  0.53 &  0.52 & 0.18 & 0.18 & 0.980 \\
		& $\tilde \gamma_0^5$ &  0.31 &  0.31 &  0.31 & 0.14 & 0.14 & 0.968 \\
		& $\tilde \gamma_0^6$ &  0.65 &  0.66 &  0.67 & 0.20 & 0.20 & 0.968 \\
		& $\gamma_1$   &  0.30 &  0.30 &  0.32 & 0.58 & 0.63 & 0.964 \\
		\hline
		\multirow{7}{*}{Frank PQD}
		& $\tilde \gamma_0^1$ & -1.17 & -1.27 & -1.18 & 0.52 & 0.41 & 0.935 \\
		& $\tilde \gamma_0^2$ &  0.20 &  0.20 &  0.20 & 0.10 & 0.11 & 0.956 \\
		& $\tilde \gamma_0^3$ &  0.08 &  0.08 &  0.09 & 0.07 & 0.07 & 0.947 \\
		& $\tilde \gamma_0^4$ &  0.29 &  0.30 &  0.29 & 0.12 & 0.13 & 0.962 \\
		& $\tilde \gamma_0^5$ &  0.14 &  0.15 &  0.15 & 0.10 & 0.10 & 0.960 \\
		& $\tilde \gamma_0^6$ &  0.36 &  0.37 &  0.37 & 0.14 & 0.14 & 0.960 \\
		& $\gamma_1$   &  0.30 &  0.39 &  0.29 & 0.75 & 0.63 & 0.956 \\
		\hline
		\multirow{7}{*}{Clayton PQD}
		& $\tilde \gamma_0^1$ & -1.78 & -1.89 & -1.80 & 0.49 & 0.41 & 0.934 \\
		& $\tilde \gamma_0^2$ &  0.21 &  0.21 &  0.21 & 0.14 & 0.15 & 0.948 \\
		& $\tilde \gamma_0^3$ &  0.24 &  0.24 &  0.24 & 0.10 & 0.10 & 0.946 \\
		& $\tilde \gamma_0^4$ &  0.47 &  0.47 &  0.46 & 0.15 & 0.16 & 0.954 \\
		& $\tilde \gamma_0^5$ &  0.42 &  0.41 &  0.41 & 0.12 & 0.13 & 0.964 \\
		& $\tilde \gamma_0^6$ &  0.66 &  0.65 &  0.65 & 0.16 & 0.17 & 0.968 \\
		& $\gamma_1$   &  0.30 &  0.41 &  0.30 & 0.73 & 0.65 & 0.956 \\
		\hline
	\end{tabular}
\end{table}

To assess misspecification, we consider data from a log-normal distribution, under which both the Lehmann and copula-based models are misspecified. Details are given in supplementary material Appendix~\ref{app:G_miss}, and Figure~\ref{fig:MAEs_LN} reports the corresponding MAEs. All models accurately estimate the bivariate survival function despite misspecification.

%To assess misspecification, we consider data generated from a covariate-adjusted log-normal distribution, under which both the Lehmann and copula-based models are misspecified. Details are given in supplementary material Appendix~\ref{app:G_miss}, where Figure~\ref{fig:MAEs_LN} reports the corresponding MAEs. Overall, all models accurately predict the bivariate survival function despite misspecification.
%To assess the impact of misspecification, we consider data generated from a covariate-adjusted log-normal distribution, under which both the Lehmann and copula-based models are misspecified. Details are given in supplementary material Appendix~\ref{app:G_miss}, where Figure~\ref{fig:MAEs_LN} reports the MAEs for this setting. Overall, all models provide accurate predictions of the bivariate survival function despite misspecification.

\section{Data analysis}\label{sec:data}

We applied our methods to bilateral cases from the GROS dataset. Previous analyses \citep{global_retinoblastoma_study_group_global_2020,fabian_global_2022,nishath_retinoblastoma_2025} focused mainly on univariate outcomes, such as overall mortality or time to first enucleation. Here, we study times to enucleation in both eyes and their association with patient characteristics. Patients enter the study at first presentation to the treatment center, when baseline covariates are recorded. Among the 1,255 bilateral cases, involvement of both eyes was almost always identified at presentation, so the two event times share a common origin. Patients with missing or implausible dates were excluded. The inclusion process is summarized in Figure~\ref{fig:flowchart.png} of the supplementary material, yielding 977 patients in the final analytic sample.

%We applied our methods to patients with bilateral disease in the GROS dataset. Previous analyses \citep{global_retinoblastoma_study_group_global_2020,fabian_global_2022,nishath_retinoblastoma_2025} have focused mainly on univariate outcomes, such as overall mortality or time to first enucleation. Here, we study the times to enucleation in both eyes and their association with patient characteristics. Patients enter the study at first presentation to the treatment center, when baseline covariates are recorded. Among the 1,255 bilateral cases, involvement of both eyes was almost always identified at presentation, so the two event times are measured from a common origin. Patients with missing or implausible dates were excluded.  The inclusion process is summarized in Figure~\ref{fig:flowchart.png} of the supplementary material, yielding a final analytic sample of 977 patients.

Let $T_1$ and $T_2$ denote the times to enucleation of the right and left eyes, measured from presentation. We include six baseline covariates measured at presentation: age, income level, tumor stage, an indicator for age above four years, sex, and family history of retinoblastoma. The age indicator was included based on clinical recommendations and prior analyses of this dataset \citep{global_retinoblastoma_study_group_global_2020,fabian_global_2022}. The covariate vector is $Z=(Z_1,\ldots,Z_6)^T$, with definitions given in Appendix~\ref{app:H} of the supplementary material. 

Because the two clinically similar eyes share the same disease process, we primarily assume exchangeability: 
%Because the two eyes are clinically similar and arise from the same underlying disease process, we adopt exchangeability as the primary modeling approach, that is,
$S(t_1,t_2\mid Z)=S(t_2,t_1\mid Z)$.
%We fit both the simple and generalized Lehmann models under this constraint (Section~\ref{sec:exchange}).
Details of the constrained estimation procedure are given in Appendix~\ref{app:H} of the supplementary material, which also presents an unconstrained analysis; results are broadly similar.

Our goal is to estimate the covariate-adjusted joint survival probability at clinically relevant time points. We estimate the probability that both eyes remain free from enucleation beyond 6, 12, 18, and 24 months, as well as the conditional probabilities $\Pr(T_i>24 \mid T_j\leq 12, Z)$ and $\Pr(T_i>24 \mid T_j>12, Z)$, for $i,j=1,2$.

\begin{figure}[!p]
	\centering
	
	\subfloat[Bivariate survival probabilities\label{fig:fig-b_exchange}]{%
		\includegraphics[width=0.99\textwidth]{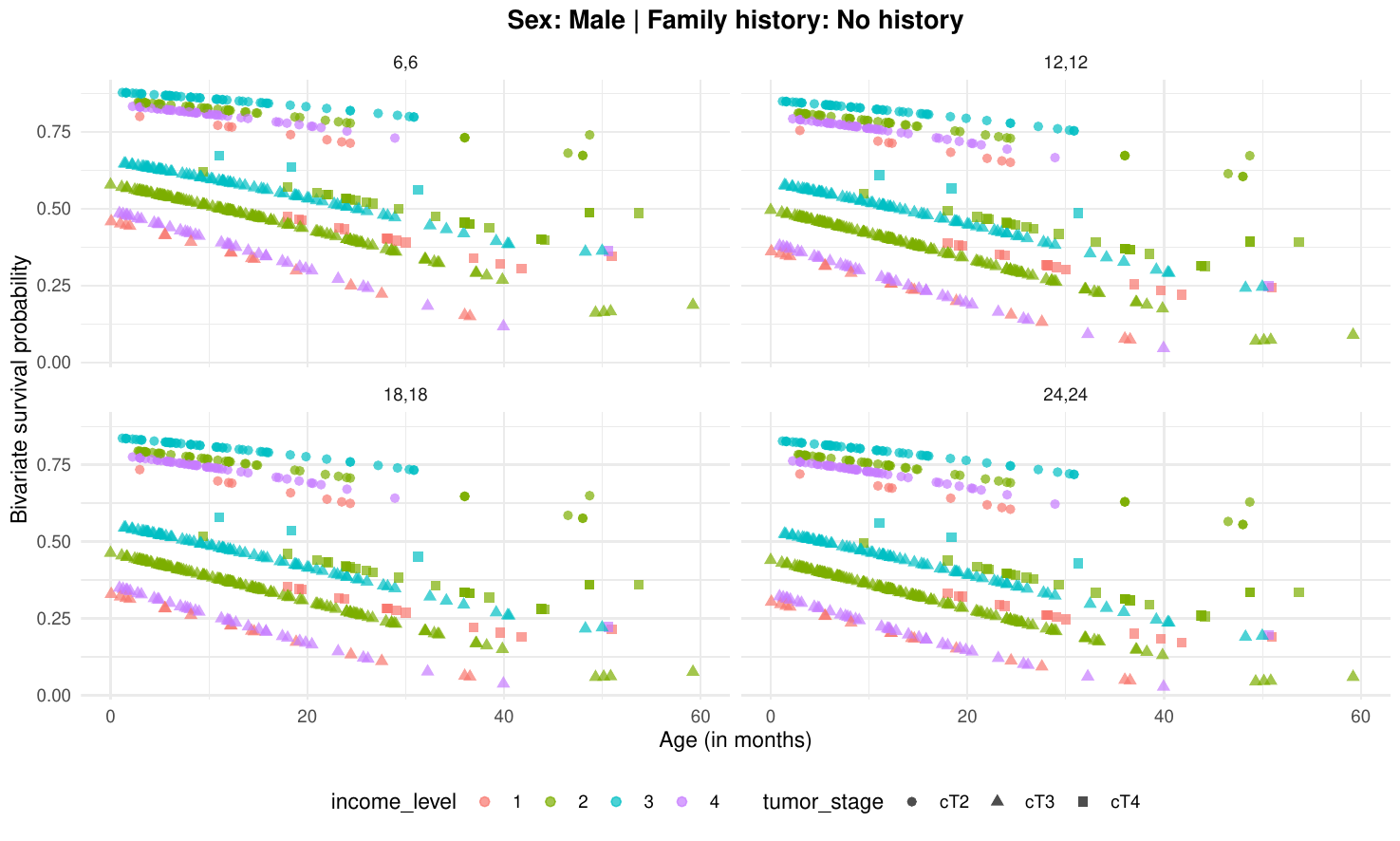}%
	}
		\vspace{0.5em}
		\subfloat[Conditional survival probabilities\label{fig:cond_probs_males_no_hist_exchange}]{%
		\includegraphics[width=0.99\textwidth]{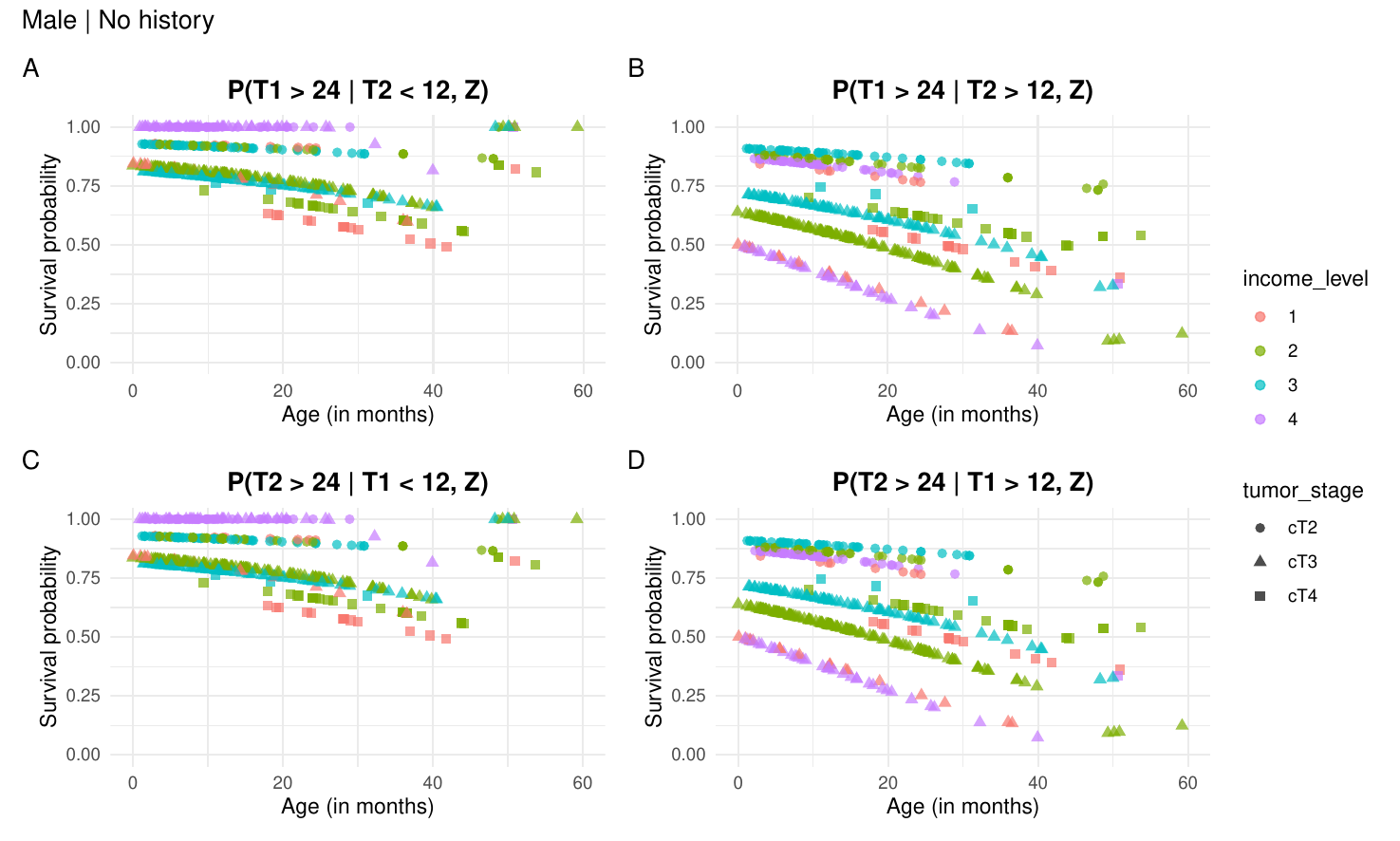}%
	}
	
\caption{Generalized Lehmann model. Top: predicted bivariate survival probabilities at 6, 12, 18, and 24 months. Bottom: predicted conditional survival probabilities. All probabilities are shown as functions of age at presentation, income level, and tumor stage for males without a family history of retinoblastoma, under the exchangeability assumption.}
\label{fig:GROS_generalized_exchange}
	
\end{figure}

Tables \ref{tab:sim_Lehmann_exchangeable}--\ref{tab:lehmann_gamma_exchangeable} in Appendix~\ref{app:H} of the supplementary material report the estimated regression parameters and their interpretation. Figure~\ref{fig:fig-b_exchange} shows predicted bivariate survival under the generalized Lehmann model at $(6,6),(12,12),(18,18),(24,24)$, by age, income, and tumor stage, for males without family history; similar patterns held in other subgroups. The simple Lehmann results in Figure~\ref{fig:fig-a_exchange} are nearly identical, and diagnostics in Figure~\ref{fig:GOF_exchange} support adequate fit. Survival decreases from 6 to 24 months and, among children diagnosed before age four, declines with age at presentation. Survival is highest for cT2 and lower for cT3–cT4, with cT4 slightly exceeding cT3, possibly due to deaths before enucleation or palliative chemotherapy only. Survival also varies by income, with level 3 exceeding level 2 and level 4; this may reflect lower mortality among higher-income patients, possibly due to broader treatment access \citep{fabian_global_2022}. Death before enucleation precludes observation of enucleation; addressing this competing risk is beyond the present scope.

%Tables~\ref{tab:sim_Lehmann_exchangeable}--\ref{tab:lehmann_gamma_exchangeable} in Appendix~\ref{app:H} of the supplementary material report the estimated regression parameters; their interpretation is provided there. Figure~\ref{fig:fig-b_exchange} displays the predicted bivariate survival probabilities under the generalized Lehmann model at the time points $(6,6)$, $(12,12)$, $(18,18)$, and $(24,24)$, as functions of age (x-axis), income level (color), and tumor stage (shape). Results are shown for males without a family history of retinoblastoma; similar patterns were observed across other subgroups. The corresponding results for the simple Lehmann model are presented in Figure~\ref{fig:fig-a_exchange} of the supplementary material. The two Lehmann models yield nearly identical predictions. Graphical diagnostics (Figure~\ref{fig:GOF_exchange}) support an adequate fit.

%As expected, survival decreases across the time points (6, 12, 18, and 24 months). Among patients diagnosed before age four, predicted survival declines with increasing age at presentation at each time point. Survival is highest for tumor stage $cT2$, with lower probabilities for $cT3$ and $cT4$; notably, $cT4$ shows slightly higher survival than $cT3$, possibly reflecting deaths prior to enucleation or treatment with palliative chemotherapy alone. Survival also varies by income level: upper-middle income (level 3) exceeds lower-middle (level 2), which in turn exceeds high income (level 4). Although this pattern may appear counterintuitive at first glance, it may reflect lower mortality among higher-income patients \citep{fabian_global_2022}, potentially due to broader access to treatment. Importantly, death prior to enucleation represents a competing risk, which is beyond the scope of this study.

Next, we estimated the survival probability of one eye conditional on failure or survival of the other eye. Specifically, we computed $\Pr(T_1>24 \mid T_2\leq 12, Z)$, $\Pr(T_1>24 \mid T_2>12, Z)$. These quantities represent the probability that one eye remains free from enucleation beyond two years, conditional on whether the other eye failed or survived during the first year.
Under exchangeability, these capture all distinct conditional survival probabilities, with the remaining obtained by symmetry. In particular, 
\begin{align*}
	\Pr(T_1>24 \mid T_2\leq 12, Z)
	&=\frac{S(24,0 \mid Z)-S(24,12 \mid Z)}{1-S(0,12 \mid Z)}
	=\frac{S_{T_1}(24 \mid Z)-S(24,12 \mid Z)}{1-S_{T_1}(12 \mid Z)}, \\
	\Pr(T_1>24 \mid T_2>12, Z)
	&=\frac{S(24,12 \mid Z)}{S(0,12 \mid Z)}
	=\frac{S(24,12 \mid Z)}{S_{T_1}(12 \mid Z)}.
\end{align*}
%with the remaining probabilities obtained by symmetry.

Figure~\ref{fig:cond_probs_males_no_hist_exchange} shows the estimated conditional survival probabilities under the generalized Lehmann model for males without a family history of retinoblastoma. Under exchangeability, panels A and C (and B and D) coincide, confirming the imposed symmetry. Conditional survival decreases with age at presentation, and separation across tumor stages and income levels persists. Corresponding results for the simple Lehmann model (Figure~\ref{fig:fig-cond_exchange} in the supplementary material) are similar.

Conditioning on enucleation of one eye (panels A and C in Figure~\ref{fig:cond_probs_males_no_hist_exchange}), the generalized Lehmann model suggests that high-income patients diagnosed before age two have near-zero probability of enucleation in the second eye. In other income groups, patients with low tumor stage ($cT2$) show higher survival probabilities than those with advanced stages, consistent with the lower likelihood of bilateral surgery in milder cases. Patients with stage $cT3$ diagnosed around age four show elevated survival probabilities, possibly reflecting instability in the estimated level change.
%When conditioning on enucleation of one eye (panels A and C in Figure~\ref{fig:cond_probs_males_no_hist_exchange}), the generalized Lehmann model suggests that high-income patients diagnosed before age two have an essentially zero probability of enucleation in the second eye. For other income groups, patients with low tumor stage ($cT2$) exhibit higher survival probabilities than those with more advanced stages, consistent with the lower likelihood of bilateral surgery in milder cases. Notably, patients with stage $cT3$ diagnosed around age four show elevated survival probabilities, possibly reflecting instability in the estimated level change.

When one eye has been enucleated, clinicians make substantial efforts to preserve the remaining eye. This is reflected in the finding that for $i\neq j$, $\Pr(T_i>24\mid T_j\leq 12,Z)>\Pr(T_i>24\mid T_j>12,Z)$,
indicating that the probability of long-term survival in one eye is higher when the other eye fails early. This inequality is equivalent to $S_{T_i}(24\mid Z)S_{T_j}(12\mid Z)>S_{ij}(24,12\mid Z)$,
which implies NQD. % between the two times.

Finally, we compared these conditional probabilities to those from the copula-based approach of \cite{marra_copula_2020}. As in Section~\ref{sec:simulations}, we considered four copulas—Frank, Clayton, Gaussian, and Gumbel—and selected the model by BIC. The Clayton copula was selected, implying PQD between event times, and was strongly favored within the copula class (BIC 9906 vs.\ 10087 for Gumbel and 10100 for Gaussian and Frank). Despite this, the implied PQD structure conflicts with both clinical expectations and empirical evidence. Nonparametric estimates based on the Dabrowska and Kaplan--Meier estimators yield $\nicefrac{\hat S(t_1,t_2)}{\hat S_{T_1}(t_1)\hat S_{T_2}(t_2)}<1$ at clinically relevant time points, indicating NQD. The discrepancy is also evident in Figure~\ref{fig:cond_prob_cop}: panels A and C show systematically lower conditional survival probabilities than panels B and D. The fitted copula model also fails to preserve exchangeability, as panels A and C and panels B and D are not identical. By contrast, the unconstrained Lehmann models (Figure~\ref{fig:cond_probs_males_no_hist} in the supplementary material) yield approximately exchangeable conditional survival probabilities.

\section{Discussion}
A key advantage of the proposed framework is that it avoids specifying a parametric dependence structure. This is particularly relevant in our data application, where the copula-based approach selected a Clayton copula implying PQD, despite clinical reasoning and nonparametric estimates indicating NQD. This discrepancy is consequential for conditional survival probabilities, which depend directly on the assumed dependence structure. The copula-based fit also fails to preserve exchangeability, whereas the unconstrained generalized Lehmann model yields approximately exchangeable conditional survival probabilities.

At the estimation stage, an additional advantage of the pseudo-observations approach is that assumptions are imposed only at selected time points rather than across the entire time scale. A limitation is computational cost, because the survival estimator must be recomputed \(n+1\) times, although the method remains feasible for moderate sample sizes. Replacing pseudo-observations with infinitesimal jackknife pseudo-values is a promising direction for future work \citep{parner_regression_2023}.

Several limitations merit consideration. Efficiency may be lower than for correctly specified likelihood-based methods, although the robustness of the GEE approach partly mitigates this. Developing likelihood-based estimators for the proposed models is an important direction for future work, with potential to improve efficiency and accommodate time-dependent covariates.

Second, the framework assumes independent censoring, as is standard in the pseudo-observations literature; violations of this assumption may induce bias \citep{andersen_pseudo-observations_2010}. A natural extension is to allow conditional independence given covariates, for example via IPCW estimators of the bivariate survival function \citep{van_der_laan_locally_2002}. As with other pseudo-observations methods, reliable inference also requires sufficiently large samples.

%Second, the framework assumes independent censoring, as is standard in the pseudo-observations literature; see \citet{andersen_pseudo-observations_2010}, who demonstrate that violations of this assumption can induce bias. A natural extension is to relax this to conditional independence given covariates, for example via inverse probability of censoring weighting (IPCW) estimators of the bivariate survival function \citep{van_der_laan_locally_2002}. As with other pseudo-observations--based methods, valid inference requires sufficiently large samples, since the approach relies on stable estimation of the survival function and repeated leave-one-out calculations, which may be unstable in small samples. %More generally, alternative estimators of the joint survival function and the use of shrinkage methods for high-dimensional covariates merit further study.
Finally, extending the framework to higher dimensions would broaden its applicability. While the simple Lehmann model extends naturally, the generalized model becomes increasingly complex, requiring \(2^d-1\) parameter vectors in $d$ dimensions.
%Finally, extending the proposed models to trivariate or higher-dimensional data would broaden their applicability to settings such as multiple organs or family data. While the simple Lehmann model extends naturally, the generalized Lehmann model requires a separate regression parameter for each derivative of the \(d\)-dimensional log survival function, yielding \(2^d-1\) parameter vectors. This grows rapidly with dimension and may become infeasible even in moderate-dimensional settings. %While the simple Lehmann model extends naturally, the generalized Lehmann model requires a separate regression parameter for each derivative of the \(d\)-dimensional log survival function, yielding \(2^d-1\) parameter vectors. This grows rapidly with dimension and may become infeasible even in moderate-dimensional settings.

%In conclusion, we introduced a new class of Lehmann-type semiparametric regression models for bivariate survival data together with a tractable pseudo-observations--based estimation framework. The proposed models extend the Cox paradigm to the bivariate setting while preserving interpretable regression effects on the survival scale. Several methodological challenges remain open, including likelihood-based estimation, higher-dimensional extensions, dependent censoring, and conditions ensuring validity of the joint survival distribution.
\section{Acknowledgments}
This work is supported in part by the Israel Science Foundation (ISF: grants Nos. 1147/20 and 767/21), and by the Kashket Memorial Fellowship.

\FloatBarrier
\clearpage
\begin{appendices}
	
	\renewcommand{\thefigure}{S\arabic{figure}}
	\renewcommand{\thetable}{S\arabic{table}}
	\renewcommand{\theproposition}{S\arabic{proposition}}
	\renewcommand{\theremark}{S\arabic{remark}}
	\renewcommand{\thelemma}{S\arabic{lemma}}
	
	\setcounter{figure}{0}
	\setcounter{table}{0}
	
	%\section{Appendix A}\label{Appendix A}
	\section{The factorization of the bivariate survival function}\label{Appendix A}
	Define $A(t_1,t_2)=\log(S(t_1,t_2))$. Using rules of double integration, for $(t_1,t_2)\in [0,\tau_1]\times [0,\tau_2]$ such that $S(\tau_1,\tau_2)>0$, we have that
	\[
	\int_{0}^{t_1}\int_{0}^{t_2} A(du,dv)=A(t_1,t_2)-A(t_1,0)-A(0,t_2)+A(0,0).
	\]
	
	Next, since $A(t_1,t_2)=\log(S(t_1,t_2))$, we have $A(0,0)=\log(S(0,0))=\log(1)=0$, $A(t_1,0)=\log(S(t_1,0))$, and $A(0,t_2)=\log(S(0,t_2))$. Consequently,
	
	\begin{equation*}
		\begin{aligned}
			S(t_1,t_2)&=\exp\{A(t_1,t_2)\}\\
			&=\exp\left\{\int_{0}^{t_1}\int_{0}^{t_2}A(du,dv)+A(t_1,0)+A(0,t_2)-A(0,0)\right\}\\
			&=\exp\left\{\int_{0}^{t_1}\int_{0}^{t_2}A(du,dv)+A(t_1,0)+A(0,t_2)\right\}\\
			&=\exp\{A(t_1,0)\}\exp\{A(0,t_2)\}\exp\left\{\int_{0}^{t_1}\int_{0}^{t_2}A(du,dv)\right\}\\
			&=S_{T_1}(t_1)S_{T_2}(t_2)\exp\left\{\int_{0}^{t_1}\int_{0}^{t_2}A(du,dv)\right\},
		\end{aligned}
	\end{equation*}
	which explains in detail the identities in Equation~\eqref{eq1}.
	
	%	\section{Appendix B}\label{app:B}
	\section{The bivariate hazard functions}\label{app:B}
	Note that if $S$ has second order partial derivatives such that
	$\frac{\partial^2S(u,v)}{\partial u\partial v}=f(u,v)$ exists, then we can write
	\[
	\frac{\partial^2A(u,v)}{\partial u\partial v}=\frac{\partial^2 \log S(u,v)}{\partial u\partial v}=\frac{f(u,v)}{S(u,v)}-\frac{\frac{\partial S(u,v)}{\partial u}}{S(u,v)}\frac{\frac{\partial S(u,v)}{\partial v}}{S(u,v)},
	\]
	which can be rewritten in terms of the bivariate hazard functions as
	\[
	\frac{\partial^2A(u,v)}{\partial u\partial v}=\frac{\partial^2 \Lambda_{11}(u,v)}{\partial u \partial v}-\frac{\partial\Lambda_{10}(u,v)}{\partial u }\frac{\partial\Lambda_{01}(u,v)}{ \partial v}.
	\]
	
	More generally, even if $S$ does not have second order partial derivatives, we obtain that
	\[
	A(du,dv)=\frac{\Lambda_{11}(du,dv)-\Lambda_{10}(du,v^-)\Lambda_{01}(u^-,dv)}{[1-\Lambda_{10}(\Delta u,v^-)][1-\Lambda_{01}(u^-,\Delta v)]},
	\]
	where $\Lambda_{10}(\Delta u,v^-)=\Lambda_{10}(u,v^-)-\Lambda_{10}(u^-,v^-)$ and where $\Lambda_{01}(u^-,\Delta v)=\Lambda_{01}(u^-, v)-\Lambda_{01}(u^-,v^-)$.
	
	\section{Additional details on the Lehmann models}\label{app:Lehmann_details}
	
	\subsection{Connection between the simple and generalized Lehmann models}
	
	By Equation~\eqref{eq1}, the baseline bivariate survival function admits the decomposition
	\[
	S^0(t_1,t_2)
	=
	S_{T_1}^0(t_1)\,S_{T_2}^0(t_2)\,
	\exp\Big\{\int_0^{t_1}\int_0^{t_2}A^0(du,dv)\Big\},
	\]
	where $S_{T_1}^0(t_1)=S^0(t_1,0)$, $S_{T_2}^0(t_2)=S^0(0,t_2)$, and $A^0(t_1,t_2)=\log S^0(t_1,t_2)$.
	
	Substituting this representation into the simple Lehmann model,
	\[
	S(t_1,t_2 \mid Z) = \left[S^0(t_1,t_2)\right]^{\exp(b^T Z)},
	\]
	yields
	\begin{align*}
		S(t_1,t_2 \mid Z)
		&=
		\left[
		S_{T_1}^0(t_1)\,S_{T_2}^0(t_2)\,
		\exp\Big\{\int_0^{t_1}\int_0^{t_2}A^0(du,dv)\Big\}
		\right]^{\exp(b^T Z)} \\
		&=
		\left[S_{T_1}^0(t_1)\right]^{\exp(b^T Z)}
		\left[S_{T_2}^0(t_2)\right]^{\exp(b^T Z)}
		\left[
		\exp\Big\{\int_0^{t_1}\int_0^{t_2}A^0(du,dv)\Big\}
		\right]^{\exp(b^T Z)}.
	\end{align*}
	
	Thus, the simple Lehmann model can be expressed as the product of three components:
	\begin{align*}
		S_{T_1}(t_1\mid Z)&= \left[S_{T_1}^0(t_1)\right]^{\exp(b^T Z)}, \\
		S_{T_2}(t_2\mid Z)&= \left[S_{T_2}^0(t_2)\right]^{\exp(b^T Z)}, \\
		\exp\Big\{\int_0^{t_1}\int_0^{t_2}A(du,dv; Z)\Big\}
		&=
		\left[
		\exp\Big\{\int_0^{t_1}\int_0^{t_2}A^0(du,dv)\Big\}
		\right]^{\exp(b^T Z)}.
	\end{align*}
	
	All three components share the same regression parameter $b$. The generalized Lehmann model arises by allowing distinct regression parameters in these three components, thereby extending the simple model while preserving its multiplicative structure.
	
	\subsection{Detailed covariate effects in the generalized Lehmann model}
	
	To better understand the effect of covariates on joint survival, consider a scalar covariate $Z$ with value $z$. Increasing $z$ by one unit yields
	\begin{align*}
		S(t_1,t_2 \mid z+1)
		&=
		[S_{T_1}(t_1 \mid z)]^{e^\alpha}
		[S_{T_2}(t_2 \mid z)]^{e^\beta}
		\left[
		\frac{S(t_1,t_2 \mid z)}{S_{T_1}(t_1 \mid z)\,S_{T_2}(t_2 \mid z)}
		\right]^{e^\gamma} \\
		&=
		[S_{T_1}(t_1 \mid z)]^{e^\alpha - e^\gamma}
		[S_{T_2}(t_2 \mid z)]^{e^\beta - e^\gamma}
		[S(t_1,t_2 \mid z)]^{e^\gamma}.
	\end{align*}
	
	This expression highlights that the overall covariate effect on joint survival is determined jointly by the marginal components and the dependence component. In particular:
	\begin{itemize}
		\item The parameters $\alpha$ and $\beta$ govern the marginal survival behavior of $T_1$ and $T_2$, respectively.
		\item The parameter $\gamma$ controls how the covariate amplifies or attenuates the dependence between $T_1$ and $T_2$.
	\end{itemize}
	
	When $\exp(\gamma)>1$, the dependence component is strengthened relative to the baseline, whereas when $\exp(\gamma)<1$, it is weakened. The overall direction of the covariate effect on $S(t_1,t_2 \mid Z)$ depends on the combined influence of these three components, and may therefore reflect competing marginal and dependence effects.
	
	\section{Hazard functions and validity constraints}\label{app:C}
	\subsection{Structural constraints on the hazard functions}
	When $\alpha=\beta=\gamma=b$, the generalized Lehmann model reduces to the simple Lehmann model. In this case, the corresponding hazard models coincide, yielding the identities  
	\[
	-\int_{0}^{t_2} a^0(t_1,v)\, dv \;=\; \lambda_{10}^0(t_1,t_2)-\lambda_{10}^0(t_1,0), 
	\quad
	-\int_{0}^{t_1} a^0(u,t_2)\, du \;=\; \lambda_{01}^0(t_1,t_2)-\lambda_{01}^0(0,t_2).
	\]
	Equivalently,  
	\[
	-\frac{\partial \lambda_{01}^0(t_1,t_2)}{\partial t_1}
	= a^0(t_1,t_2)
	= -\frac{\partial \lambda_{10}^0(t_1,t_2)}{\partial t_2}.
	\]

	These identities follow from the continuity of $S^0(t_1,t_2)$, which implies equality of mixed partial derivatives:
	\[
	\frac{\partial}{\partial t_2}\left(\frac{\partial \log S^0(t_1,t_2)}{\partial t_1}\right)
	= \frac{\partial}{\partial t_1}\left(\frac{\partial \log S^0(t_1,t_2)}{\partial t_2}\right).
	\]  
	The same identities hold for any covariate $Z$:
	\begin{equation*}%\label{eq:haz_constraint}
		-\frac{\partial\lambda_{01}(t_1,t_2\mid Z)}{\partial t_1 }=a(t_1,t_2; Z)=-\frac{\partial\lambda_{10}(t_1,t_2\mid Z)}{\partial t_2}.
	\end{equation*}
	
	\subsubsection{On the incompatibility of three proportional hazards models}\label{sec:3PH_not_valid}
	At first sight, it seems natural to extend the Cox proportional hazards model to bivariate survival data by specifying three bivariate proportional hazard models:
	\begin{equation*}%\label{eq:3PH}
		\begin{aligned}
			\lambda_{10}(t_1,t_2\mid Z)&= e^{\alpha^TZ}\lambda_{10}^0(t_1,t_2),\\
			\lambda_{01}(t_1,t_2\mid Z)&= e^{\beta^TZ}\lambda_{01}^0(t_1,t_2),\\
			\lambda_{11}(t_1,t_2\mid Z)&= e^{\gamma^TZ}\lambda_{11}^0(t_1,t_2),
		\end{aligned}
	\end{equation*}
	where $\lambda_{10}^0$, $\lambda_{01}^0$, and $\lambda_{11}^0$ are valid baseline bivariate hazard functions.  
	
	However, these models cannot all hold. The continuity constraint $-\frac{\partial \lambda_{10}(t_1,t_2\mid Z)}{\partial t_2}=a(t_1,t_2 ; Z)=-\frac{\partial \lambda_{01}(t_1,t_2\mid Z)}{\partial t_1}$, together with $a(t_1,t_2 ; Z)=\lambda_{11}(t_1,t_2\mid Z)-\lambda_{10}(t_1,t_2\mid Z)\lambda_{01}(t_1,t_2\mid Z)$, means that any valid hazards model needs to satisfy
	\begin{align*}
		\lambda_{10}(t_1,t_2\mid Z)&=-\int_{0}^{t_2}a(t_1,v;Z)dv+C_1(t_1,Z)\\
		\lambda_{01}(t_1,t_2\mid Z)&=-\int_{0}^{t_1}a(u,t_2; Z)du+C_2(t_2,Z)\\
		\lambda_{11}(t_1,t_2\mid Z)&=a(t_1,t_2 ; Z)+\lambda_{10}(t_1,t_2\mid Z)\lambda_{01}(t_1,t_2\mid Z),
	\end{align*}
	where $C_1(t_1,Z)=\lambda_{10}(t_1,0\mid Z)=\lambda_1(t_1\mid Z)$ and $C_2(t_2,Z)=\lambda_{01}(0,t_2\mid Z)=\lambda_2(t_2\mid Z)$ are the two marginal hazard functions. Specifically, if we use two marginal Cox models such that $C_1(t_1,Z)=\lambda_1(t_1\mid Z)=\lambda_{T_1}^0(t_1)e^{\alpha^TZ}$ and $C_2(t_2,Z)=\lambda_2(t_2\mid Z)=\lambda_{T_2}^0(t_2)e^{\beta^TZ}$, then a consistent model for the double-failure rate must satisfy
	\begin{align*}
		&\lambda_{11}(t_1,t_2\mid Z)\\
		=&a(t_1,t_2 ; Z)+\left(\lambda_{T_1}^0(t_1)e^{\alpha^TZ}-\int_{0}^{t_2}a(t_1,v;Z)dv\right)\left(\lambda_{T_2}^0(t_2)e^{\beta^TZ}-\int_{0}^{t_1}a(u,t_2; Z)du\right)
	\end{align*}
	which does not admit a multiplicative form such as $\lambda_{11}(t_1,t_2\mid Z)=e^{\gamma^TZ}\lambda_{11}^0(t_1,t_2)$. Thus, the three proportional hazards models proposed in \cite{pons_nonparametric_1992,tien_proportional_2002, prentice_regression_2021} do not correspond to a valid bivariate survival distribution.
	
	\subsection{Proof of Proposition~\ref{prop:hazards}}
	\begin{proof}%[Proof of Proposition~\ref{prop:hazards}]
		\textbf{Simple Lehmann model:}  
		The simple Lehmann model takes the form
		\[
		S(t_1,t_2\mid Z)=\left[S^0(t_1,t_2)\right]^{e^{b^TZ}}.
		\]
		By the chain rule, it's derivative with respect to $t_1$ satisfies
		\begin{align*}
			\frac{\partial S(t_1,t_2\mid Z)}{\partial t_1}=e^{b^TZ}\left[S^0(t_1,t_2)\right]^{e^{b^TZ}-1}\frac{\partial S^0(t_1,t_2)}{\partial t_1}.
		\end{align*}
		Consequently,
		\[
		\frac{\frac{\partial S(t_1,t_2\mid Z)}{\partial t_1}}{S(t_1,t_2\mid Z)}=\frac{e^{b^TZ}\left[S^0(t_1,t_2)\right]^{e^{b^TZ}-1}\frac{\partial S^0(t_1,t_2)}{\partial t_1}}{\left[S^0(t_1,t_2)\right]^{e^{b^TZ}}}=
		e^{b^TZ}\frac{\frac{\partial S^0(t_1,t_2)}{\partial t_1}}{S^0(t_1,t_2)}=-\lambda_{10}^0(t_1,t_2)e^{b^TZ}.
		\]
		Similarly,
		\[
		\frac{\frac{\partial S(t_1,t_2\mid Z)}{\partial t_2}}{S(t_1,t_2\mid Z)}=\frac{e^{b^TZ}\left[S^0(t_1,t_2)\right]^{e^{b^TZ}-1}\frac{\partial S^0(t_1,t_2)}{\partial t_2}}{\left[S^0(t_1,t_2)\right]^{e^{b^TZ}}}=
		e^{b^TZ}\frac{\frac{\partial S^0(t_1,t_2)}{\partial t_2}}{S^0(t_1,t_2)}=-\lambda_{01}^0(t_1,t_2)e^{b^TZ}.
		\]
		Hence,
		\[
		\lambda_{10}(t_1,t_2\mid Z)=\frac{-\frac{\partial S(t_1,t_2\mid Z)}{\partial t_1}}{S(t_1,t_2\mid Z)}=\lambda_{10}^0(t_1,t_2)e^{b^TZ} \ \text{and} \ \lambda_{01}(t_1,t_2\mid Z)=\frac{-\frac{\partial S(t_1,t_2\mid Z)}{\partial t_2}}{S(t_1,t_2\mid Z)}=\lambda_{01}^0(t_1,t_2)e^{b^TZ},
		\]
		which proves the first two equalities in Equation~\ref{eq:cond_hazard_simple}.
		
		The cross partial derivative is
		\begin{align*}
			&\frac{\partial^2 S(t_1,t_2\mid Z)}{\partial t_1 \partial t_2}=\frac{\partial}{\partial t_2}\left[e^{b^TZ}\left[S^0(t_1,t_2)\right]^{e^{b^TZ}-1}\frac{\partial S^0(t_1,t_2)}{\partial t_1}\right]\\
			=&e^{b^TZ}(e^{b^TZ}-1)\left[S^0(t_1,t_2)\right]^{e^{b^TZ}-2}\frac{\partial S^0(t_1,t_2)}{\partial t_1}\frac{\partial S^0(t_1,t_2)}{\partial t_2}+e^{b^TZ}\left[S^0(t_1,t_2)\right]^{e^{b^TZ}-1}\frac{\partial^2 S^0(t_1,t_2)}{\partial t_1 \partial t_2}.
		\end{align*}
		Consequently,
		\begin{align*}
			\lambda_{11}(t_1,t_2\mid Z)&=\frac{\frac{\partial^2 S(t_1,t_2\mid Z)}{\partial t_1\partial t_2}}{S(t_1,t_2\mid Z)}\\
			&=\frac{e^{b^TZ}(e^{b^TZ}-1)\left[S^0(t_1,t_2)\right]^{e^{b^TZ}-2}\frac{\partial S^0(t_1,t_2)}{\partial t_1}\frac{\partial S^0(t_1,t_2)}{\partial t_2}+e^{b^TZ}\left[S^0(t_1,t_2)\right]^{e^{b^TZ}-1}\frac{\partial^2 S^0(t_1,t_2)}{\partial t_1 \partial t_2}}{\left[S^0(t_1,t_2)\right]^{e^{b^TZ}}}\\
			&=\frac{e^{b^TZ}(e^{b^TZ}-1)}{[ S^0(t_1,t_2)]^2}\frac{\partial S^0(t_1,t_2)}{\partial t_1}\frac{\partial S^0(t_1,t_2)}{\partial t_2}+\frac{e^{b^TZ}}{S^0(t_1,t_2)}\frac{\partial^2 S^0(t_1,t_2)}{\partial t_1\partial t_2}\\
			&=e^{b^TZ}(e^{b^TZ}-1)\lambda_{10}^0(t_1,t_2)\lambda_{01}^0(t_1,t_2)+e^{b^T Z}\lambda_{11}^0(t_1,t_2)\\
			&=\lambda_{11}^0(u,v)e^{b^TZ}-\lambda_{10}^0(u,v)\lambda_{01}^0(u,v)\left[e^{b^TZ}-e^{2b^TZ}\right].
		\end{align*}
		
		\textbf{Generalized Lehmann model:}  
		The generalized Lehmann model takes the form
		\begin{align*}
			S(t_1,t_2\mid Z)=\left[S_1^0(t_1)\right]^{e^{\alpha^TZ}}\left[S_2^0(t_2)\right]^{e^{\beta^TZ}}\left[\exp\left\{\int_{0}^{t_1}\int_{0}^{t_2}A^0(du,dv)\right\}\right]^{e^{\gamma^TZ}}.
		\end{align*}
		To find the corresponding covariate-adjusted bivariate hazard functions, we need to differentiate the covariate-adjusted bivariate survival function.
		\begin{align*}
			\frac{\partial S(t_1,t_2\mid Z)}{\partial t_1}&=\frac{\partial\left[S_1^0(t_1)\right]^{e^{\alpha^TZ}}}{\partial t_1}\frac{ S(t_1,t_2\mid Z)}{\left[S_1^0(t_1)\right]^{e^{\alpha^TZ}}}+S(t_1,t_2\mid Z)\frac{\partial \int_{0}^{t_1}\int_{0}^{t_2}e^{\gamma^TZ}a^0(u,v)dudv}{\partial t_1}\\
			&=e^{\alpha^TZ}\left[S_1^0(t_1)\right]^{e^{\alpha^TZ}-1}\frac{\partial S_1^0(t_1)}{\partial t_1}\frac{ S(t_1,t_2\mid Z)}{\left[S_1^0(t_1)\right]^{e^{\alpha^TZ}}}+S(t_1,t_2\mid Z)e^{\gamma^TZ}\int_{0}^{t_2}a^0(t_1,v)dv\\
			&=e^{\alpha^TZ}\frac{\frac{\partial S_1^0(t_1)}{\partial t_1}}{S_1^0(t_1)}S(t_1,t_2\mid Z)
			+S(t_1,t_2\mid Z)e^{\gamma^TZ}\int_{0}^{t_2}a^0(t_1,v)dv\\
			&=S(t_1,t_2\mid Z)\left[-e^{\alpha^TZ}\lambda_{10}^0(t_1,0)+e^{\gamma^TZ}\int_{0}^{t_2}a^0(t_1,v)dv\right].
		\end{align*}
		Consequently,
		\[
		\lambda_{10}(t_1,t_2\mid Z)=\frac{-\frac{\partial S(t_1,t_2\mid Z)}{\partial t_1}}{S(t_1,t_2\mid Z)}=
		e^{\alpha^TZ}\lambda_{10}^0(t_1,0)-e^{\gamma^TZ}\int_{0}^{t_2}a^0(t_1,v)dv.
		\]
		Similarly,
		\[
		\lambda_{01}(t_1,t_2\mid Z)=\frac{-\frac{\partial S(t_1,t_2\mid Z)}{\partial t_2}}{S(t_1,t_2\mid Z)}=
		e^{\beta^TZ}\lambda_{01}^0(0,t_2)-e^{\gamma^TZ}\int_{0}^{t_1}a^0(u,t_2)du.
		\]
		Next,
		\begin{align*}
			&\frac{\partial^2 S(t_1,t_2\mid Z)}{\partial t_1\partial t_2}=
			\frac{\partial}{\partial t_2}\left(S(t_1,t_2\mid Z)\left[-e^{\alpha^TZ}\lambda_{10}^0(t_1,0)+e^{\gamma^TZ}\int_{0}^{t_2}a^0(t_1,v)dv\right]\right)\\
			&=\frac{\partial S(t_1,t_2\mid Z)}{\partial t_2}\left[-e^{\alpha^TZ}\lambda_{10}^0(t_1,0)+e^{\gamma^TZ}\int_{0}^{t_2}a^0(t_1,v)dv\right]+S(t_1,t_2\mid Z)e^{\gamma^TZ}a^0(t_1,t_2)\\
			&=S(t_1,t_2\mid Z)\left[-e^{\beta^TZ}\lambda_{01}^0(0,t_2)+e^{\gamma^TZ}\int_{0}^{t_1}a^0(u,t_2)du\right]\left[-e^{\alpha^TZ}\lambda_{10}^0(t_1,0)+e^{\gamma^TZ}\int_{0}^{t_2}a^0(t_1,v)dv\right]\\
			&+S(t_1,t_2\mid Z)e^{\gamma^TZ}a^0(t_1,t_2).
		\end{align*}
		Hence,
		\begin{align*}
			&\lambda_{11}(t_1,t_2\mid Z)=\frac{\frac{\partial^2 S(t_1,t_2\mid Z)}{\partial t_1\partial t_2}}{S(t_1,t_2\mid Z)}\\
			&=\left[-e^{\beta^TZ}\lambda_{01}^0(0,t_2)+e^{\gamma^TZ}\int_{0}^{t_1}a^0(u,t_2)du\right]\left[-e^{\alpha^TZ}\lambda_{10}^0(t_1,0)+e^{\gamma^TZ}\int_{0}^{t_2}a^0(t_1,v)dv\right]+e^{\gamma^TZ}a^0(t_1,t_2)\\
			&=\left[e^{\beta^TZ}\lambda_{01}^0(0,t_2)-e^{\gamma^TZ}\int_{0}^{t_1}a^0(u,t_2)du\right]\left[e^{\alpha^TZ}\lambda_{10}^0(t_1,0)-e^{\gamma^TZ}\int_{0}^{t_2}a^0(t_1,v)dv\right]+e^{\gamma^TZ}a^0(t_1,t_2).
		\end{align*}
		
	\end{proof}

	\section{The validity of both Lehmann models}\label{app:E}
	
	Here we check if and when the proposed models correspond to a valid bivariate survival function. A valid bivariate survival function of a pair of non-negative random variables needs to satisfy 
	\begin{enumerate}
		\item $S(t_1,t_2\mid Z)\geq 0$, $\forall t_1,t_2 \geq 0$,
		\item $S(0,0\mid Z)=1$, 
		\item $S_{T_1}(t_1\mid Z)=S(t_1,0\mid Z)$ and $S_{T_2}(t_2\mid Z)=S(0,t_2\mid Z)$,
		\item $\lim_{t_1,t_2 \rightarrow \infty}S(t_1,t_2\mid Z)=0$, $\lim_{t_1 \rightarrow \infty}S(t_1,t_2\mid Z)=0$, and $\lim_{t_2 \rightarrow \infty}S(t_1,t_2\mid Z)=0$,
		\item $S(t_1,t_2\mid Z)$ is monotonically decreasing with respect to each coordinate,
		\item $S(b,d\mid Z)-S(b,c\mid Z)-S(a,d\mid Z)+S(a,c\mid Z)\geq 0$ for $a\leq b$ and $c\leq d$. 
	\end{enumerate}
	
	Properties (1)--(3) are relatively easy to check in these generalized Lehmann models. Property (4) holds trivially for the simple Lehmann model; the generalized Lehmann model is only defined on a rectangle $[0,\tau_1]\times [0,\tau_2]$ where the factorization in Equation~\eqref{eq1} holds. Hence, it only needs to satisfy the validity conditions locally on this rectangle, and does not need to satisfy Property (4) globally. This means that the survival function is valid for all times within the range where the model is defined, which is sufficient for practical applications where the observation window is finite. 
	
	Property (5) holds when 
	\[
	\frac{\partial S(t_1,t_2\mid Z)}{\partial t_1}, \ \frac{\partial S(t_1,t_2\mid Z)}{\partial t_2} < 0,
	\] 
	and Property (6) holds when 
	\[
	f(t_1,t_2\mid Z)=\frac{\partial^2 S(t_1,t_2\mid Z)}{\partial t_1 \partial t_2} \geq 0.
	\] 
	Consequently, it is enough to check the derivatives of $S(t_1,t_2\mid Z)$. Specifically, we need to verify that
	\begin{enumerate}
		\item $f(t_1,t_2\mid Z)=\frac{\partial^2 S(t_1,t_2\mid Z)}{\partial t_1 \partial t_2}\geq 0$ (positivity over rectangles),
		\item $\frac{\partial S(t_1,t_2\mid Z)}{\partial t_1}, \ \frac{\partial S(t_1,t_2\mid Z)}{\partial t_2} \leq 0$ (monotonicity with respect to each coordinate).
	\end{enumerate}
	
	Note that verifying that the bivariate density function $f(t_1,t_2\mid Z)$ is non-negative is equivalent to verifying that the double-failure rate 
	\[
	\lambda_{11}(t_1,t_2\mid Z)=\frac{f(t_1,t_2\mid Z)}{S(t_1,t_2\mid Z)}
	\] 
	is non-negative. Similarly, due to the relations 
	\[
	\lambda_{10}(t_1,t_2\mid Z) = -\frac{\partial S(t_1,t_2\mid Z)/\partial t_1}{S(t_1,t_2\mid Z)}, \quad
	\lambda_{01}(t_1,t_2\mid Z) = -\frac{\partial S(t_1,t_2\mid Z)/\partial t_2}{S(t_1,t_2\mid Z)},
	\] 
	checking that the two partial derivatives of the survival function are non-positive is equivalent to checking that $\lambda_{10}(t_1,t_2\mid Z)$ and $\lambda_{01}(t_1,t_2\mid Z)$ are both non-negative. That is, for our model to be valid, we need to verify that 
	\[
	\lambda_{10}(t_1,t_2\mid Z) \geq 0, \quad
	\lambda_{01}(t_1,t_2\mid Z) \geq 0, \quad
	\lambda_{11}(t_1,t_2\mid Z) \geq 0,
	\] 
	for all $(t_1,t_2)$ where our models are defined. Below we present conditions that guarantee the validity of both the simple and generalized Lehmann models.
	
	\subsection{Proofs of propositions~\ref{prop:suff_cond_simple} and \ref{prop:suff_cond_generalized}}\label{app:E_proofs}
	\begin{proof}[Proof of Proposition~\ref{prop:suff_cond_simple}]
		By Proposition~\ref{prop:hazards}, the bivariate hazard functions corresponding to the simple Lehmann model satisfy
		\begin{equation*}
			\begin{aligned}
				\lambda_{10}(u,v\mid Z)&=\lambda_{10}^0(u,v)e^{b^TZ},\\		
				\lambda_{01}(u,v\mid Z)&=\lambda_{01}^0(u,v)e^{b^TZ},\\
				\lambda_{11}(u,v\mid Z)&=\lambda_{11}^0(u,v)e^{b^TZ}-\lambda_{10}^0(u,v)\lambda_{01}^0(u,v)\left[e^{b^TZ}-e^{2b^TZ}\right].
			\end{aligned}
		\end{equation*}
		Since the baseline bivariate hazard functions correspond to a valid baseline bivariate distribution, they must satisfy
		\[
		\lambda_{10}^0(t_1,t_2) \geq 0, \quad
		\lambda_{01}^0(t_1,t_2) \geq 0, \quad
		\lambda_{11}^0(t_1,t_2) \geq 0.
		\] 
		Consequently, 
		\[
		\lambda_{10}(u,v\mid Z)=\lambda_{10}^0(u,v)e^{b^TZ}	\geq 0 \ \text{and} \ 
		\lambda_{01}(u,v\mid Z)=\lambda_{01}^0(u,v)e^{b^TZ} \geq 0.
		\]
		As for the double-failure rate, recall that $a^0(t_1,t_2)=\lambda_{11}^0(t_1,t_2)-\lambda_{10}^0(t_1,t_2)\lambda_{01}^0(t_1,t_2)$. Hence,
		\begin{enumerate}
			\item If $a^0(t_1,t_2)\geq 0$ then
			\begin{align*}
				\lambda_{11}(u,v\mid Z)&=\lambda_{11}^0(u,v)e^{b^TZ}-\lambda_{10}^0(u,v)\lambda_{01}^0(u,v)\left[e^{b^TZ}-e^{2b^TZ}\right]\\
				&=e^{b^TZ}a^0(t_1,t_2)+e^{2b^TZ}\lambda_{10}^0(u,v)\lambda_{01}^0(u,v) \geq 0,
			\end{align*}
			as the sum of two non-negative terms.
			\item If $b^TZ \geq 0$ for all $Z$ then $\left[e^{2b^TZ}-e^{b^TZ}\right]\geq 0$. Hence,
			\begin{align*}
				\lambda_{11}(u,v\mid Z)&=\lambda_{11}^0(u,v)e^{b^TZ}-\lambda_{10}^0(u,v)\lambda_{01}^0(u,v)\left[e^{b^TZ}-e^{2b^TZ}\right]\\
				&=\lambda_{11}^0(u,v)e^{b^TZ}+\lambda_{10}^0(u,v)\lambda_{01}^0(u,v)\left[e^{2b^TZ}-e^{b^TZ}\right] \geq 0,
			\end{align*}
			as the sum of two non-negative terms.
		\end{enumerate}
	\end{proof}
	
	\begin{proof}[Proof of Proposition~\ref{prop:suff_cond_generalized}]
		By Proposition~\ref{prop:hazards}, the bivariate hazard functions corresponding to the generalized Lehmann model satisfy
		\begin{equation*}
			\begin{aligned}
				\lambda_{10}(t_1,t_2\mid Z)& = e^{\alpha^TZ}\lambda_{10}^0(t_1,0)-e^{\gamma^TZ}\int_{0}^{t_2}a^0(t_1,v)dv,\\
				\lambda_{01}(t_1,t_2\mid Z)& = e^{\beta^TZ}\lambda_{01}^0(0,t_2)-e^{\gamma^TZ}\int_{0}^{t_1}a^0(u,t_2)du,\\
				\lambda_{11}(t_1,t_2\mid Z)& = \lambda_{10}(t_1,t_2\mid Z)\lambda_{01}(t_1,t_2\mid Z)+e^{\gamma^TZ}a^0(t_1,t_2),
			\end{aligned}
		\end{equation*}
		
		Recall the equality constraints 
		\[
		-\frac{\partial \lambda_{01}^0(t_1,t_2)}{\partial t_1}
		= a^0(t_1,t_2)
		= -\frac{\partial \lambda_{10}^0(t_1,t_2)}{\partial t_2},
		\]
		or equivalently,
		\[
		-\int_{0}^{t_2} a^0(t_1,v)\, dv \;=\; \lambda_{10}^0(t_1,t_2)-\lambda_{10}^0(t_1,0), 
		\quad
		-\int_{0}^{t_1} a^0(u,t_2)\, du \;=\; \lambda_{01}^0(t_1,t_2)-\lambda_{01}^0(0,t_2).
		\]
		
		Hence,
		\begin{align*}
			\lambda_{10}(t_1,t_2\mid Z)&\geq 0 \\
			& \iff \\
			e^{\alpha^TZ}\lambda_{10}^0(t_1,0)-e^{\gamma^TZ}[\lambda_{10}^0(t_1,0)-\lambda_{10}^0(t_1,t_2)]&\geq0\\
			& \iff \\
			e^{\alpha^TZ}\lambda_{10}^0(t_1,0)&\geq e^{\gamma^TZ}[\lambda_{10}^0(t_1,0)-\lambda_{10}^0(t_1,t_2)]\\
			& \iff \\
			e^{(\alpha-\gamma)^TZ}&\geq \frac{\lambda_{10}^0(t_1,0)-\lambda_{10}^0(t_1,t_2)}{\lambda_{10}^0(t_1,0)}=1-\frac{\lambda_{10}^0(t_1,t_2)}{\lambda_{10}^0(t_1,0)},
		\end{align*}
		where we assume everywhere that $\lambda_{10}^0(t_1,0)\neq 0$ and that $\lambda_{01}^0(0,t_2)\neq 0$.
		That is,
		\begin{equation}\label{eq:sing_fail_1_iff}
			\lambda_{10}(t_1,t_2\mid Z)\geq 0 \iff e^{(\alpha-\gamma)^TZ}\geq 1-\frac{\lambda_{10}^0(t_1,t_2)}{\lambda_{10}^0(t_1,0)}.
		\end{equation}
		Similarly, for the second single-failure rate we obtain that 
		\begin{equation}\label{eq:sing_fail_2_iff}
			\lambda_{01}(t_1,t_2\mid Z)\geq 0 \iff 	e^{(\beta-\gamma)^TZ}\geq 1-\frac{\lambda_{01}^0(t_1,t_2)}{\lambda_{01}^0(0,t_2)}.
		\end{equation}
		If $(\alpha-\gamma)^TZ$ and $(\beta-\gamma)^TZ$ are both non-negative then Equations~\eqref{eq:sing_fail_1_iff}-\eqref{eq:sing_fail_2_iff} hold for any baseline dependency structure and all $(t_1,t_2)$. That is, when $\gamma^TZ \leq \min\{\alpha^TZ, \beta^TZ\}$, the two single-failure rates are non-negative.
		
		the conditional double-failure rate can be written as
		\begin{equation}\label{eq:double_failure}
			\begin{aligned}
				&\lambda_{11}(t_1,t_2\mid Z)\\
				=&\left[e^{\beta^TZ}\lambda_{01}^0(0,t_2)-e^{\gamma^TZ}\int_{0}^{t_1}a^0(u,t_2)du\right]\left[e^{\alpha^TZ}\lambda_{10}^0(t_1,0)-e^{\gamma^TZ}\int_{0}^{t_2}a^0(t_1,v)dv\right]+e^{\gamma^TZ}a^0(t_1,t_2)\\
				=&\left[e^{\beta^TZ}\lambda_{01}^0(0,t_2)+e^{\gamma^TZ}\left(\lambda_{01}^0(t_1,t_2)-\lambda_{01}^0(0,t_2)\right)\right]\left[e^{\alpha^TZ}\lambda_{10}^0(t_1,0)+e^{\gamma^TZ}\left(\lambda_{10}^0(t_1,t_2)-\lambda_{10}^0(t_1,0)\right)\right]\\
				+&e^{\gamma^TZ}\left(\lambda_{11}^0(t_1,t_2)-\lambda_{10}^0(t_1,t_2)\lambda_{01}^0(t_1,t_2)\right)\\
				=& e^{(\alpha+\beta)^TZ}\lambda_{10}^0(t_1,0)\lambda_{01}^0(0,t_2) -e^{(\alpha+\gamma)^TZ}\lambda_{10}^0(t_1,0)\lambda_{01}^0(0,t_2)+e^{(\alpha+\gamma)^TZ}\lambda_{10}^0(t_1,0)\lambda_{01}^0(t_1,t_2)\\
				-&e^{(\beta+\gamma)^TZ}\lambda_{10}^0(t_1,0)\lambda_{01}^0(0,t_2)+e^{(\beta+\gamma)^TZ}\lambda_{10}^0(t_1,t_20)\lambda_{01}^0(0,t_2)+e^{\gamma^TZ}\lambda_{11}^0(t_1,t_2)-e^{\gamma^TZ}\lambda_{10}^0(t_1,t_2)\lambda_{01}^0(t_1,t_2)\\
				+&e^{2\gamma^TZ}\left[\lambda_{10}^0(t_1,0)\lambda_{01}^0(0,t_2)-\lambda_{10}^0(t_1,0)\lambda_{01}^0(t_1,t_2)-\lambda_{10}^0(t_1,t_2)\lambda_{01}^0(0,t_2)+\lambda_{10}^0(t_1,t_2)\lambda_{01}^0(t_1,t_2)\right]\\
				=&\lambda_{10}^0(t_1,0)\lambda_{01}^0(0,t_2)\left[e^{(\alpha+\beta)^TZ}-e^{(\alpha+\gamma)^TZ}-e^{(\beta+\gamma)^TZ}+e^{2\gamma^TZ}\right]
				+\lambda_{10}^0(t_1,0)\lambda_{01}^0(t_1,t_2)\left[e^{(\alpha+\gamma)^TZ}-e^{2\gamma^TZ}\right]\\
				+&\lambda_{10}^0(t_1,t_2)\lambda_{01}^0(0,t_2)\left[e^{(\beta+\gamma)^TZ}-e^{2\gamma^TZ}\right]+\lambda_{10}^0(t_1,t_1)\lambda_{01}^0(t_1,t_2)\left[e^{2\gamma^TZ}-e^{\gamma^TZ}\right]+e^{\gamma^TZ}\lambda_{11}^0(t_1,t_2)\\
				\equiv&a+b+c+d+e,
			\end{aligned}
		\end{equation}
		
		where \begin{align*}
			a&=\lambda_{10}^0(t_1,0)\lambda_{01}^0(0,t_2)\left[e^{(\alpha+\beta)^TZ}-e^{(\alpha+\gamma)^TZ}-e^{(\beta+\gamma)^TZ}+e^{2\gamma^TZ}\right]\\
			b&=\lambda_{10}^0(t_1,0)\lambda_{01}^0(t_1,t_2)\left[e^{(\alpha+\gamma)^TZ}-e^{2\gamma^TZ}\right]\\
			c&=\lambda_{10}^0(t_1,t_2)\lambda_{01}^0(0,t_2)\left[e^{(\beta+\gamma)^TZ}-e^{2\gamma^TZ}\right]\\
			d&=\lambda_{10}^0(t_1,t_1)\lambda_{01}^0(t_1,t_2)\left[e^{2\gamma^TZ}-e^{\gamma^TZ}\right]\\
			e&=\lambda_{11}^0(t_1,t_2)e^{\gamma^TZ}.
		\end{align*}
		
		The condition $\gamma^TZ \leq \min\{\alpha^TZ, \beta^TZ\}$ guarantees that
		\begin{align*}
			&\left[e^{(\alpha+\beta)^TZ}-e^{(\alpha+\gamma)^TZ}-e^{(\beta+\gamma)^TZ}+e^{2\gamma^TZ}\right]=\left[e^{\alpha^TZ}-e^{\gamma^TZ}\right]\left[e^{\beta^TZ}-e^{\gamma^TZ}\right]\geq 0\\
			&\left[e^{(\alpha+\gamma)^TZ}-e^{2\gamma^TZ}\right]=e^{\gamma^TZ}\left[e^{\alpha^TZ}-e^{\gamma^TZ}\right]\geq 0\\
			&\left[e^{(\beta+\gamma)^TZ}-e^{2\gamma^TZ}\right]=e^{\gamma^TZ}\left[e^{\beta^TZ}-e^{\gamma^TZ}\right]\geq 0.
		\end{align*}
		
		Hence, $a, b,c$ in Equation~\eqref{eq:double_failure} are all non-negative as the product of the two single-failure rates and a non-negative value. Next, note that 
		\[
		d+e=e^{\gamma^TZ}\left[\lambda_{11}^0(t_1,t_2)-\lambda_{10}^0(t_1,t_2)\lambda_{01}^0(t_1,t_2)\right]+e^{2\gamma^TZ}\lambda_{10}^0(t_1,t_2)\lambda_{01}^0(t_1,t_2)
		\]
		which is non-negative when:
		\begin{enumerate}
			\item $0\leq a^0(t_1,t_2)=\lambda_{11}^0(t_1,t_2)-\lambda_{10}^0(t_1,t_2)\lambda_{01}^0(t_1,t_2)$ and then 
			\[
			d+e=e^{\gamma^TZ}a^0(t_1,t_2)+e^{2\gamma^TZ}\lambda_{10}^0(t_1,t_2)\lambda_{01}^0(t_1,t_2)\geq 0
			\]
			as the sum of two non-negative terms.
			\item $e^{2\gamma^TZ}-e^{\gamma^TZ}\geq 0$ (which can be written as $\gamma^TZ\geq 0$), and thus
			\[
			d+e=e^{\gamma^TZ}\lambda_{11}^0(t_1,t_2)+[e^{2\gamma^TZ}-e^{\gamma^TZ}]\lambda_{10}^0(t_1,t_2)\lambda_{01}^0(t_1,t_2) \geq 0,
			\]
			as the sum of two non-negative terms.
		\end{enumerate}
		In summary, the condition $\gamma^TZ \leq \min\{\alpha^TZ, \beta^TZ\}$ guarantees non-negativity of the two single-failure rates. The additional conditions (1) or (2) above  guarantee non-negativity of the double-failure rate, which concludes the proof.
	\end{proof}
	
	\subsection{Gumbel–Barnett copula: a valid model beyond the sufficient conditions}\label{sec:exp2D_valid}
	\begin{example}\label{ex:Gumbel}
		The bivariate exponential distribution of \citet{gumbel_bivariate_1960}, corresponding to a special case of the Gumbel–Barnett copula, is defined by
		\[
		S(t_1,t_2)=\exp\{-(t_1+t_2+t_1t_2)\}.
		\]
		Its density function is
		\[
		f(t_1,t_2)=\exp\{-(t_1+t_2+t_1t_2)\}(t_1+t_2+t_1t_2).
		\]
		The corresponding hazard components are
		\begin{align*}
			\lambda_{11}(s,t) &= \frac{f(s,t)}{S(s^-,t^-)} = s+t+st, \\
			\lambda_{10}(s,t) &= \frac{\int_{t}^{\infty} f(s,v)\,dv}{S(s^-,t)} = t+1, \\
			\lambda_{01}(s,t) &= \frac{\int_{s}^{\infty} f(u,t)\,du}{S(s,t^-)} = s+1.
		\end{align*}
		%		Integrating, we obtain
		%		\begin{align*}
			%			\Lambda_{11}(t_1,t_2) &= \int_{0}^{t_1}\int_{0}^{t_2} (s+t+st)\,ds\,dt
			%			= \tfrac{1}{4}\big(2t_1^2 t_2 + 2t_1 t_2^2 + t_1^2 t_2^2\big), \\
			%			\Lambda_{10}(t_1,t_2) &= \int_{0}^{t_1} (t_2+1)\,ds = t_1 t_2 + t_1, \\
			%			\Lambda_{01}(t_1,t_2) &= \int_{0}^{t_2} (t_1+1)\,dt = t_1 t_2 + t_2.
			%		\end{align*}
		In particular,
		\[
		a(s,t) = \lambda_{11}(s,t) - \lambda_{10}(s,t)\lambda_{01}(s,t) = -1 < 0,
		\]
		so this distribution exhibits negative quadrant dependence (NQD).
	\end{example}
	
	We assume that the baseline distribution corresponds to Example~\ref{ex:Gumbel}. Then
	\[
	\lambda_{11}^0(s,t)=s+t+st, \quad
	\lambda_{10}^0(s,t)=t+1, \quad
	\lambda_{01}^0(s,t)=s+1, \quad
	a^0(s,t)=-1.
	\]
	Combining this with Proposition~\ref{prop:hazards}, the three hazard components under the generalized Lehmann model are
	\begin{equation*}
		\begin{aligned}
			\lambda_{10}(t_1,t_2\mid Z)
			&= e^{\alpha^T Z}\lambda_{10}^0(t_1,0)
			+ e^{\gamma^T Z}\big(\lambda_{10}^0(t_1,t_2)-\lambda_{10}^0(t_1,0)\big)
			= e^{\alpha^T Z} + t_2 e^{\gamma^T Z} \ge 0, \\
			\lambda_{01}(t_1,t_2\mid Z)
			&= e^{\beta^T Z}\lambda_{01}^0(0,t_2)
			+ e^{\gamma^T Z}\big(\lambda_{01}^0(t_1,t_2)-\lambda_{01}^0(0,t_2)\big)
			= e^{\beta^T Z} + t_1 e^{\gamma^T Z} \ge 0, \\
			\lambda_{11}(t_1,t_2\mid Z)
			&= \lambda_{10}(t_1,t_2\mid Z)\lambda_{01}(t_1,t_2\mid Z)
			+ e^{\gamma^T Z} a^0(t_1,t_2) \\
			&= \big(e^{\alpha^T Z} + t_2 e^{\gamma^T Z}\big)
			\big(e^{\beta^T Z} + t_1 e^{\gamma^T Z}\big)
			- e^{\gamma^T Z} \\
			&= e^{(\alpha+\beta)^T Z}
			+ t_1 e^{(\alpha+\gamma)^T Z}
			+ t_2 e^{(\beta+\gamma)^T Z}
			- e^{\gamma^T Z}.
		\end{aligned}
	\end{equation*}
	If $\gamma = \alpha + \beta$, then $\lambda_{11}(t_1,t_2 \mid Z) \ge 0$ for all $t_1,t_2$ and $Z$, and the resulting model defines a valid bivariate survival function despite violating Proposition~\ref{prop:suff_cond_generalized}.
	
	\section{Dependency measures}\label{app:D}
	%	\cite{clayton_model_1978}, \cite{oakes_bivariate_1989}, and \cite{hsu_assessing_1996} define the \emph{cross ratio function} by
	%	\[
	%	c(t_1,t_2)=\frac{S(dt_1,dt_2)S(t_1,t_2)}{S(dt_1,t_2)S(t_1,dt_2)}.
	%	\]
	%	The value of $c(t_1,t_2)$ indicates positive association ($>1$), no association ($=1$), or negative local association ($<1$) between the two failure times at $(t_1,t_2)$.
	%	
	%	For continuous survival times, the cross ratio function can also be expressed in terms of the bivariate hazard functions:
	\citet{clayton_model_1978}, \citet{oakes_bivariate_1989}, and \citet{hsu_assessing_1996} define the \emph{cross-ratio function}, which for continuous survival times can be expressed as
	\[
	c(t_1,t_2)=\frac{\frac{\partial^2S(t_1,t_2)}{\partial t_1\partial t_2}S(t_1,t_2)}{\frac{\partial S(t_1,t_2)}{\partial t_1}\frac{\partial S(t_1,t_2)}{\partial t_2}}
	=\frac{\lambda_{11}(t_1,t_2)}{\lambda_{10}(t_1,t_2)\lambda_{01}(t_1,t_2)}.
	\]

	We propose the \emph{covariate-adjusted cross ratio function}:
	\[
	c(t_1,t_2\mid Z)=\frac{\frac{\partial^2 S(t_1,t_2\mid Z)}{\partial t_1 \partial t_2} S(t_1,t_2\mid Z)}
	{\frac{\partial S(t_1,t_2\mid Z)}{\partial t_1} \frac{\partial S(t_1,t_2\mid Z)}{\partial t_2}}
	= \frac{\lambda_{11}(t_1,t_2\mid Z)}{\lambda_{10}(t_1,t_2\mid Z)\lambda_{01}(t_1,t_2\mid Z)}
	= 1 + \frac{a(t_1,t_2;Z)}{\lambda_{10}(t_1,t_2\mid Z)\lambda_{01}(t_1,t_2\mid Z)},
	\]
	where $a(t_1,t_2;Z)=\lambda_{11}(t_1,t_2\mid Z)-\lambda_{10}(t_1,t_2\mid Z)\lambda_{01}(t_1,t_2\mid Z)$.
	\begin{proposition}\label{prop:cross-ratio}
		Assume either Lehmann model. The value of the covariate-adjusted cross ratio function $c(t_1,t_2\mid Z)$ is determined from the sign of $a^0(t_1,t_2) $:
		\begin{enumerate}
			\item If $a^0(t_1,t_2)>0$, then $c(t_1,t_2\mid Z) > 1$.
			\item If $a^0(t_1,t_2)=0$, then $c(t_1,t_2\mid Z) = 1$.
			\item If $a^0(t_1,t_2)<0$, then $c(t_1,t_2\mid Z) < 1$.
		\end{enumerate}
	\end{proposition}
	The proof the proposition follows directly from the definition of the covariate-adjusted cross ratio term and from the fact that $a(t_1,t_2;Z)=a^0(t_1,t_2)e^{b^TZ}$ in the simple Lehmann model and $a(t_1,t_2;Z)=a^0(t_1,t_2)e^{\gamma^TZ}$ in the generalized Lehmann model.
	
	The corresponding weighted average of the covariate-adjusted cross ratio term is
	\begin{equation}\label{eq:weighted_CR}
		C(t_1,t_2\mid Z)=\frac{\int_{0}^{t_1}\int_{0}^{t_2} \lambda_{11}(s_1,s_2\mid Z) ds_1 ds_2}{\int_{0}^{t_1}\int_{0}^{t_2} \lambda_{10}(s_1,s_2\mid Z)\lambda_{01}(s_1,s_2\mid Z) ds_1 ds_2},
	\end{equation}
	which is a special case of the general weighted class
	\[
	D(t_1,t_2\mid Z)=\frac{\int_{0}^{t_1}\int_{0}^{t_2} \phi\{c(s_1,s_2\mid Z)\} w_Z(ds_1,ds_2)}{\int_{0}^{t_1}\int_{0}^{t_2} w_Z(ds_1,ds_2)},
	\]
	by setting $\phi(c)=c$ and 
	\[
	w_Z(ds_1,ds_2)=\frac{S(ds_1,s_2\mid Z)S(s_1,ds_2\mid Z)}{[S(s_1,s_2\mid Z)]^2} = \lambda_{10}(s_1,s_2\mid Z)\lambda_{01}(s_1,s_2\mid Z) ds_1 ds_2.
	\]
	
	\cite{prentice_regression_2021} proposed a similar weighted average:
	\[
	\tilde{C}(t_1,t_2\mid Z) = \frac{\int_{0}^{t_1}\int_{0}^{t_2} S(s_1,s_2\mid Z) \lambda_{11}(s_1,s_2\mid Z) ds_1 ds_2}{\int_{0}^{t_1}\int_{0}^{t_2} S(s_1,s_2\mid Z) \lambda_{10}(s_1,s_2\mid Z) \lambda_{01}(s_1,s_2\mid Z) ds_1 ds_2},
	\]
	which is a special case of $D(t_1,t_2\mid Z)$ using $\phi(c)=c$ and weights 
	\[
	w_Z(ds_1,ds_2)=\frac{S(ds_1,s_2\mid Z)S(s_1,ds_2\mid Z)}{S(s_1,s_2\mid Z)} = S(s_1,s_2\mid Z)\lambda_{10}(s_1,s_2\mid Z)\lambda_{01}(s_1,s_2\mid Z) ds_1 ds_2.
	\]
	
	\begin{remark}\label{rem:weighted_average}
		The weighted average $D(t_1,t_2\mid Z) $  with $\phi\{c\}=c$ can be re-written in terms of the function $a(t_1,t_2; Z)$:
		\[
		D(t_1,t_2\mid Z) = 1 + \frac{\int_0^{t_1}\int_0^{t_2} \frac{a(s_1,s_2;Z)}{\lambda_{10}(s_1,s_2\mid Z)\lambda_{01}(s_1,s_2\mid Z)} w_Z(ds_1,ds_2)}{\int_0^{t_1}\int_0^{t_2} w_Z(ds_1,ds_2)}.
		\]
		
		Specifically, for the special case $w_Z(ds_1,ds_2) = \lambda_{10}(s_1,s_2\mid Z)\lambda_{01}(s_1,s_2\mid Z) ds_1 ds_2$, the covariate-adjusted weighted average in \eqref{eq:weighted_CR} becomes
		\[
		C(t_1,t_2\mid Z) = 1 + \frac{\int_0^{t_1}\int_0^{t_2} a(s_1,s_2;Z) ds_1 ds_2}{\int_0^{t_1}\int_0^{t_2} \lambda_{10}(s_1,s_2\mid Z)\lambda_{01}(s_1,s_2\mid Z) ds_1 ds_2},
		\]
		and its value is determined by the dependency structure:
		\[
		\int_0^{t_1}\int_0^{t_2} a(s_1,s_2;Z) ds_1 ds_2 = \log\frac{S(t_1,t_2\mid Z)}{S_{T_1}(t_1\mid Z) S_{T_2}(t_2\mid Z)}.
		\]
		
	\end{remark}
	
	\begin{proposition}\label{prop:weighted_cross-ratio}
		Assume either Lehmann model. The value of the covariate-adjusted weighted average $C(t_1,t_2\mid Z)$ is determined from the sign of the log baseline dependency $\int_0^{t_1}\int_0^{t_2} a^0(s_1,s_2) ds_1 ds_2=\log \left(\frac{S^0(t_1,t_2)}{S_1^0(t_1)S_2^0(t_2)}\right)$:
		\begin{enumerate}
			\item If $\int_0^{t_1}\int_0^{t_2} a^0(s_1,s_2) ds_1 ds_2>0$, then $C(t_1,t_2\mid Z) > 1$.
			\item If $\int_0^{t_1}\int_0^{t_2} a^0(s_1,s_2) ds_1 ds_2=0$, then $C(t_1,t_2\mid Z) = 1$.
			\item If $\int_0^{t_1}\int_0^{t_2} a^0(s_1,s_2) ds_1 ds_2<0$, then $C(t_1,t_2\mid Z) < 1$.
		\end{enumerate}
	\end{proposition}
	The proof follows directly from the presentation of the weighted average in Remark~\ref{rem:weighted_average}.
	
	%	\subsection{old}
	%		For a continuous bivariate survival function $S(t_1,t_2)$ we have that $$a(t_1,t_2)=\frac{\partial^2A(t_1,t_2)}{\partial t_1 \partial t_2}=\lambda_{11}(t_1,t_2)-\lambda_{10}(t_1,t_2)\lambda_{01}(t_1,t_2)$$ where $A(t_1,t_2)=\log S(t_1,t_2)$.
	%		Interestingly, the sign of $a(t_1,t_2)$ determines whether the cross ratio term $c(t_1,t_2)$ is greater than 1, 1 or less than 1:
	%		\begin{enumerate}
		%			\item If $a(t_1,t_2)>0$ then 
		%			%the  dependency structure is PQD and we have that 
		%			\[
		%			c(t_1,t_2)=\frac{\lambda_{11}(t_1,t_2)}{\lambda_{10}(t_1,t_2)\lambda_{01}(t_1,t_2)}>1.
		%			\]
		%			\item If $a(t_1,t_2)=0$ then 
		%			%the dependency structure is independent, and 
		%			\[
		%			c(t_1,t_2)=\frac{\lambda_{11}(t_1,t_2)}{\lambda_{10}(t_1,t_2)\lambda_{01}(t_1,t_2)}=1.
		%			\]
		%			\item If $a(t_1,t_2)<0$ then
		%			%the baseline dependency structure is NQD and we have that 
		%			\[
		%			c(t_1,t_2)=\frac{\lambda_{11}(t_1,t_2)}{\lambda_{10}(t_1,t_2)\lambda_{01}(t_1,t_2)}<1.
		%			\]
		%		\end{enumerate}
	
	\section{Additional theoretical details}\label{app:F}
	\subsection{Regularity conditions}
	
	We express the regularity conditions in terms of the full parameter vector $\tilde b$ and the augmented covariates $Z_i^*$ introduced in Section~\ref{sec:fit_Lehmann}.
	
	Let
	\[
	A(\tilde b;Z_i^*)
	=
	\left(\frac{\partial}{\partial \tilde b} g^{-1}(\tilde b^T Z_i^*)\right)^T
	V_i^{-1},
	\]
	where $V_i$ is the working covariance matrix. In particular, if
	\[
	g(x)=\log\{-\log(x)\},
	\]
	then
	\[
	g^{-1}(u)=\exp\{-e^u\},
	\qquad
	\frac{\partial}{\partial \tilde b} g^{-1}(\tilde b^T Z_i^*)
	=
	- e^{\tilde b^T Z_i^*}\exp\{-e^{\tilde b^T Z_i^*}\} Z_i^*,
	\]
	so that
	\[
	A(\tilde b;Z_i^*)
	=
	- e^{\tilde b^T Z_i^*}\exp\{-e^{\tilde b^T Z_i^*}\} (Z_i^*)^T V_i^{-1}.
	\]
	Assume that the following conditions hold \citep{overgaard_asymptotic_2017}:
	
	\begin{enumerate}
		\item[(C1)] The functions $g^{-1}(\tilde b^T z^*)$ and $A(\tilde b;z^*)$ are continuously differentiable in $\tilde b$ for (almost) all $z^*$.
		
		\item[(C2)] $A(\tilde b^*,Z^*)$ has finite second moment.
		
		\item[(C3)] The functions
		\[
		\frac{\partial}{\partial \tilde b} A(\tilde b;Z^*)\, g^{-1}(\tilde b^T Z^*)
		\quad \text{and} \quad
		A(\tilde b;Z^*)\, \frac{\partial}{\partial \tilde b} g^{-1}(\tilde b^T Z^*)
		\]
		are dominated by integrable random variables in a neighborhood of $\tilde b^*$.
		
		\item[(C4)] $\left\lvert \frac{\partial}{\partial \tilde b} A(\tilde b;Z^*) \right\rvert$
		is dominated by an integrable random variable in a neighborhood of $\tilde b^*$.
		
		\item[(C5)] The matrix
		\[
		M
		=
		E\!\left[
		A(\tilde b^*;Z^*)\,
		\left.\frac{\partial}{\partial \tilde b} g^{-1}(\tilde b^T Z^*)\right|_{\tilde b=\tilde b^*}
		\right]
		\]
		has full rank.
	\end{enumerate}
	
	\subsection{The covariance matrix $\Sigma_{\tilde b}$}
	
	Denote by $O_i=(\tilde T_{1i},\Delta_{1i},\tilde T_{2i}, \Delta_{2i})$ the observed data (excluding covariates). Then
	\begin{equation*}
		\Sigma_{\tilde b}
		=
		\mathrm{Var}\!\left[
		\left.\frac{\partial g^{-1}(\tilde b^T Z_i^*)}{\partial \tilde b}\right|_{\tilde b=\tilde b^*}
		V_i^{-1}
		\left\{
		\phi(F) + \dot{\phi}(O_i) - g^{-1}\big((\tilde b^*)^T Z_i^*\big)
		\right\}
		+ h_1(O_i)
		\right],
	\end{equation*}
	where $\phi$ is the estimating functional corresponding to the Dabrowska estimator, 
	$\dot{\phi}$ is its first-order influence function, and
	\[
	h_1(o)
	=
	E\!\left[
	\left.\frac{\partial g^{-1}(\tilde b^T Z^*)}{\partial \tilde b}\right|_{\tilde b=\tilde b^*}
	V^{-1}
	\,\ddot{\phi}(o, O)
	\right],
	\]
	where $\ddot{\phi}$ is the second-order influence function.
	\subsection{Two-step pseudo-observation estimator}
	\begin{lemma}\label{lem}
		Under independent censoring, the pseudo-observations
		$\hat\theta_{ji}$, $j=1,2,3$, corresponding to the Kaplan--Meier
		and Dabrowska estimators, admit the expansion
		\[
		\hat\theta_{ji}
		=
		\theta_j+\dot\phi_j(O_i)
		+\frac{1}{n-1}\sum_{\ell\neq i}\ddot\phi_j(O_i,O_\ell)
		+r_{n,ji},
		\]
		where $\dot\phi_j$ and $\ddot\phi_j$ denote the first- and second-order influence functions, $O_i=(\tilde T_{1i},\Delta_{1i},\tilde T_{2i},\Delta_{2i})$, and $\max_i |r_{n,ji}| = o_p(n^{-1/2})$. Moreover,
		\[
		E[\dot\phi_1(O)\mid Z]
		=
		S_{T_1}(t_1^0\mid Z)-S_{T_1}(t_1^0), \quad
		E[\dot\phi_2(O)\mid Z]
		=
		S_{T_2}(t_2^0\mid Z)-S_{T_2}(t_2^0),
		\]
		and $E[\dot\phi_3(O)\mid Z]
		=
		S(t_1^0,t_2^0\mid Z)-S(t_1^0,t_2^0)$.
		
	\end{lemma}

	%		\begin{proof}
		%			For $j=1,2$, Proposition~3.1 of \citet{overgaard_asymptotic_2017} applied to the Kaplan--Meier functional gives the expansion, and the conditional expectation identity follows from the Kaplan--Meier influence-function result (see \citealt[Eq.~4.7]{overgaard_asymptotic_2017}). For $j=3$, the expansion follows from the von Mises representation of the Dabrowska pseudo-observations and the conditional expectation identity from the corresponding influence-function calculations (see Web Appendix~A of \citet{travis2024pseudo}).
		%		\end{proof}
	\begin{proof}
		For $j=1,2$, the pseudo-observations are the usual univariate Kaplan--Meier pseudo-observations for the two margins. Hence, Proposition~3.1 of \citet{overgaard_asymptotic_2017}, applied separately to each margin, yields the stated expansion. The corresponding conditional expectation identities follow from the Kaplan--Meier influence-function result; see \citealt[Eq.~4.7]{overgaard_asymptotic_2017}. For $j=3$, the expansion follows from the von Mises representation of the Dabrowska pseudo-observations, and the conditional expectation identity follows from the corresponding influence-function calculations; see Web Appendix~A of \citet{travis2024pseudo}.
	\end{proof}

	\begin{proof}[Proof of Theorem~\ref{thm:2-step}]
		By Lemma~\ref{lem}, the pseudo-observations in the first-step estimating equation admit the required asymptotic linear representation and conditional expectation identities. Hence, standard pseudo-observation GEE arguments imply
		\[
		\hat\xi_1 \xrightarrow{p} \xi_1^* .
		\]
		
		For the second step, let
		\[
		Y_i^\circ=
		\frac{\hat{\theta}_{3i}}
		{S_{T_1}(t_1^0\mid Z_i)S_{T_2}(t_2^0\mid Z_i)}
		\]
		denote the oracle response obtained by evaluating the marginal survival probabilities at their true values.
		
		By Lemma~\ref{lem},
		\[
		\hat{\theta}_{3i}
		=
		\theta_3+\dot\phi_3(O_i)
		+\frac{1}{n-1}\sum_{\ell\neq i}\ddot\phi_3(O_i,O_\ell)
		+r_{n,3i},
		\]
		where $\theta_3=S(t_1^0,t_2^0)$ and $\max_i|r_{n,3i}|=o_p(n^{-1/2})$. Since
		$S_{T_1}(t_1^0\mid Z_i)$ and $S_{T_2}(t_2^0\mid Z_i)$ are measurable with respect to $Z_i$, Lemma~\ref{lem} gives
		\begin{align*}
			E[Y_i^\circ\mid Z_i]
			&=
			\frac{\theta_3+E\{\dot\phi_3(O_i)\mid Z_i\}}
			{S_{T_1}(t_1^0\mid Z_i)S_{T_2}(t_2^0\mid Z_i)}+o_p(1) \\
			&=
			\frac{S(t_1^0,t_2^0\mid Z_i)}
			{S_{T_1}(t_1^0\mid Z_i)S_{T_2}(t_2^0\mid Z_i)}+o_p(1)
			=
			\mu_i(\xi_2^*)+o_p(1).
		\end{align*}
		%	Thus,
		%	\[
		%	E(Y_i^\circ)=E\{\mu_i(\xi_2^*)\}.
		%	\]
		%	
		It remains to relate $Y_i^\circ$ to the actual response
		\[
		Y_i=
		\frac{\hat{\theta}_{3i}}
		{\hat S_{T_1}(t_1^0\mid Z_i)\hat S_{T_2}(t_2^0\mid Z_i)}.
		\]
		The denominator is a continuously differentiable function of the first-step parameter \(\xi_1\). A first-order Taylor expansion around $\xi_1^*$ yields, uniformly in $i$,
		\[
		\hat S_{T_j}(t_j^0\mid Z_i)
		=
		S_{T_j}(t_j^0\mid Z_i)
		+
		\nabla_{\xi_1} S_{T_j}(t_j^0\mid Z_i)^T(\hat\xi_1-\xi_1^*)
		+ o_p(n^{-1/2}), \qquad j=1,2.
		\]
		Since the map $(a,b)\mapsto (ab)^{-1}$ is continuously differentiable on sets bounded away from zero, and the marginal survival probabilities are bounded away from zero,
		\[
		\frac{1}
		{\hat S_{T_1}(t_1^0\mid Z_i)\hat S_{T_2}(t_2^0\mid Z_i)}
		=
		\frac{1}
		{S_{T_1}(t_1^0\mid Z_i)S_{T_2}(t_2^0\mid Z_i)}
		+O_p(n^{-1/2}),
		\]
		uniformly in $i$. Since $\hat\theta_{3i}=O_p(1)$ under the regularity conditions,
		\[
		Y_i=Y_i^\circ+\rho_{n,i},
		\qquad
		\max_{1\le i\le n}|\rho_{n,i}|=O_p(n^{-1/2}).
		\]
		
		That is, we showed that
		\[
		E(Y_i^\circ\mid Z_i)=\mu_i(\xi_2^*)+o_p(1)
		\]
		
		and that, for consistency, replacing the true marginal survival probabilities
		by their fitted counterparts changes the second-step estimating equation only
		by an asymptotically negligible term.
		Therefore, the second-step estimating equation has limiting expectation
		\[
		E\{U_{2i}(\xi_1^*,\xi_2)\},
		\]
		which vanishes at \(\xi_2=\xi_2^*\). Standard estimating-equation arguments then give
		\[
		\hat\xi_2 \xrightarrow{p} \xi_2^* .
		\]
		Consequently,
		\[
		n^{-1/2}
		\begin{pmatrix}
			U_1(\xi_1^*)\\
			U_2(\xi_1^*,\xi_2^*)
		\end{pmatrix}
		\xrightarrow{d}N(0,\Sigma).
		\]

		It remains to derive the limiting distribution of \(\hat\xi_2\). Let
		\[
		A_{1n}(\xi_1)
		=
		-\frac{\partial U_1(\xi_1)}{\partial \xi_1},
		\qquad
		B_{1n}(\xi_1,\xi_2)
		=
		-\frac{\partial U_2(\xi_1,\xi_2)}{\partial \xi_1},
		\qquad
		B_{2n}(\xi_1,\xi_2)
		=
		-\frac{\partial U_2(\xi_1,\xi_2)}{\partial \xi_2}.
		\]
		\[
		A_1
		=
		-E\left[
		\frac{\partial U_{1i}(\xi_1)}{\partial \xi_1}
		\bigg|_{\xi_1=\xi_1^*}
		\right],
		\qquad
		B_1
		=
		-E\left[
		\frac{\partial U_{2i}(\xi_1,\xi_2)}{\partial \xi_1}
		\bigg|_{(\xi_1,\xi_2)=(\xi_1^*,\xi_2^*)}
		\right],
		\]
		\[
		B_2
		=
		-E\left[
		\frac{\partial U_{2i}(\xi_1,\xi_2)}{\partial \xi_2}
		\bigg|_{(\xi_1,\xi_2)=(\xi_1^*,\xi_2^*)}
		\right],
		\]
		where \(U_{1i}\) and \(U_{2i}\) denote the individual contributions to
		\(U_1\) and \(U_2\), respectively.
		
		By the regularity conditions, together with the standard pseudo-observation
		GEE arguments of \citet{overgaard_asymptotic_2017} and a uniform law of large numbers,
		\[
		n^{-1}A_{1n}(\tilde\xi_1)\xrightarrow{p}A_1,\qquad
		n^{-1}B_{1n}(\bar\xi_1,\bar\xi_2)\xrightarrow{p}B_1,\qquad
		n^{-1}B_{2n}(\bar\xi_1,\bar\xi_2)\xrightarrow{p}B_2,
		\]
		where \(\tilde\xi_1\) lies between \(\hat\xi_1\) and \(\xi_1^*\), and
		\((\bar\xi_1,\bar\xi_2)\) lies between
		\((\hat\xi_1,\hat\xi_2)\) and \((\xi_1^*,\xi_2^*)\).

		A first-order Taylor expansion of \(U_1\) around \(\hat\xi_1\) gives
		\[
		0
		=
		U_1(\hat\xi_1)
		=
		U_1(\xi_1^*)
		-
		A_{1n}(\tilde\xi_1)(\hat\xi_1-\xi_1^*),
		\]
		and therefore
		\[
		\sqrt n(\hat\xi_1-\xi_1^*)
		=
		\left\{n^{-1}A_{1n}(\tilde\xi_1)\right\}^{-1}
		n^{-1/2}U_1(\xi_1^*)
		=
		A_1^{-1}n^{-1/2}U_1(\xi_1^*)+o_p(1).
		\]
		
		Similarly, a first-order Taylor expansion of \(U_2\) around
		\((\hat\xi_1,\hat\xi_2)\) gives
		\[
		0
		=
		U_2(\hat\xi_1,\hat\xi_2)
		=
		U_2(\xi_1^*,\xi_2^*)
		-
		B_{1n}(\bar\xi_1,\bar\xi_2)(\hat\xi_1-\xi_1^*)
		-
		B_{2n}(\bar\xi_1,\bar\xi_2)(\hat\xi_2-\xi_2^*).
		\]
		Hence
		\[
		\sqrt n(\hat\xi_2-\xi_2^*)
		=
		\left\{n^{-1}B_{2n}(\bar\xi_1,\bar\xi_2)\right\}^{-1}
		\left[
		n^{-1/2}U_2(\xi_1^*,\xi_2^*)
		-
		\{n^{-1}B_{1n}(\bar\xi_1,\bar\xi_2)\}
		\sqrt n(\hat\xi_1-\xi_1^*)
		\right].
		\]
		Substituting the expansion for \(\sqrt n(\hat\xi_1-\xi_1^*)\) yields
		\[
		\sqrt n(\hat\xi_2-\xi_2^*)
		=
		B_2^{-1}
		\left[
		n^{-1/2}U_2(\xi_1^*,\xi_2^*)
		-
		B_1A_1^{-1}n^{-1/2}U_1(\xi_1^*)
		\right]
		+o_p(1).
		\]
		Since
		\[
		n^{-1/2}
		\begin{pmatrix}
			U_1(\xi_1^*)\\
			U_2(\xi_1^*,\xi_2^*)
		\end{pmatrix}
		\xrightarrow{d}
		N\left\{
		0,
		\begin{pmatrix}
			\Sigma_1 & \Sigma_{12}\\
			\Sigma_{21} & \Sigma_2
		\end{pmatrix}
		\right\},
		\]
		and the preceding expansion expresses
		\(\sqrt n(\hat\xi_2-\xi_2^*)\) as a continuous linear transformation of
		\[
		\left(
		n^{-1/2}U_1(\xi_1^*),
		\,n^{-1/2}U_2(\xi_1^*,\xi_2^*)
		\right),
		\]
		up to an \(o_p(1)\) term, the continuous mapping theorem and Slutsky's theorem give
		\[
		\sqrt n(\hat\xi_2-\xi_2^*)
		\xrightarrow{d}
		N(0,\Omega),
		\]
		where
		\[
		\Omega
		=
		B_2^{-1}
		\left(
		\Sigma_2
		+
		B_1A_1^{-1}\Sigma_1A_1^{-T}B_1^T
		-
		\Sigma_{21}A_1^{-T}B_1^T
		-
		B_1A_1^{-1}\Sigma_{12}
		\right)
		B_2^{-T}.
		\]

	\end{proof}

	\section{Estimation of parameters under exchangeability}\label{sec:sup_exchange}
	
	In some applications, the two failure times are exchangeable, as in the retinoblastoma example in Section~\ref{sec:data}. Exchangeability means that $S(t_1,t_2\mid Z)=S(t_2,t_1\mid Z)$. In the simple Lehmann model, this holds whenever the baseline survival function is symmetric, $S^0(t_1,t_2)=S^0(t_2,t_1)$. In the generalized Lehmann model, exchangeability further requires identical marginals, $S_{T_1}(t\mid Z)=S_{T_2}(t\mid Z)$.
	
	The estimation procedure can be adapted to incorporate these constraints. In the simple Lehmann model, symmetry is enforced by assigning a common intercept to symmetric time points $(t_1^0,t_2^0)$ and $(t_2^0,t_1^0)$, which is achieved by computing pseudo-observations at both points and fitting them jointly.
	
	For the generalized Lehmann model, exchangeability is imposed by combining the marginal pseudo-observations for $S_{T_1}$ and $S_{T_2}$ into a single regression, yielding common slope parameters and two time-specific intercepts corresponding to $t_1^0$ and $t_2^0$. The dependence component is then estimated by jointly fitting the modified responses $Y_i$ at $(t_1^0,t_2^0)$ and $(t_2^0,t_1^0)$, resulting in a single intercept $\gamma_0$ shared across symmetric time points.
	
	\section{Graphical goodness-of-fit}\label{app:GOF}

	For univariate survival data, \cite{perme_checking_2008} and \cite{andersen_pseudo-observations_2010} propose to use pseudo-residuals and pseudo-scatterplots for checking model assumptions. Specifically, They consider the case where $h(T)=1(T>t_0)\in \{0,1\}$ and $\theta=E[h(t)]=S_T(t_0)$. They then calculate the pseudo-observations $\hat{\theta}_i$, and treat them as the outcome. Denote by $\hat{S}_T(t_0|Z_i)$ the predicted value that is based on the model. The raw pseudo-residual are $\hat{\theta}_i-\hat{S}_T(t_0|Z_i)$, and the standardized 
	(Pearson) residuals are
	\[
	\frac{\hat{\theta}_i-\hat{S}_T(t_0|Z_i)}{\sqrt{\hat{S}_T(t_0|Z_i)(1-\hat{S}_T(t_0|Z_i))}}.
	\]
	
	For the bivariate setting, consider the function $h_3(T_1,T_2)=1(T_1>t_1^0,T_2>t_2^0)\in \{0,1\}$ and $\theta_3=E[h_3(T_1,T_2)]=S(t_1^0,t_2^0)$. For the pseudo-observations $\hat{\theta}_{3_i}$ and the predicted probabilities $\hat{S}(t_1^0,t_2^0|Z_i)$, the raw pseudo-residual are $\hat{\theta}_{3_i}-\hat{S}(t_1^0,t_2^0|Z_i)$, and the standardized 
	(Pearson) residuals are
	\[
	\frac{\hat{\theta}_{3_i}-\hat{S}(t_1^0,t_2^0|Z_i)}{\sqrt{\hat{S}(t_1^0,t_2^0|Z_i)(1-\hat{S}(t_1^0,t_2^0|Z_i))}}.
	\]
	We shall use these residuals as a graphical diagnostic tool to evaluate the model's fit. If the model fits the data well, no trends should be seen in the residuals when we plot
	them against a covariate. In practice, we plot the residuals
	against the covariate at a few chosen time points $(t_1^0,t_2^0)\in\{(t_1^1,t_2^1),\ldots,(t_1^K,t_2^K)\}$ (in which case the pseudo-values are $K$-dimensional). Finally, for the simple Lehmann model, one can also define pseudo-scatterplots where the cloglog link-transformed pseudo-values and predicted survival probabilities are both plotted against a covariate value. However, for the generalized Lehmann model there is no single link function that when applied to the pseudo-values should return a linear combination of the covariate. 
	
	\section{Simulation Design and Additional Results}\label{app:G}
	\subsection{Detailed description of simulation settings}
	We consider four data-generating mechanisms (DGMs) under the generalized Lehmann model.
	Specifically, the baseline bivariate survival function is specified via the following copulas: Frank (NQD), Frank (PQD), Clayton (PQD), and Gumbel–Barnett (NQD; bivariate exponential). The four DGMs differ primarily in their baseline dependence structure, covering both negative and positive dependence induced by Frank, Clayton, and Gumbel–Barnett copulas. 
	Across all DGMs, the marginal baseline distributions are unit exponentials, censoring times follow a univariate exponential distribution with rate $\lambda=0.3$, and a single covariate $Z$ is drawn from a $\mathrm{Uniform}(0,1)$ distribution.

	To generate bivariate failure times from the generalized Lehmann model in \eqref{eq:overall_survival},
	we first sample $T_1\sim F_{T_1}(t_1\mid Z)$
	from the marginal cumulative distribution function of $T_1$. We then sample $T_2$ conditional on $T_1=t_1$ using the conditional cumulative distribution function $F_{T_2|T_1=t_1}(\cdot\mid Z)$. For all DGMs, we set the regression parameters to $\alpha_1=1$, $\beta_1=0.7$, and $\gamma_1=0.3$, corresponding to the three slope parameters in our generalized Lehmann model. 
	
	In DGM 1 (Frank NQD), the baseline bivariate distribution is given by the Frank copula
	$$C_\eta(u,v)
	=
	-\frac{1}{\eta}
	\ln\left(
	1 + \frac{(e^{-\eta u}-1)(e^{-\eta v}-1)}{e^{-\eta}-1}
	\right),
	\quad u,v \in [0,1], \ \eta \neq 0,$$
	with copula parameter $\eta=-5$, corresponding to moderate negative baseline dependence ($\tau=-0.46$). Covariate-adjusted bivariate failure times are then generated using the inverse transform method.
	Independent exponential censoring is applied to each margin, yielding average censoring proportions of 16\% for $T_1$ and 17\% for $T_2$. As a result, about 74\% of observations are fully observed (i.e., both event times are uncensored).
	
	In DGM 2 (Frank PQD), the baseline distribution is defined by the Frank copula with parameter $\eta=5$, corresponding to moderate positive baseline dependence ($\tau=0.46$). Covariate-adjusted bivariate failure times are then generated using the inverse transform method.
	Independent exponential censoring is applied to each margin, yielding similar marginal censoring proportions to DGM 1. As a result, about 78\% of observations are fully observed. 
	For both DGM 1 and DGM 2, pseudo-observations are computed at six pre-specified bivariate time points,
	\[
	(t_1,t_2)\in\{(0.5,0.6),(0.5,0.7),(0.5,0.8),(0.7,0.6),(0.7,0.7),(0.7,0.8)\}.
	\] 
	This fixed time point selection approach was found to work well in the pseudo-observations literature, for both univariate \citep{furberg_bivariate_2023} and bivariate \cite{travis2024pseudo} settings. Since both DGMs share the same marginal distributions, the marginal event proportions by time are similar across settings, with differences arising only from sampling variability. In contrast, the joint event proportions differ due to the underlying dependence structure: for DGM 1 (Frank NQD), the proportion of double events ranges from 0.22 to 0.35, whereas for DGM 2 (Frank PQD), it ranges from 0.35 to 0.46.

	In DGM 3 (Clayton PQD), the baseline bivariate distribution is given by the Clayton copula
	$$
	C_\eta(u,v)
	=
	\left(u^{(1-\eta)} + v^{(1-\eta)} - 1\right)^{1/(1-\eta)},
	\quad u,v \in [0,1], \ \eta > 1.
	$$
	with copula parameter $\eta=3$, corresponding to moderate positive baseline dependence ($\tau=0.5$). Covariate-adjusted bivariate failure times are then generated using the inverse transform method. As in the previous settings, independent exponential censoring is applied to each margin, yielding similar marginal censoring proportions. Consequently, approximately 77\% of observations are fully observed.
	
	In DGM 4 (Gumbel--Barnett NQD), the baseline bivariate distribution is given by the Gumbel--Barnett copula
	$$
	C_\eta(u,v)
	=
	uv\exp\{-\eta \log(u)\log(v)\},
	\quad u,v \in [0,1], \ 0 \le \eta \le 1,
	$$
	with copula parameter $\eta = 1$, corresponding to moderate negative baseline dependence ($\tau = -0.36$).The corresponding baseline survival function under unit exponential margins is
	$$
	S^0(t_1,t_2)=\exp\big(-(t_1+t_2+\eta t_1 t_2)\big),
	$$
	which for $\eta=1$ reduces to $S^0(t_1,t_2)=\exp(-(t_1+t_2+t_1 t_2))$, a commonly used bivariate exponential model. Covariate-adjusted bivariate failure times are then generated using the inverse transform method. As in the previous settings, independent exponential censoring is applied to each margin, yielding similar marginal censoring proportions. As a result, approximately 72\% of observations are fully observed (i.e., both event times are uncensored).
	
	For both DGM 3 and DGM 4, pseudo-observations are computed at six pre-specified bivariate time points,
	$$
	\{(0.5,0.3),(0.5,0.4),(0.5,0.5),(0.7,0.3),(0.7,0.4),(0.7,0.5)\}.
	$$
	Since both DGMs share the same marginal distributions, the marginal event proportions by time are similar across settings, with differences arising only from sampling variability. In contrast, the joint event proportions differ due to the underlying dependence structure: for DGM 3 (Clayton PQD), the proportion of double events ranges from 0.21 to 0.38, whereas for DGM 4 (Gumbel--Barnett NQD), it ranges from 0.14 to 0.24.

	In addition to the main analyses, we consider two supplementary estimation settings based on DGM 4 (Gumbel--Barnett NQD) under the simple Lehmann model, in which pseudo-observations are computed at four bivariate time points,
	$$
	\{(0.3,0.3), (0.5,0.3),(0.3,0.5),(0.5,0.5)\},
	$$
	thereby reducing the dimensionality of the estimating equations. For this reduced time grid, we implement both an unconstrained fit, in which all marginal parameters are estimated separately, and an exchangeable version, in which equality constraints are imposed across selected parameters. Importantly, both analyses are based on the same data-generating mechanism, differing only in the imposed parameter constraints at the estimation stage. These additional settings allow us to assess the impact of model parsimony on estimation accuracy and variability. 
	
	\subsection{Simulation diagnostics and filtering criteria}\label{sec:filtering}
	We considered two sample sizes $n\in \{400,800\}$, with 500 replications for each. Simulation runs exhibiting numerical instability were excluded.
	We first removed runs with extremely large variance estimates ($>1000$) or parameter estimates (absolute value $>100$), as these indicated model non-convergence. Among the remaining runs, additional outliers were excluded if either the estimate of the slope parameter $\gamma_1$ or its variance deviated by more than four standard deviations from the sample-size--specific mean. Overall, these exclusions corresponded to approximately 10 (2\%) and 5 (1\%) replications for $n=400$ and $n=800$, respectively, on average across DGMs.
	
	\subsection{Interpretation of intercept parameters}\label{sec:app_intercept_params}
	For each DGM, the first intercept parameter $\tilde{\gamma}_0^1$ corresponds to the link-transformed value of the baseline quantity
	\[
	\exp\left\{\int_{0}^{t_1^0}\int_{0}^{t_2^0} A^0(du,dv)\right\}
	=
	\frac{S^0(t_1^0,t_2^0)}{S_{T_1}^0(t_1^0)\,S_{T_2}^0(t_2^0)},
	\]
	evaluated at the first bivariate time point $(t_1^0,t_2^0)=(t_1^1,t_2^1)$. For NQD baseline distributions, this yields
	\[
	\tilde{\gamma}_0^1 = \log\!\left(-\log\!\left(\frac{S^0(t_1^1,t_2^1)}{S_{T_1}^0(t_1^1)\,S_{T_2}^0(t_2^1)}\right)\right),
	\]
	whereas for PQD distributions,
	\[
	\tilde{\gamma}_0^1 = \log\!\left(\log\!\left(\frac{S^0(t_1^1,t_2^1)}{S_{T_1}^0(t_1^1)\,S_{T_2}^0(t_2^1)}\right)\right).
	\]
	The remaining parameters $\tilde{\gamma}_0^2, \ldots, \tilde{\gamma}_0^6$ represent contrasts relative to this reference level across the remaining bivariate time points.

	\subsection{Prediction accuracy and MAE}
	For each time point and Monte Carlo replication, we estimated the joint survival probability $\hat{S}(t_1^j,t_2^j\mid Z_i)$ by plugging the parameter estimates into Equation~\eqref{eq:overall_survival}, and compared it to the corresponding true survival probability. We then computed the average absolute difference between $\hat{S}(t_1^j,t_2^j\mid Z_i)$ and $S(t_1^j,t_2^j\mid Z_i)$ across all covariate values $Z_i$, $i=1,\ldots,n$. That is, for each time point and replication, we report the mean absolute error (MAE),
	$$
	{\rm MAE}=\frac{1}{n}\sum_{i=1}^{n}\left|\hat{S}(t_1^j,t_2^j\mid Z_i)-S(t_1^j,t_2^j\mid Z_i)\right|.
	$$
	
\begin{figure}[h]
		\centering
		
		\subfloat[]{%
			\includegraphics[width=0.95\textwidth]{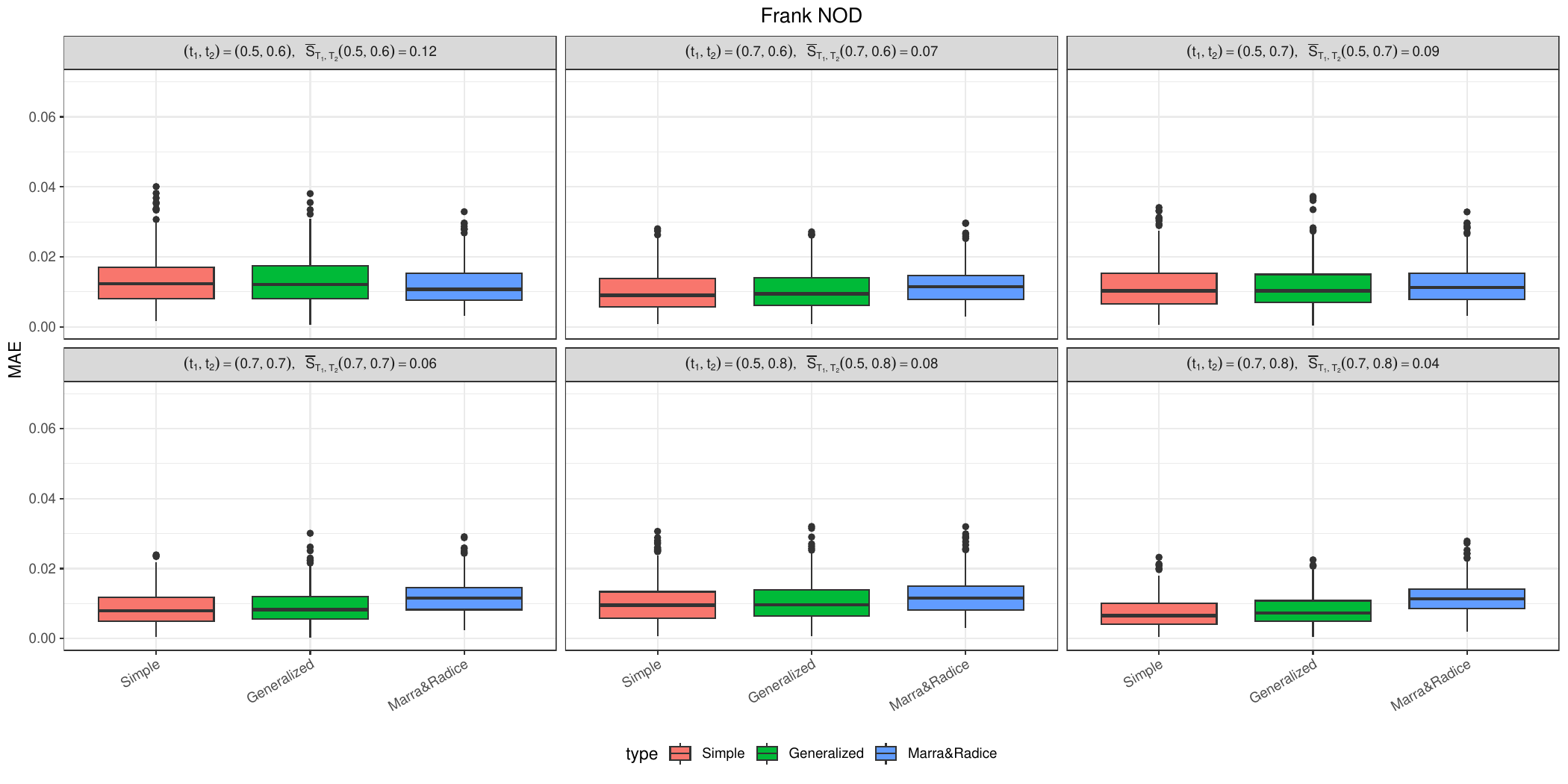}%
		}
		
		\vspace{0.5em}
		
		\subfloat[]{%
			\includegraphics[width=0.95\textwidth]{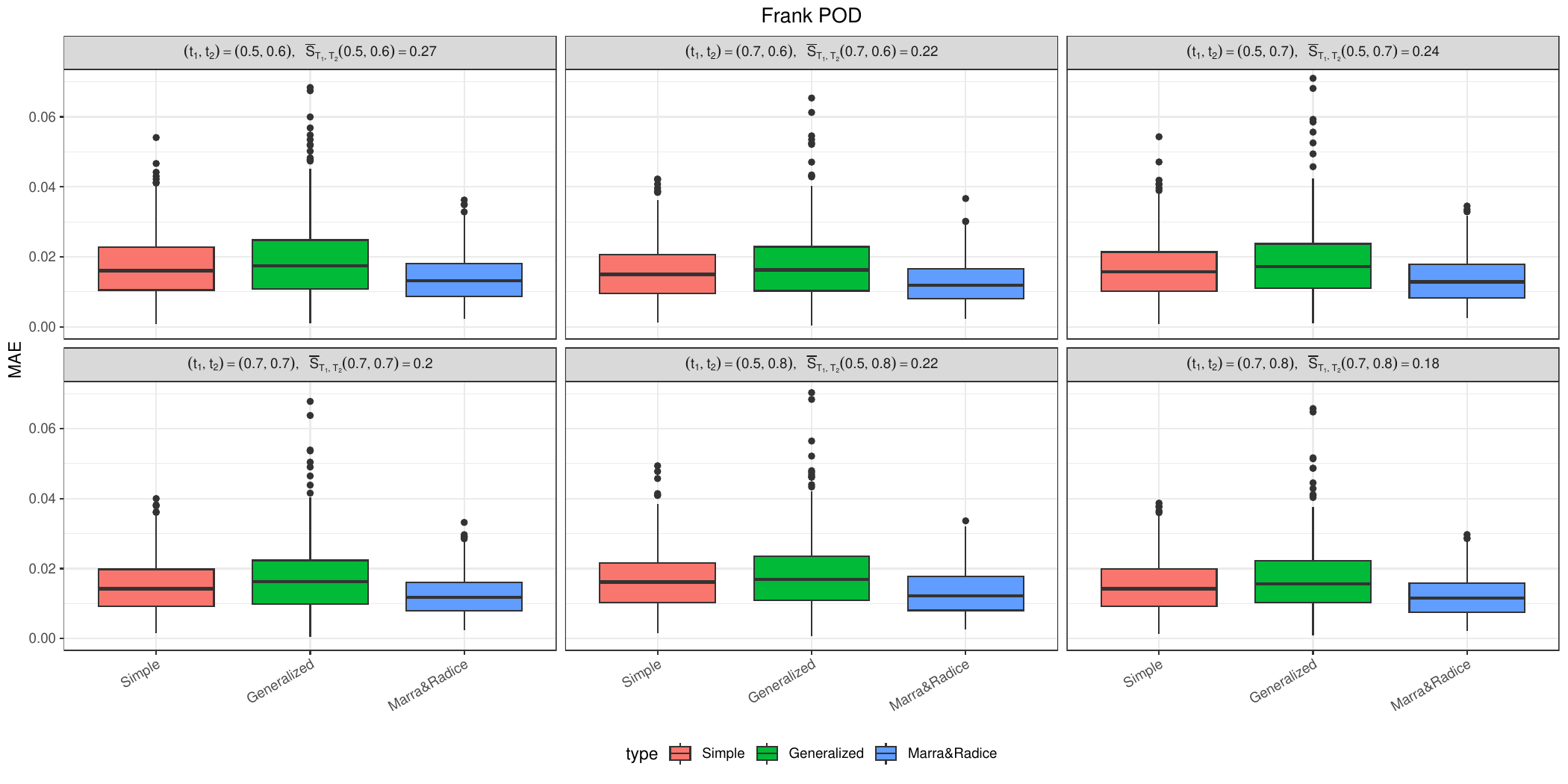}%
		}
		
		\caption{Comparison of MAEs for the simple Lehmann model (red), the generalized Lehmann model (green), and the copula-based approach of \citealt{marra_copula_2020} (blue). Results are shown for DGM~1 (Frank NQD, top panel) and DGM~2 (Frank PQD, bottom panel), with sample size $n=800$ and $m=500$ Monte Carlo replications. MAEs are computed between the true bivariate survival probability and its estimate at the six time points $\{(0.5,0.6),(0.5,0.7),(0.5,0.8),(0.7,0.6),(0.7,0.7),(0.7,0.8)\}$. The strip labels indicate the time point $(t_1^j,t_2^j)$, $j=1,\ldots,6$, along with the corresponding average true joint survival probability $\bar{S}(t_1^j,t_2^j)=\frac{1}{nm}\sum_{i=1}^{n}\sum_{l=1}^{m} S(t_1^j,t_2^j \mid Z_i)$.}
		\label{fig:MAEs}
		
	\end{figure}

	Figure~\ref{fig:MAEs} presents boxplots of the MAEs for DGM~1 (Frank NQD, top panel) and DGM~2 (Frank PQD, bottom panel), with sample size 
	n=800. Results are shown for the simple and generalized Lehmann models, as well as for the copula-based approach of \cite{marra_copula_2020}.
	
	\subsection{Additional simulation results}
	Table~\ref{tab:sim_marginal} presents the regression estimates for the marginal parameters in the three benchmark scenarios: Frank NQD, Frank PQD, and Clayton PQD, and for sample sizes $n=400,800$. The intercept parameters $\alpha_0^j$ and $\beta_0^j$ correspond to the link-transformed baseline marginal survival functions, given by $\log\!\left(-\log\!\big(S_{T_1}(t_1^j)\big)\right)$ and $\log\!\left(-\log\!\big(S_{T_2}(t_2^j)\big)\right)$, respectively. As can be seen, the univariate pseudo-observations approach accurately recovers the marginal regression parameters.
	
	\begin{table}[ht]
		\centering
		\small
		\caption{Simulation results for the marginal parameters in the three benchmark scenarios. For each parameter, the table reports the true value, Monte Carlo mean, median, and empirical standard deviation (SD), for sample sizes \(n=400\) and \(n=800\).}
		\label{tab:sim_marginal}
		\begin{tabular}{llrrrrr}
			\toprule
			DGM & $n$ & Parameter & True & Mean & Median & SD \\
			\midrule
			
			\multirow{14}{*}{Frank NQD}
			& \multirow{7}{*}{400}
			& $\alpha_1$      &  1.00 &  0.99 &  0.99 & 0.25 \\
			& & $\beta_1$     &  0.70 &  0.73 &  0.73 & 0.24 \\
			& & $\alpha_0^1$& -0.69 & -0.69 & -0.69 & 0.15 \\
			& & $\alpha_0^2$& -0.36 & -0.35 & -0.35 & 0.15 \\
			& & $\beta_0^1$ & -0.51 & -0.53 & -0.53 & 0.15 \\
			& & $\beta_0^2$ & -0.36 & -0.37 & -0.38 & 0.14 \\
			& & $\beta_0^3$ & -0.22 & -0.24 & -0.24 & 0.14 \\
			\cmidrule(lr){2-7}
			& \multirow{7}{*}{800}
			& $\alpha_1$      &  1.00 &  1.00 &  1.00 & 0.16 \\
			& & $\beta_1$     &  0.70 &  0.71 &  0.71 & 0.16 \\
			& & $\alpha_0^1$& -0.69 & -0.69 & -0.69 & 0.10 \\
			& & $\alpha_0^2$& -0.36 & -0.35 & -0.36 & 0.10 \\
			& & $\beta_0^1$ & -0.51 & -0.52 & -0.52 & 0.10 \\
			& & $\beta_0^2$ & -0.36 & -0.36 & -0.36 & 0.10 \\
			& & $\beta_0^3$ & -0.22 & -0.23 & -0.23 & 0.10 \\
			
			\midrule
			
			\multirow{14}{*}{Frank PQD}
			& \multirow{7}{*}{400}
			& $\alpha_1$      &  1.00 &  0.99 &  0.99 & 0.25 \\
			& & $\beta_1$     &  0.70 &  0.70 &  0.71 & 0.24 \\
			& & $\alpha_0^1$& -0.69 & -0.69 & -0.69 & 0.15 \\
			& & $\alpha_0^2$& -0.36 & -0.35 & -0.35 & 0.15 \\
			& & $\beta_0^1$ & -0.51 & -0.52 & -0.51 & 0.15 \\
			& & $\beta_0^2$ & -0.36 & -0.36 & -0.36 & 0.15 \\
			& & $\beta_0^3$ & -0.22 & -0.23 & -0.23 & 0.15 \\
			\cmidrule(lr){2-7}
			& \multirow{7}{*}{800}
			& $\alpha_1$      &  1.00 &  1.00 &  1.00 & 0.16 \\
			& & $\beta_1$     &  0.70 &  0.71 &  0.71 & 0.17 \\
			& & $\alpha_0^1$& -0.69 & -0.70 & -0.69 & 0.10 \\
			& & $\alpha_0^2$& -0.36 & -0.36 & -0.36 & 0.10 \\
			& & $\beta_0^1$ & -0.51 & -0.52 & -0.52 & 0.10 \\
			& & $\beta_0^2$ & -0.36 & -0.36 & -0.36 & 0.10 \\
			& & $\beta_0^3$ & -0.22 & -0.23 & -0.23 & 0.10 \\
			
			\midrule
			
			\multirow{14}{*}{Clayton PQD}
			& \multirow{7}{*}{400}
			& $\alpha_1$      &  1.00 &  0.99 &  0.98 & 0.25 \\
			& & $\beta_1$     &  0.70 &  0.71 &  0.71 & 0.27 \\
			& & $\alpha_0^1$& -0.69 & -0.69 & -0.69 & 0.15 \\
			& & $\alpha_0^2$& -0.36 & -0.35 & -0.35 & 0.15 \\
			& & $\beta_0^1$ & -1.20 & -1.22 & -1.22 & 0.17 \\
			& & $\beta_0^2$ & -0.92 & -0.93 & -0.93 & 0.17 \\
			& & $\beta_0^3$ & -0.69 & -0.71 & -0.71 & 0.16 \\
			\cmidrule(lr){2-7}
			& \multirow{7}{*}{800}
			& $\alpha_1$      &  1.00 &  1.00 &  1.00 & 0.16 \\
			& & $\beta_1$     &  0.70 &  0.71 &  0.70 & 0.18 \\
			& & $\alpha_0^1$& -0.69 & -0.70 & -0.69 & 0.10 \\
			& & $\alpha_0^2$& -0.36 & -0.36 & -0.36 & 0.10 \\
			& & $\beta_0^1$ & -1.20 & -1.21 & -1.21 & 0.12 \\
			& & $\beta_0^2$ & -0.92 & -0.93 & -0.92 & 0.12 \\
			& & $\beta_0^3$ & -0.69 & -0.70 & -0.70 & 0.11 \\
			
			\bottomrule
		\end{tabular}
	\end{table}
	
	Table~\ref{tab:sim_main_n400} presents the main benchmark scenarios for sample size $n=400$. Compared to Table~\ref{tab:sim_main_n800}, this smaller sample size exhibits higher bias and reduced coverage, highlighting the importance of sufficiently large samples for stable estimation.
	\begin{table}[ht]
		\centering
		\small
		\caption{Simulation results for the main benchmark scenarios with sample size $n=400$. For each parameter, the table reports the true value, Monte Carlo mean, median, empirical standard deviation (SD), average estimated standard error for the two-step approach (SE), and empirical coverage of the two-step 95\% confidence interval.}
		\label{tab:sim_main_n400}
		\begin{tabular}{llrrrrrr}
			\hline
			DGM & Parameter & True & Mean & Median & SD & SE & Cov. \\
			\hline
			\multirow{7}{*}{Frank NQD}
			& $\tilde \gamma_0^1$ & -0.84 & -0.92 & -0.89 & 0.47 & 0.44 & 0.852 \\
			& $\tilde \gamma_0^2$ &  0.36 &  0.39 &  0.38 & 0.23 & 0.23 & 0.860 \\
			& $\tilde \gamma_0^3$ &  0.17 &  0.17 &  0.17 & 0.14 & 0.15 & 0.893 \\
			& $\tilde \gamma_0^4$ &  0.52 &  0.56 &  0.55 & 0.26 & 0.26 & 0.868 \\
			& $\tilde \gamma_0^5$ &  0.31 &  0.32 &  0.31 & 0.20 & 0.21 & 0.881 \\
			& $\tilde \gamma_0^6$ &  0.65 &  0.69 &  0.68 & 0.30 & 0.29 & 0.864 \\
			& $\gamma_1$   &  0.30 &  0.37 &  0.41 & 0.95 & 0.92 & 0.836 \\
			\hline
			\multirow{7}{*}{Frank PQD}
			& $\tilde \gamma_0^1$ & -1.17 & -1.49 & -1.15 & 1.34 & 1.21 & 0.831 \\
			& $\tilde \gamma_0^2$ &  0.20 &  0.19 &  0.21 & 0.19 & 0.43 & 0.872 \\
			& $\tilde \gamma_0^3$ &  0.08 &  0.08 &  0.08 & 0.10 & 0.11 & 0.856 \\
			& $\tilde \gamma_0^4$ &  0.29 &  0.30 &  0.30 & 0.20 & 0.23 & 0.864 \\
			& $\tilde \gamma_0^5$ &  0.14 &  0.14 &  0.15 & 0.18 & 0.23 & 0.868 \\
			& $\tilde \gamma_0^6$ &  0.36 &  0.37 &  0.36 & 0.23 & 0.32 & 0.858 \\
			& $\gamma_1$   &  0.30 &  0.60 &  0.23 & 1.63 & 1.44 & 0.840 \\
			\hline
			\multirow{7}{*}{Clayton PQD}
			& $\tilde \gamma_0^1$ & -1.78 & -2.15 & -1.82 & 1.57 & 1.45 & 0.842 \\
			& $\tilde \gamma_0^2$ &  0.21 &  0.23 &  0.22 & 0.23 & 0.26 & 0.859 \\
			& $\tilde \gamma_0^3$ &  0.24 &  0.24 &  0.24 & 0.14 & 0.15 & 0.881 \\
			& $\tilde \gamma_0^4$ &  0.47 &  0.48 &  0.47 & 0.22 & 0.25 & 0.885 \\
			& $\tilde \gamma_0^5$ &  0.42 &  0.42 &  0.42 & 0.18 & 0.19 & 0.861 \\
			& $\tilde \gamma_0^6$ &  0.66 &  0.66 &  0.66 & 0.24 & 0.26 & 0.887 \\
			& $\gamma_1$   &  0.30 &  0.66 &  0.33 & 1.89 & 1.69 & 0.857 \\
			\hline
		\end{tabular}
	\end{table}

	Tables~\ref{tab:sim_exp2D_n400} and \ref{tab:sim_exp2D_n800} present the simulation results for the Gumbel--Barnett scenarios with sample sizes $n=400$ and $n=800$, respectively. These scenarios are based on a baseline bivariate survival function induced by the Gumbel--Barnett copula and are considered under three modeling frameworks: (i) the generalized Lehmann model, (ii) the simple Lehmann model, and (iii) the simple Lehmann model analyzed under an exchangeability constraint. The exchangeable analysis imposes equality constraints on parameters that are estimated separately in the unconstrained setting. Importantly, scenarios (ii) and (iii) are generated from the simple Lehmann model, which is a special case of the generalized Lehmann model. This allows us to directly assess the performance of the simple model under both unconstrained and constrained estimation.
	
	Overall, the results are consistent with those reported in Tables~\ref{tab:sim_main_n400} and \ref{tab:sim_main_n800}. In particular, the smaller sample size ($n=400$) exhibits larger variability and somewhat reduced coverage, whereas for $n=800$ the estimators show good finite-sample performance, with small bias and coverage close to the nominal level.
	
	Table~\ref{tab:sim_exp2D_marginal} summarizes the estimation of the marginal regression parameters for the Gumbel--Barnett scenarios. The univariate pseudo-observations approach accurately recovers these parameters across all scenarios, with bias decreasing and variability shrinking as the sample size increases.
	
	Table~\ref{tab:sim_exp2D_simple} presents the results for the simple Lehmann model under both the unconstrained and exchangeable analyses in scenarios (ii) and (iii) of the Gumbel--Barnett scenarios, where the simple model is correctly specified. In both cases, the regression parameters are estimated accurately, and coverage probabilities are close to the nominal level. The exchangeable analysis yields comparable performance to the unconstrained approach, indicating that the latter can be reliably used even when exchangeability is plausible, thereby avoiding the need to impose additional structural assumptions.
	
	%Tables~\ref{tab:sim_exp2D_n400} and \ref{tab:sim_exp2D_n800} present the simulation results for the Gumbel--Barnett scenarios with sample sizes $n=400$ and $n=800$, respectively. The three Gumbel--Barnett scenarios correspond to a baseline bivariate survival function from the Gumbel--Barnett copula used in: (1) the generalized Lehmann model, (2) the simple Lehmann model, and (3) the simple Lehmann model analyzed under the exchangeability constraint. This exchangeable analysis imposes equality constraints on parameters that are estimated separately under the unconstrained analysis. The results are similar to those reported in Tables~\ref{tab:sim_main_n400} and \ref{tab:sim_main_n800}, where the smaller sample size ($n=400$) exhibits higher bias and reduced coverage, and the larger sample size ($n=800$) corresponds to good estimation performance.
	%Table~\ref{tab:sim_exp2D_marginal} presents the regression estimates for the marginal parameters in these three Gumbel--Barnett scenarios, for sample sizes $n=400,800$. As can be seen, the univariate pseudo-observations approach accurately recovers the marginal regression parameters. Finally, Table X presents the results for the simple Lehmann model under both the constrained exchangeable analysis and its unconstrained version. Importantly, all the constrained exchangeable analyses and their unconstrained versions accurately estimate the regression parameters, showing that the unconstrained fits can be used also in exchangeable settings, thereby further reducing the assumptions that need to be made on the data.

	\begin{table}[ht]
		\centering
		\small
		\caption{Simulation results for the Gumbel--Barnett scenarios with sample size $n=400$. For each parameter, the table reports the true value, Monte Carlo mean, median, empirical standard deviation (SD), average estimated standard error for the two-step approach (SE), and empirical coverage of the two-step 95\% confidence interval.}
		\label{tab:sim_exp2D_n400}
		\begin{tabular}{llrrrrrr}
			\hline
			DGM & Parameter & True & Mean & Median & SD & SE & Cov. \\
			\hline
			
			\multirow{7}{*}{Generalized}
			& $\tilde \gamma_0^1$ & -1.90 & -2.13 & -1.95 & 0.87 & 0.79 & 0.812 \\
			& $\tilde \gamma_0^2$ &  0.34 &  0.36 &  0.35 & 0.40 & 1.31 & 0.870 \\
			& $\tilde \gamma_0^3$ &  0.29 &  0.31 &  0.30 & 0.27 & 0.32 & 0.852 \\
			& $\tilde \gamma_0^4$ &  0.62 &  0.69 &  0.65 & 0.41 & 0.44 & 0.845 \\
			& $\tilde \gamma_0^5$ &  0.51 &  0.54 &  0.52 & 0.34 & 0.38 & 0.849 \\
			& $\tilde \gamma_0^6$ &  0.85 &  0.92 &  0.88 & 0.45 & 0.47 & 0.841 \\
			& $\gamma_1$          &  0.30 &  0.44 &  0.42 & 1.46 & 1.43 & 0.858 \\
			
			\hline
			
			\multirow{5}{*}{Simple (unconstrained)}
			& $\tilde \gamma_0^1$ & -2.41 & -2.66 & -2.51 & 0.83 & 0.76 & 0.814 \\
			& $\tilde \gamma_0^2$ &  0.51 &  0.52 &  0.51 & 0.33 & 0.38 & 0.867 \\
			& $\tilde \gamma_0^3$ &  0.51 &  0.54 &  0.52 & 0.32 & 0.37 & 0.857 \\
			& $\tilde \gamma_0^4$ &  1.02 &  1.07 &  1.04 & 0.39 & 0.44 & 0.878 \\
			& $\gamma_1$          &  1.00 &  1.23 &  1.14 & 1.21 & 1.15 & 0.837 \\
			
			\hline
			
			\multirow{4}{*}{Simple (exchangeable)}
			& $\tilde \gamma_0^1$ & -2.41 & -2.62 & -2.48 & 0.81 & 0.71 & 0.795 \\
			& $\tilde \gamma_0^2$ &  0.51 &  0.54 &  0.51 & 0.24 & 0.26 & 0.854 \\
			& $\tilde \gamma_0^3$ &  1.02 &  1.06 &  1.04 & 0.38 & 0.40 & 0.864 \\
			& $\gamma_1$          &  1.00 &  1.21 &  1.14 & 1.19 & 1.11 & 0.827 \\
			
			\hline
		\end{tabular}
	\end{table}
	
	\begin{table}[ht]
		\centering
		\small
		\caption{Simulation results for the Gumbel--Barnett scenarios with sample size $n=800$. For each parameter, the table reports the true value, Monte Carlo mean, median, empirical standard deviation (SD), average estimated standard error for the two-step approach (SE), and empirical coverage of the two-step 95\% confidence interval.}
		\label{tab:sim_exp2D_n800}
		\begin{tabular}{llrrrrrr}
			\hline
			DGM & Parameter & True & Mean & Median & SD & SE & Cov. \\
			\hline
			
			\multirow{7}{*}{Generalized}
			& $\tilde \gamma_0^1$ & -1.90 & -1.99 & -1.93 & 0.49 & 0.44 & 0.942 \\
			& $\tilde \gamma_0^2$ &  0.34 &  0.35 &  0.33 & 0.23 & 0.23 & 0.950 \\
			& $\tilde \gamma_0^3$ &  0.29 &  0.29 &  0.27 & 0.17 & 0.17 & 0.952 \\
			& $\tilde \gamma_0^4$ &  0.62 &  0.64 &  0.63 & 0.26 & 0.26 & 0.950 \\
			& $\tilde \gamma_0^5$ &  0.51 &  0.51 &  0.51 & 0.20 & 0.22 & 0.962 \\
			& $\tilde \gamma_0^6$ &  0.85 &  0.87 &  0.85 & 0.28 & 0.28 & 0.968 \\
			& $\gamma_1$          &  0.30 &  0.35 &  0.29 & 0.85 & 0.87 & 0.958 \\
			
			\hline
			
			\multirow{5}{*}{Simple (unconstrained)}
			& $\tilde \gamma_0^1$ & -2.41 & -2.51 & -2.46 & 0.47 & 0.43 & 0.946 \\
			& $\tilde \gamma_0^2$ &  0.51 &  0.52 &  0.53 & 0.22 & 0.22 & 0.948 \\
			& $\tilde \gamma_0^3$ &  0.51 &  0.52 &  0.52 & 0.22 & 0.22 & 0.950 \\
			& $\tilde \gamma_0^4$ &  1.02 &  1.05 &  1.03 & 0.27 & 0.27 & 0.948 \\
			& $\gamma_1$          &  1.00 &  1.06 &  1.03 & 0.73 & 0.70 & 0.960 \\
			
			\hline
			
			\multirow{4}{*}{Simple (exchangeable)}
			& $\tilde \gamma_0^1$ & -2.41 & -2.50 & -2.44 & 0.46 & 0.40 & 0.932 \\
			& $\tilde \gamma_0^2$ &  0.51 &  0.53 &  0.51 & 0.17 & 0.16 & 0.944 \\
			& $\tilde \gamma_0^3$ &  1.02 &  1.04 &  1.03 & 0.27 & 0.26 & 0.944 \\
			& $\gamma_1$          &  1.00 &  1.06 &  1.03 & 0.72 & 0.68 & 0.958 \\
			
			\hline
		\end{tabular}
	\end{table}
	
	\begin{table}[ht]
		\centering
		\small
		\caption{Simulation results for the marginal parameters in the Gumbel--Barnett scenarios. For each parameter, the table reports the true value, Monte Carlo mean, median, and empirical standard deviation (SD), for sample sizes \(n=400\) and \(n=800\).}
		\label{tab:sim_exp2D_marginal}
		\begin{tabular}{llrrrrr}
			\toprule
			DGM & $n$ & Parameter & True & Mean & Median & SD \\
			\midrule
			
			% ================= GENERALIZED =================
			\multirow{14}{*}{Generalized}
			& \multirow{7}{*}{400}
			& $\alpha_1$      &  1.00 &  0.99 &  0.98 & 0.26 \\
			& & $\beta_1$     &  0.70 &  0.72 &  0.72 & 0.28 \\
			& & $\alpha_0^1$  & -0.69 & -0.69 & -0.69 & 0.16 \\
			& & $\alpha_0^2$  & -0.36 & -0.35 & -0.35 & 0.15 \\
			& & $\beta_0^1$   & -1.20 & -1.22 & -1.21 & 0.17 \\
			& & $\beta_0^2$   & -0.92 & -0.93 & -0.93 & 0.17 \\
			& & $\beta_0^3$   & -0.69 & -0.71 & -0.71 & 0.17 \\
			\cmidrule(lr){2-7}
			& \multirow{7}{*}{800}
			& $\alpha_1$      &  1.00 &  1.00 &  1.00 & 0.16 \\
			& & $\beta_1$     &  0.70 &  0.71 &  0.71 & 0.18 \\
			& & $\alpha_0^1$  & -0.69 & -0.70 & -0.69 & 0.10 \\
			& & $\alpha_0^2$  & -0.36 & -0.36 & -0.36 & 0.10 \\
			& & $\beta_0^1$   & -1.20 & -1.22 & -1.21 & 0.12 \\
			& & $\beta_0^2$   & -0.92 & -0.93 & -0.93 & 0.11 \\
			& & $\beta_0^3$   & -0.69 & -0.71 & -0.70 & 0.11 \\
			
			\midrule
			
			% ================= SIMPLE =================
			\multirow{12}{*}{Simple (unconstrained)}
			& \multirow{6}{*}{400}
			& $\alpha_1$      &  1.00 &  0.99 &  0.98 & 0.26 \\
			& & $\beta_1$     &  1.00 &  1.03 &  1.02 & 0.27 \\
			& & $\alpha_0^1$  & -1.20 & -1.21 & -1.20 & 0.17 \\
			& & $\alpha_0^2$  & -0.69 & -0.69 & -0.69 & 0.16 \\
			& & $\beta_0^1$   & -1.20 & -1.22 & -1.22 & 0.17 \\
			& & $\beta_0^2$   & -0.69 & -0.71 & -0.71 & 0.17 \\
			\cmidrule(lr){2-7}
			& \multirow{6}{*}{800}
			& $\alpha_1$      &  1.00 &  0.99 &  1.01 & 0.18 \\
			& & $\beta_1$     &  1.00 &  1.01 &  1.01 & 0.17 \\
			& & $\alpha_0^1$  & -1.20 & -1.20 & -1.20 & 0.11 \\
			& & $\alpha_0^2$  & -0.69 & -0.69 & -0.70 & 0.11 \\
			& & $\beta_0^1$   & -1.20 & -1.22 & -1.22 & 0.11 \\
			& & $\beta_0^2$   & -0.69 & -0.70 & -0.70 & 0.11 \\
			
			\midrule
			
			% ================= EXCHANGEABLE =================
			
			\multirow{6}{*}{Simple (exchangeable)}
			& \multirow{3}{*}{400}
			& $\beta_1$     &  1.00 &  1.01 &  1.00 & 0.16 \\
			& & $\beta_0^1$ & -1.20 & -1.21 & -1.21 & 0.10 \\
			& & $\beta_0^2$ &  -0.69 &  -0.70 &  -0.70 & 0.09 \\
			\cmidrule(lr){2-7}
			& \multirow{3}{*}{800}
			& $\beta_1$     &  1.00 &  1.00 &  1.00 & 0.11 \\
			& & $\beta_0^1$ & -1.20 & -1.21 & -1.21 & 0.07 \\
			& & $\beta_0^2$ &  -0.69 &  -0.70 &  -0.70 & 0.07 \\
			
			\bottomrule
		\end{tabular}
	\end{table}
	
	\begin{table}[ht]
		\centering
		\small
		\caption{Simulation results for the simple Lehmann model in the Gumbel--Barnett scenarios. For each parameter, the table reports the true value, Monte Carlo mean, median, empirical standard deviation (SD), model-based standard error (SE), and coverage probability, for sample sizes \(n=400\) and \(n=800\).}
		\label{tab:sim_exp2D_simple}
		\begin{tabular}{llrrrrrrr}
			\toprule
			DGM & $n$ & Parameter & True & Mean & Median & SD & SE & Cov. \\
			\midrule
			
			% ================= SIMPLE UNCONSTRAINED =================
			\multirow{10}{*}{Simple (unconstrained)}
			& \multirow{5}{*}{400}
			& $b_0^1$ & -0.37 & -0.38 & -0.37 & 0.14 & 0.13 & 0.957 \\
			& & $b_0^2$ &  0.32 &  0.32 &  0.32 & 0.05 & 0.05 & 0.943 \\
			& & $b_0^3$ &  0.32 &  0.32 &  0.32 & 0.05 & 0.05 & 0.945 \\
			& & $b_0^4$ &  0.59 &  0.60 &  0.59 & 0.06 & 0.07 & 0.955 \\
			& & $b_1$           &  1.00 &  1.02 &  1.00 & 0.22 & 0.22 & 0.951 \\
			\cmidrule(lr){2-9}
			& \multirow{5}{*}{800}
			& $b_0^1$ & -0.37 & -0.38 & -0.38 & 0.10 & 0.09 & 0.933 \\
			& & $b_0^2$ &  0.32 &  0.32 &  0.32 & 0.03 & 0.04 & 0.954 \\
			& & $b_0^3$ &  0.32 &  0.32 &  0.32 & 0.03 & 0.04 & 0.962 \\
			& & $b_0^4$ &  0.59 &  0.60 &  0.60 & 0.05 & 0.05 & 0.954 \\
			& & $b_1$           &  1.00 &  1.00 &  1.01 & 0.15 & 0.15 & 0.948 \\
			
			\midrule
			
			% ================= SIMPLE EXCHANGEABLE =================
			\multirow{8}{*}{Simple (exchangeable)}
			& \multirow{4}{*}{400}
			& $b_0^1$ & -0.37 & -0.38 & -0.37 & 0.13 & 0.13 & 0.955 \\
			& & $b_0^2$ &  0.32 &  0.32 &  0.32 & 0.03 & 0.04 & 0.947 \\
			& & $b_0^3$ &  0.59 &  0.60 &  0.59 & 0.06 & 0.07 & 0.955 \\
			& & $b_1$           &  1.00 &  1.02 &  1.00 & 0.22 & 0.22 & 0.951 \\
			\cmidrule(lr){2-9}
			& \multirow{4}{*}{800}
			& $b_0^1$ & -0.37 & -0.38 & -0.38 & 0.10 & 0.09 & 0.934 \\
			& & $b_0^2$ &  0.32 &  0.32 &  0.32 & 0.02 & 0.03 & 0.948 \\
			& & $b_0^3$ &  0.59 &  0.60 &  0.60 & 0.05 & 0.05 & 0.956 \\
			& & $b_1$           &  1.00 &  1.00 &  1.01 & 0.15 & 0.15 & 0.948 \\
			
			\bottomrule
		\end{tabular}
	\end{table}
	
	\subsection{Model misspecification}\label{app:G_miss}
	To test for the effect of model misspecification, we consider also data that are generated from a log-normal distribution in which case both Lehmann models, and the copula-based approach, are misspecified. We generate a random vector $(Z_1,\ldots,Z_n)^T$ from a uniform $U(0,1)$ distribution. We define the following two functions: $\mu_1(Z)=\gamma_1 Z$ and  $\mu_2(Z)=\gamma_2 Z$ where we use $\gamma_1=0.2$ and $\gamma_2=1$. For each $i=1,\ldots,n$ we generate a bivariate random variable from the following Gaussian distribution
	\[
	\begin{pmatrix}
		X_{1i}\\
		X_{2i}
	\end{pmatrix}
	\sim N\left(\begin{pmatrix}
		\mu_1(Z_i) \\
		\mu_2(Z_i)
	\end{pmatrix},
	\begin{pmatrix}
		\sigma_1^2 & \rho\sigma_1\sigma_2\\
		\rho\sigma_1\sigma_2 & \sigma_2^2
	\end{pmatrix}\right),
	\]
	where $\sigma_1^2=2$, $\sigma_2^2=5$, and $\rho=0.9487$, which correspond to the covariance matrix \[
	\begin{pmatrix}
		\sigma_1^2 & \rho\sigma_1\sigma_2\\
		\rho\sigma_1\sigma_2 & \sigma_2^2
	\end{pmatrix}=
	\begin{pmatrix}
		2 & 3\\
		3 & 5
	\end{pmatrix}.
	\]
	We then define our bivariate failure times as $(T_{1},T_{2})^T=(e^{X_{1}},e^{X_{2}})^T$.  Finally we added univariate exponential censoring, which correspond to about $36\%$ censoring of $T_1$, and about $46\%$ censoring of $T_2$.

\begin{figure}[h]
		\centering
		\begin{subfigure}[b]{0.95\textwidth}
			\centering
			\includegraphics[width=0.95\linewidth]{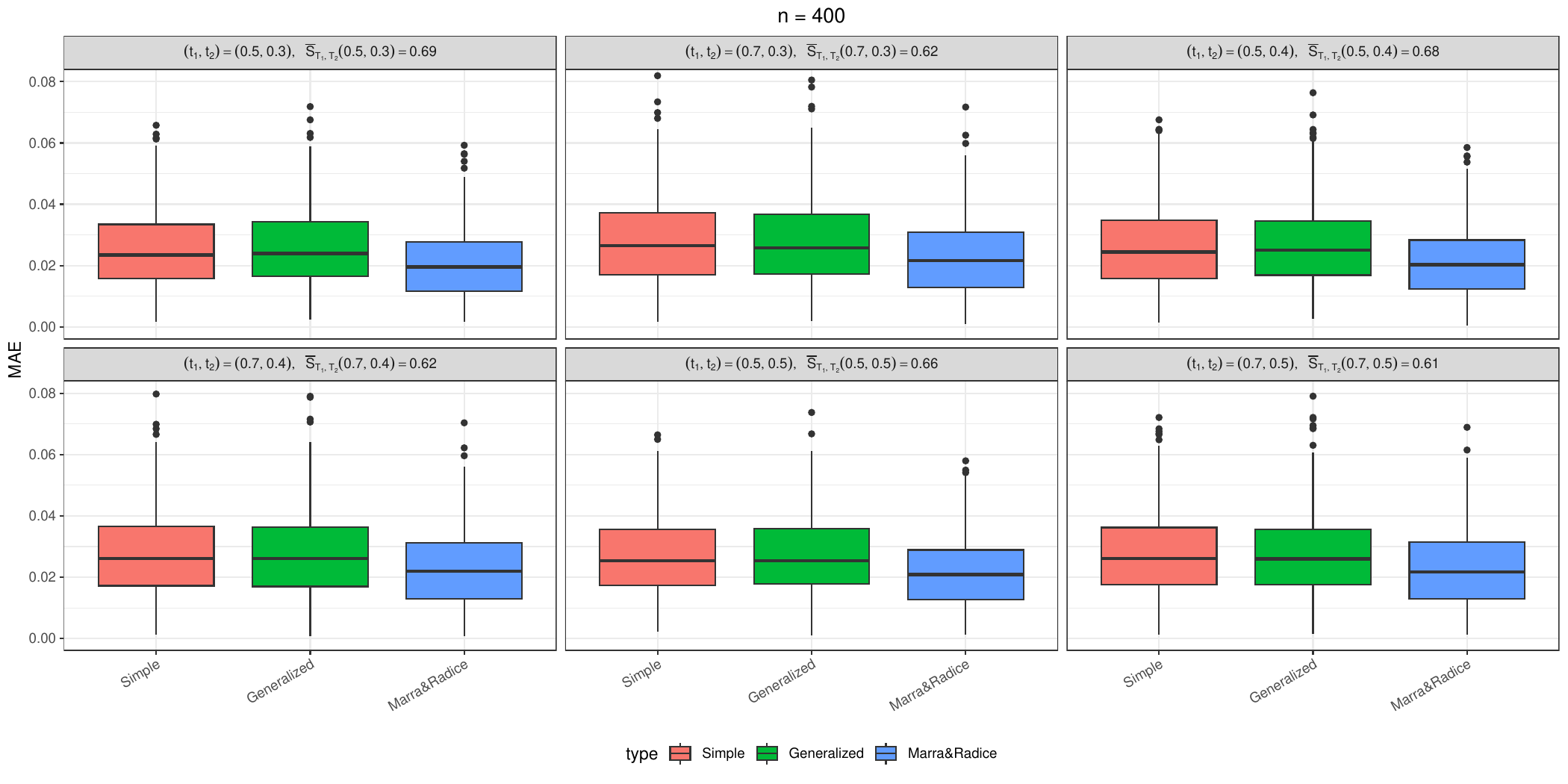}
			%				\caption{Conditional survival probability $S_{T_1| T_2\leq 36}(60\mid Z)=P(T_1>60\mid T_2 \leq 36, Z)$} \label{fig:retinopathy_1}
		\end{subfigure}
		\hfill
		\begin{subfigure}[b]{0.95\textwidth}
			\centering
			\includegraphics[width=0.95\linewidth]{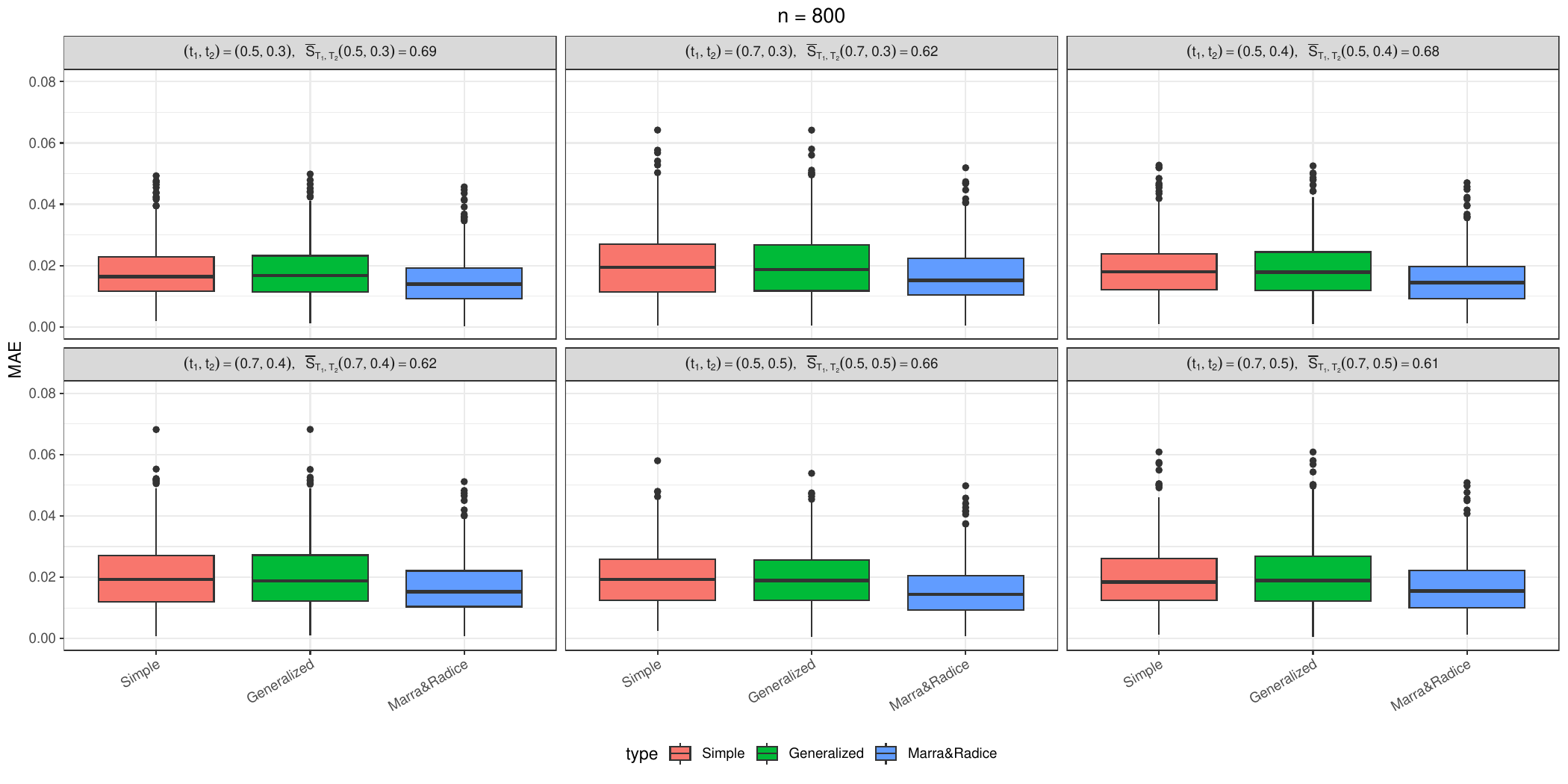}
			%				\caption{Conditional survival probability $S_{T_1| T_2> 36}(60\mid Z)=P(T_1>60\mid T_2 > 36, Z)$} \label{fig:retinopathy_2}
		\end{subfigure}
		\caption{Comparison of MAEs for the simple Lehmann model (red), the generalized Lehmann model (green), and the copula-based approach of \citealt{marra_copula_2020} (blue). Results are shown for bivariate log-normal failure times, with sample sizes $n=400$ (top panel) and $n=800$ (bottom panel), and $m=500$ Monte Carlo replications. MAEs are computed between the true bivariate survival probability and its estimate at the six time points $\{(0.5,0.3),(0.5,0.4),(0.5,0.5),(0.7,0.3),(0.7,0.4),(0.7,0.5)\}$. The strip labels indicate the time point $(t_1^j,t_2^j)$, $j=1,\ldots,6$, along with the corresponding average true joint survival probability $\bar{S}(t_1^j,t_2^j)=\frac{1}{nm}\sum_{i=1}^{n}\sum_{l=1}^{m} S(t_1^j,t_2^j \mid Z_i)$.}
		\label{fig:MAEs_LN}
	\end{figure}

	We consider the six time points $\{(0.5,0.3),(0.5,0.4),(0.5,0.5),(0.7,0.3),(0.7,0.4),(0.7,0.5)\}$. Figure~\ref{fig:MAEs_LN} presents boxplots of the MAEs for this setting, with sample sizes 
	n=400 (top) and n=800 (bottom). Results are shown for the simple and generalized Lehmann models, as well as for the copula-based approach of \cite{marra_copula_2020}, using a Gaussian copula function. As can be seen by the relatively low MAEs, all methods estimate quite well the conditional joint survival probability even under this model misspecification.

	\section{Further data analysis results}\label{app:H}
	Our analysis included the following covariates:
	
	\begin{enumerate}
		\item $Z_1$ – age at presentation (months)
		\item $Z_2$ – income level (categorical, four levels: 1 = low, 2 = lower-middle, 3 = upper-middle, 4 = high)
		\item $Z_3$ – tumor stage at presentation (categorical, three levels: $cT2$, $cT3$, $cT4$, with $cT2$ including cases originally classified as $cT1$)
		\item $Z_4$ – indicator for age at presentation over 4 years (binary)
		\item $Z_5$ – sex (binary)
		\item $Z_6$ – family history of retinoblastoma (binary)
	\end{enumerate}
	
	We also included an interaction between $Z_1$ (age) and $Z_4$ (age over 4 years) to allow the effect of age to differ for patients presenting after age 4. The full covariate vector is $Z=(Z_1,Z_2,Z_3,Z_4,Z_5,Z_6)^T$.
	
\begin{figure}[h]
		\centering
		\includegraphics[width=0.75\textwidth]{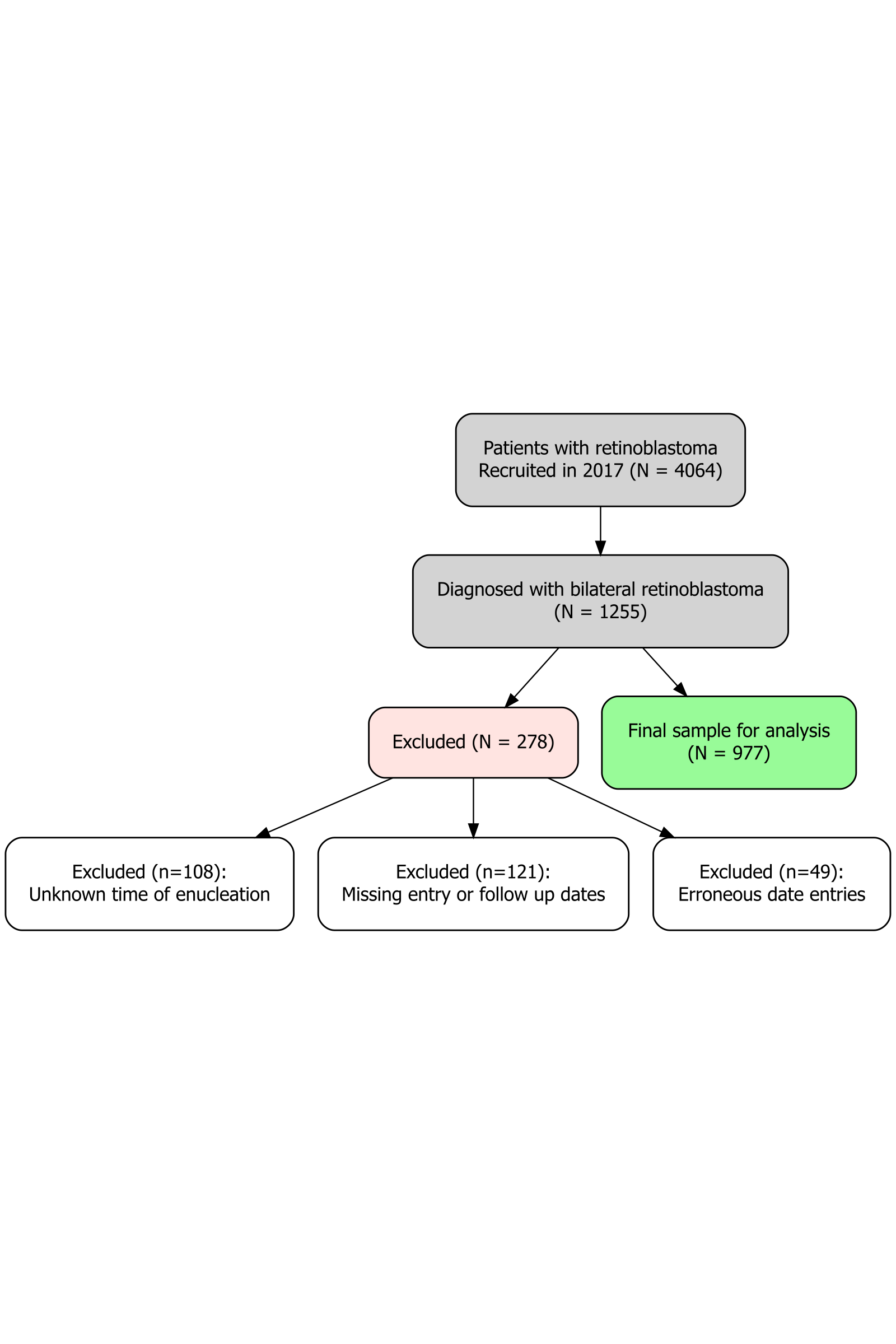}
		\caption{Flowchart of patient inclusion and exclusion.}\label{fig:flowchart.png}
	\end{figure}

	Missing covariates were imputed using the full dataset of 4,064 patients: categorical variables were replaced by the most frequent value within each economic group, and age by the group-specific median. We restricted our analysis to bilateral cases, comprising 1,255 of the 4,064 patients. After excluding patients with missing or implausible dates, the final analytic sample included 977 patients. Figure~\ref{fig:flowchart.png} summarizes the inclusion and exclusion criteria.

	\subsection{Regression estimates under exchangeability}

	\subsubsection{Exchangeability in the simple Lehmann model}
	To ensure symmetry of the baseline bivariate survival function in the simple Lehmann model, we imposed the constraint that each time point $(t_1^0,t_2^0)$ shares the same intercept as its symmetric counterpart $(t_2^0,t_1^0)$. We used the ten bivariate time points:
	$$\{(0,12), (24,0), (24,12), (12,0), (0,24), (12,24), (6,6), (12,12), (18,18), (24,24)\}.$$
	However, only seven unique intercepts were estimated because we required $b_0(0,12)=b_0(12,0)$, $b_0(0,24)=b_0(24,0)$, and $b_0(24,12)=b_0(12,24)$.
	
	Table~\ref{tab:sim_Lehmann_exchangeable} reports the regression coefficients, robust standard errors, and p-values. The first seven rows correspond to the ten bivariate time points, with the intercept at $(12,12)$ serving as the reference. Coefficients for other time points represent deviations from this baseline.
	These intercepts relate to the baseline bivariate survival function via $b_0(t_1^0,t_2^0) = \log(-\log(S^0(t_1^0,t_2^0)))$. As expected, coefficients increase with time, reflecting decreasing survival probabilities along the partial order on the bivariate time plane.
	
\begin{figure}[h]
		\centering
		
		\subfloat[Bivariate survival probabilities\label{fig:fig-a_exchange}]{%
			\includegraphics[width=0.99\textwidth]{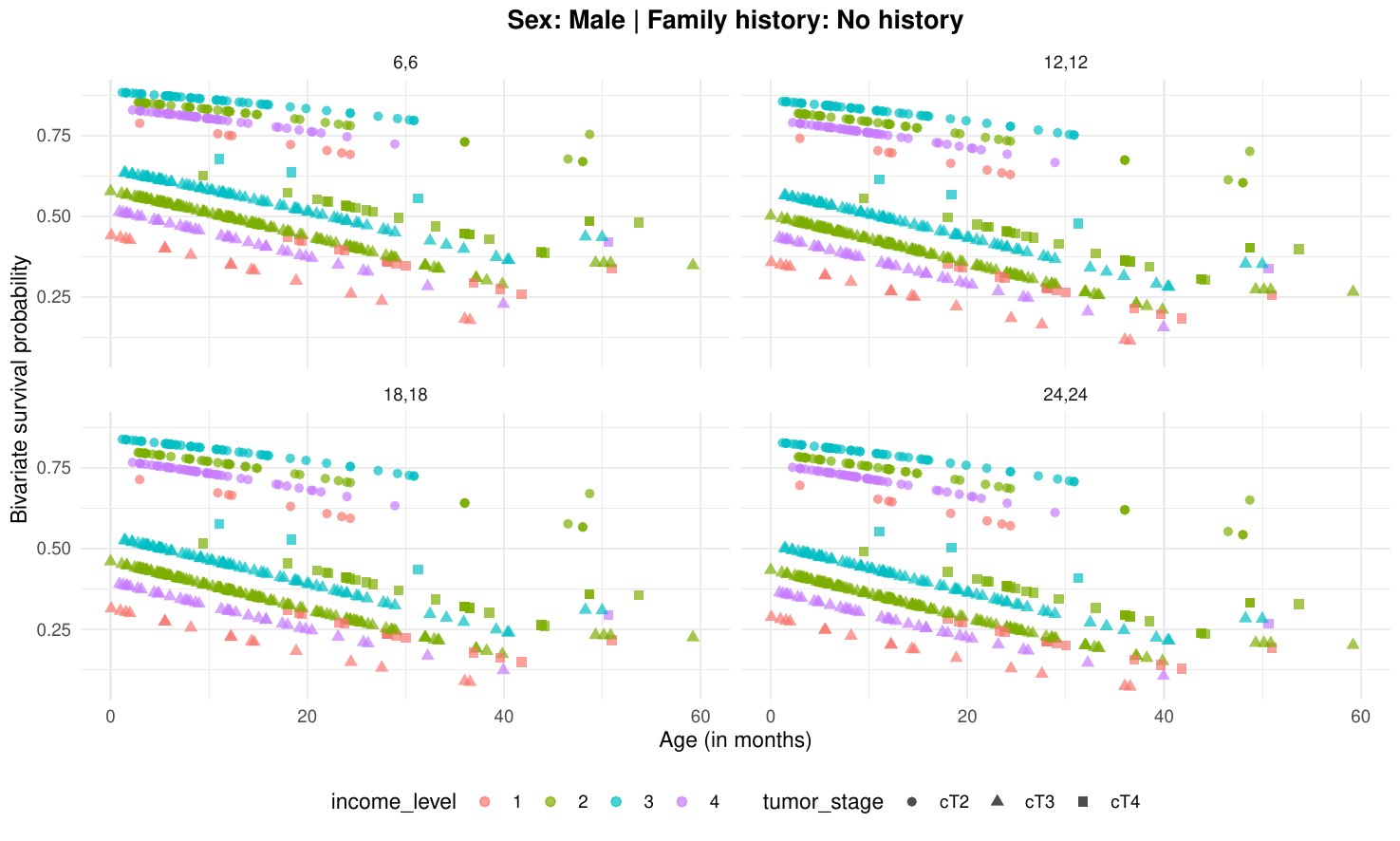}%
		}
		
		\vspace{0.5em}
		
		\subfloat[Conditional survival probabilities\label{fig:fig-cond_exchange}]{%
			\includegraphics[width=0.99\textwidth]{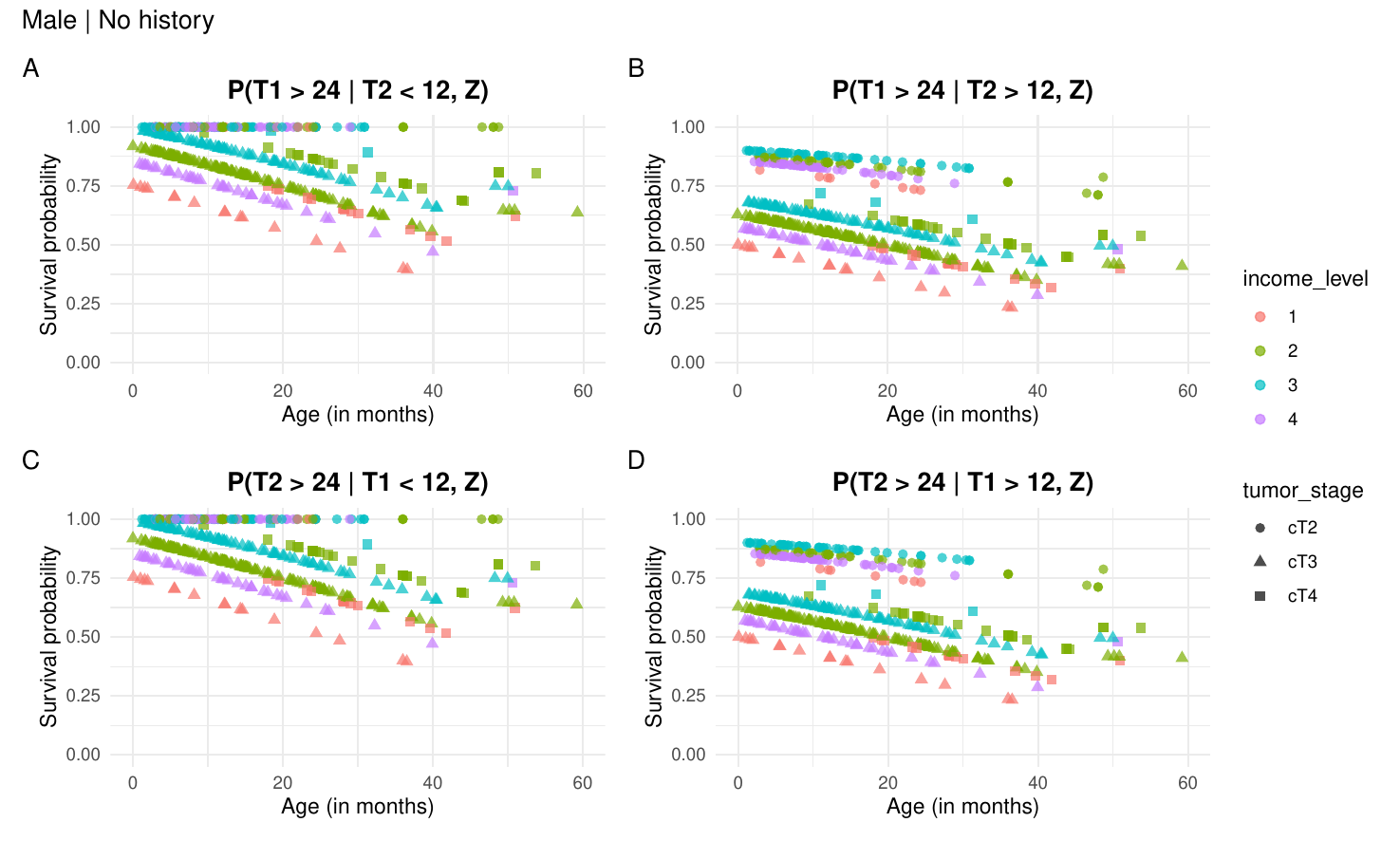}%
		}
		
		\caption{Simple Lehmann model. Top: predicted bivariate survival probabilities at 6, 12, 18, and 24 months. Bottom: predicted conditional survival probabilities. All probabilities are shown as functions of age at presentation, income level, and tumor stage for males without a family history of retinoblastoma, under the exchangeability assumption.}
		\label{fig:GROS_simple_exchange}
		
	\end{figure}

	For the covariates, a positive regression coefficient $b_j$ indicates that increasing the value of covariate $Z_j$ reduces the predicted bivariate survival probability. In particular, older age at diagnosis decreases survival probability. For patients diagnosed after age four, the interaction term with age indicates that this decrease is more gradual. Among the categorical covariates, sex and family history of retinoblastoma do not significantly affect the bivariate survival probability, whereas income level and tumor stage do. The effect of income is non-monotonic: income level 3 (upper-middle) is associated with the highest predicted survival, followed by level 2 (lower-middle), and then level 4 (high). This pattern may reflect differential access to treatment and follow-up. For tumor stage, as expected, lower stage (cT2) corresponds to the highest survival probability. Interestingly, the coefficient for cT4 is smaller than that of cT3, suggesting that an advanced tumor stage (cT4) has a higher survival probability than a medium tumor stage (cT3), which may be explained by death before enucleation and palliation by chemotherapy only. These results are generally consistent with previous studies on retinoblastoma \citep{fabian_global_2022}, which considered only univariate time to first enucleation. The effects of all covariates are also illustrated graphically in Figure~\ref{fig:fig-a_exchange}.
	%Both the intercept and slope estimates are very similar to those reported in Table~\ref{tab:simple_Lehmann_results}, and their interpretation remains unchanged.

	\begin{table}[h]
		\centering
		\caption{Exchangeable analysis. Regression estimates for the simple Lehmann model at 10 bivariate time points and seven intercepts. Positive coefficients indicate lower bivariate survival probability.}
		\begin{tabular}{lccc}
			\hline
			\textbf{Time point / Covariate} & \textbf{Estimate} & \textbf{SE} & \textbf{p-value} \\
			\hline
			$(12,12)$ (Intercept) & -1.484 & 0.239 & $<0.001$ \\
			$(18,18)$           & 0.115  & 0.021 & $<0.001$ \\
			$(24,24)$           & 0.188  & 0.027 & $<0.001$ \\
			$(24,12)$ and $(12,24)$         & 0.102  & 0.015 & $<0.001$ \\
			$(6,6)$             & -0.228 & 0.031 & $<0.001$ \\
			$(12,0)$ and $(0,12)$       & -0.834 & 0.024 & $<0.001$ \\
			$(24,0)$ and $(0,24)$        & -0.674 & 0.034 & $<0.001$ \\
			\midrule
			Age at diagnosis     & 0.020  & 0.004 & $<0.001$ \\
			Income level 2       & -0.398 & 0.143 & 0.005 \\
			Income level 3       & -0.616 & 0.165 & $<0.001$ \\
			Income level 4       & -0.224 & 0.175 & 0.196 \\
			Tumor stage cT3      & 1.294  & 0.128 & $<0.001$ \\
			Tumor stage cT4      & 0.942  & 0.194 & $<0.001$ \\
			Over 4               & 0.524  & 0.561 & 0.285 \\
			Sex (female)         & 0.044  & 0.090 & 0.624 \\
			Family history (no)  & 0.217  & 0.190 & 0.245 \\
			Age $\times$ Over 4  & -0.018 & 0.009 & 0.026 \\
			\hline
		\end{tabular}
		\label{tab:sim_Lehmann_exchangeable}
	\end{table}
	
	\subsubsection{Exchangeability in the generalized Lehmann model}
	For the generalized Lehmann model, we first fitted a single marginal Cox model using univariate pseudo-observations at the time points 
	$\{6,12,18,24\}$. Table~\ref{tab:gen_Lehmann_exchangeable_marginals} reports the regression coefficients, standard errors, and p-values. %Compared to Table~\ref{tab:gen_Lehmann_results}, this model includes exactly half the number of parameters, reflecting the imposed constraint $S_1(t\mid Z)=S_2(t\mid Z)$. As expected, each coefficient inF Table~\ref{tab:gen_Lehmann_exchangeable_marginals} lies between the corresponding estimates for the right and left eyes reported in Table~\ref{tab:gen_Lehmann_results}, since enforcing exchangeability effectively averages the two marginal effects.

	\begin{table}[h]
		\centering
		\caption{Regression estimates for the marginal Cox model at four bivariate time points under exchangeability. Positive coefficients indicate lower bivariate survival probability.}
		\begin{tabular}{lccc}
			\hline
			\textbf{Time point / Covariate} & \textbf{Estimate} & \textbf{SE} & \textbf{p-value} \\
			\hline
			$(6,6)$ (Intercept) & -2.448 & 0.216 & $<0.001$ \\
			$(12,12)$           & 0.224  & 0.028 & $<0.001$ \\
			$(18,18)$           & 0.323  & 0.033 & $<0.001$ \\
			$(24,24)$           & 0.383  & 0.035 & $<0.001$ \\
			\midrule
			Age at diagnosis     & 0.018  & 0.004 & $<0.001$ \\
			Income level 2       & -0.271 & 0.115 & 0.019 \\
			Income level 3       & -0.482 & 0.141 & $<0.001$ \\
			Income level 4       & -0.210 & 0.143 & 0.142 \\
			Tumor stage cT3      & 1.173  & 0.119 & $<0.001$ \\
			Tumor stage cT4      & 0.937  & 0.178 & $<0.001$ \\
			Over 4               & 0.240  & 0.410 & 0.558 \\
			Sex (female)         & 0.057  & 0.078 & 0.464 \\
			Family history (no)  & 0.162  & 0.176 & 0.356 \\
			Age $\times$ Over 4  & -0.014 & 0.007 & 0.044 \\
			\hline
		\end{tabular}
		\label{tab:gen_Lehmann_exchangeable_marginals}
	\end{table}
	
	The first four rows of Table~\ref{tab:gen_Lehmann_exchangeable_marginals} correspond to the four time points $\{6, 12, 18, 24\}$, with each time point having its own intercept. Under the exchangeability assumption, the two marginal survival functions coincide, so a single set of intercepts suffices. These intercepts are related to the common baseline marginal survival function via $\beta_0(t^0)=\log\left(-\log S_T^0(t^0)\right)$.
	The intercepts increase with time, reflecting the expected decrease in survival probability.
	
	Regarding the covariates, a positive regression coefficient for covariate $Z_j$ indicates that higher values of $Z_j$ are associated with lower survival probabilities. Specifically, older age at diagnosis decreases survival probability, while net age effect is reduced after age four.
	For the categorical covariates, sex and family history of retinoblastoma are not statistically significant, whereas income level and tumor stage are. As in the simple Lehmann model (Table~\ref{tab:sim_Lehmann_exchangeable}), the effect of income is non-monotonic: patients in income level 3 (upper-middle) have the highest survival probabilities, followed by income level 2 (lower-middle) and then income level 4 (high). For tumor stage, patients with low-stage tumors (cT2) have the highest survival, while the regression coefficient for cT4 is smaller than that of cT3, indicating that patients with advanced tumors (cT4) have higher survival than those with intermediate-stage tumors (cT3).
	
	Overall, these findings are broadly consistent with previous studies on retinoblastoma \citep{fabian_global_2022}, which focused on different outcomes such as time to first enucleation. In summary, age, income level, and tumor stage are the primary determinants of marginal survival probabilities, while sex and family history have negligible effects.
	
	In the second step, we estimated the remaining regression parameter $\gamma$ using modified pseudo-values (original bivariate pseudo-observations divided by the product of marginal survival probabilities), requiring that $\gamma_0(24,12)=\gamma_0(12,24)$. Estimation used time points $$(t_1,t_2) \in \{(24,12),(12,24),(6,6),(12,12),(18,18),(24,24)\}.$$ Table~\ref{tab:lehmann_gamma_exchangeable} shows the results. As can be seen, most covariates have non-significant effects on the modified pseudo-response, and the overall impact on bivariate survival depends on both marginal and joint effects. These combined effects are illustrated graphically in Figure~\ref{fig:fig-b_exchange}.
	%As can be seen, both the intercept and slope estimates are very similar to those reported in Table~\ref{tab:lehmann_gamma}.

	\begin{table}[h]
		\centering
		\caption{Regression estimates for the joint parameter $\gamma$ in the generalized Lehmann model at 6 bivariate time points under exchangeability. Positive coefficients indicate lower bivariate survival probability.}
		\begin{tabular}{lccc}
			\hline
			\textbf{Time point / Covariate} & \textbf{Estimate} & \textbf{SE} & \textbf{p-value} \\
			\hline
			(12,12) (Intercept) & -4.998 & 2.236 & 0.025 \\
			(6,6)               & -0.479 & 0.249 & 0.055 \\
			(24,12) and (12,24) & 0.085  & 0.066 & 0.200 \\
			(18,18)             & 0.047  & 0.084 & 0.577 \\
			(24,24)             & 0.162  & 0.139 & 0.245 \\
			\midrule
			Age at diagnosis     & 0.045  & 0.026 & 0.081 \\
			Income level 2       & -0.959 & 0.923 & 0.298 \\
			Income level 3       & -1.967 & 1.381 & 0.155 \\
			Income level 4       & 0.371  & 0.907 & 0.683 \\
			Tumor stage cT3      & 2.917  & 1.599 & 0.068 \\
			Tumor stage cT4      & 0.521  & 2.190 & 0.812 \\
			Over 4               & 4.073  & 1.864 & 0.029 \\
			Sex (female)         & -0.484 & 0.577 & 0.401 \\
			Family history (no)  & 0.428  & 1.365 & 0.754 \\
			Age $\times$ Over 4  & -0.064 & 0.037 & 0.082 \\
			\hline
		\end{tabular}
		\label{tab:lehmann_gamma_exchangeable}
	\end{table}

	\subsubsection{Graphical goodness-of-fit}
\begin{figure}[h]
		\centering
		\begin{subfigure}[t]{0.99\textwidth}
			\centering
			\includegraphics[width=0.99\linewidth]{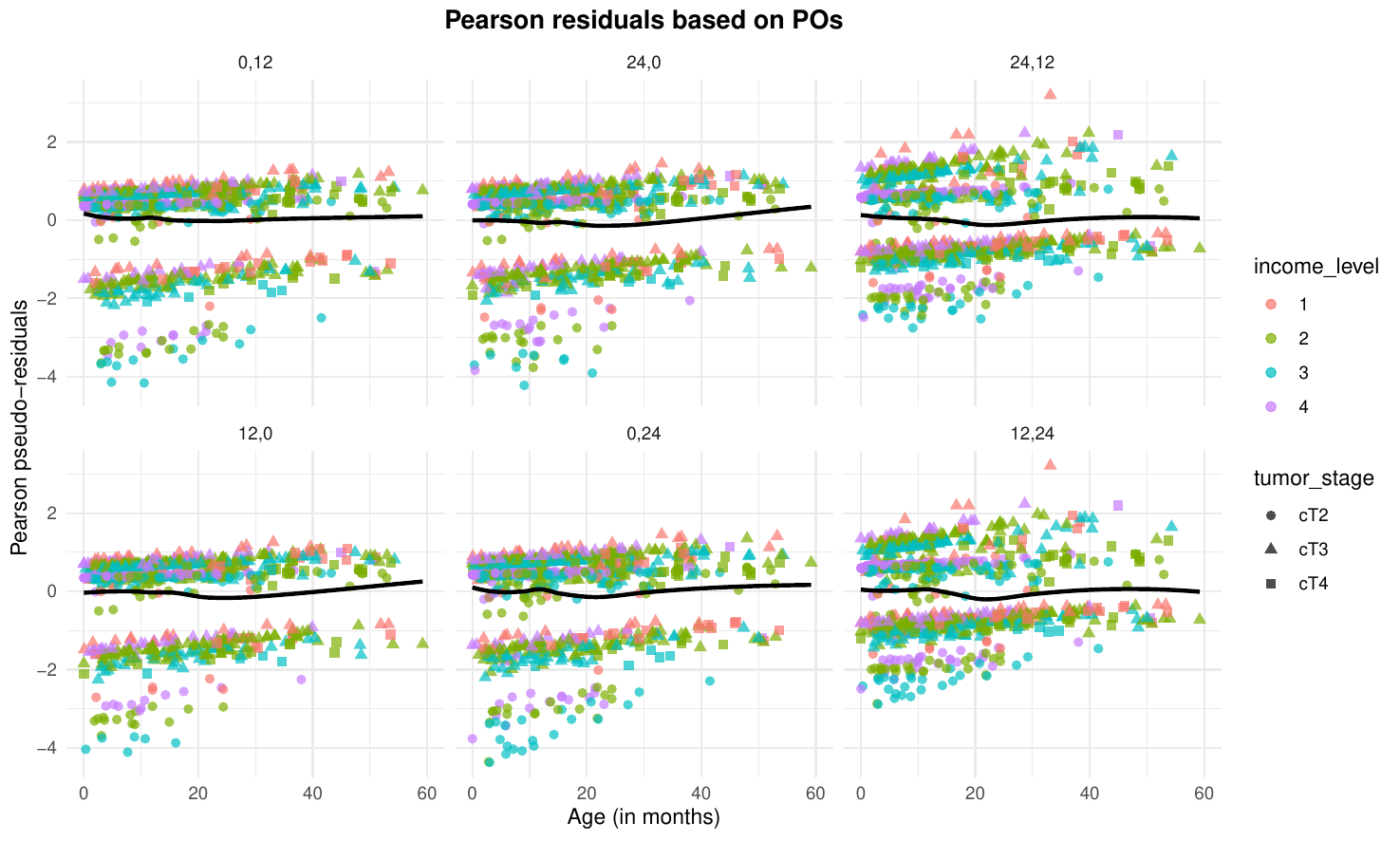}
			\caption{Simple Lehmann model}
		\end{subfigure}
		\hfill
		\begin{subfigure}[t]{0.99\textwidth}
			\centering
			\includegraphics[width=0.99\linewidth]{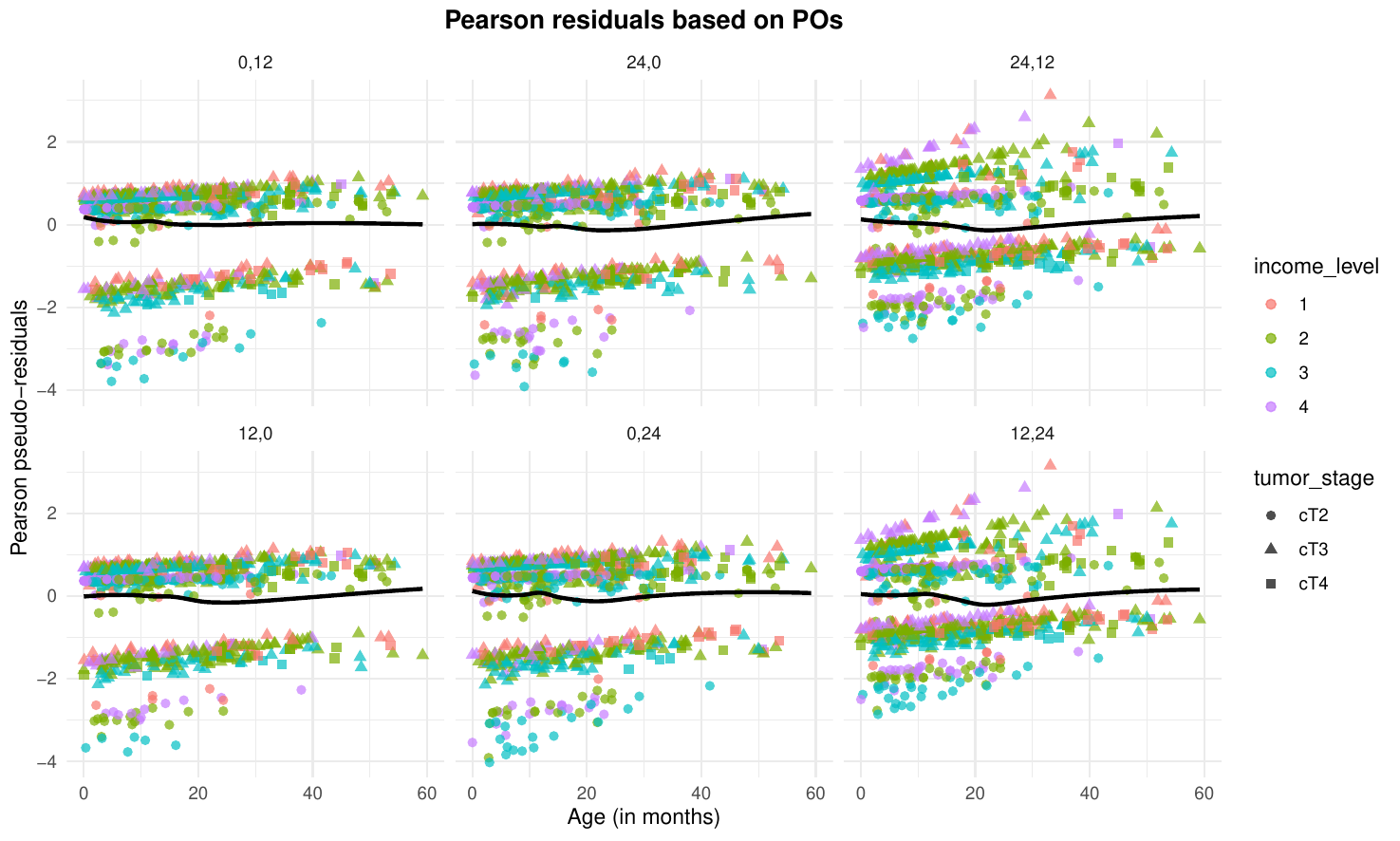}
			\caption{Generalized Lehmann model}
		\end{subfigure}
		\caption{Exchangeable analysis. Graphical goodness-of-fit based on pseudo-residuals at six time points, plotted against age at diagnosis (in months). Top: simple Lehmann model; bottom: generalized Lehmann model.}
		\label{fig:GOF_exchange}
	\end{figure}

	We use the diagnostic tool described in Section~\ref{app:GOF} to evaluate model fit. Figure~\ref{fig:GOF_exchange} displays the Pearson pseudo-residuals for both the simple (top) and generalized (bottom) Lehmann models, based on six time points. The figure focuses on ages at diagnosis up to 60 months (5 years), as only 14 of the 977 patients were diagnosed after this age. Across all six time points, the residuals are centered around zero with no apparent trend, providing support for the proportional derivatives of the log survival assumption at these bivariate time points.

	\subsection{Additional results for the copula-based analysis}
	
\begin{figure}[h]
		\centering
		\includegraphics[width=0.99\textwidth]{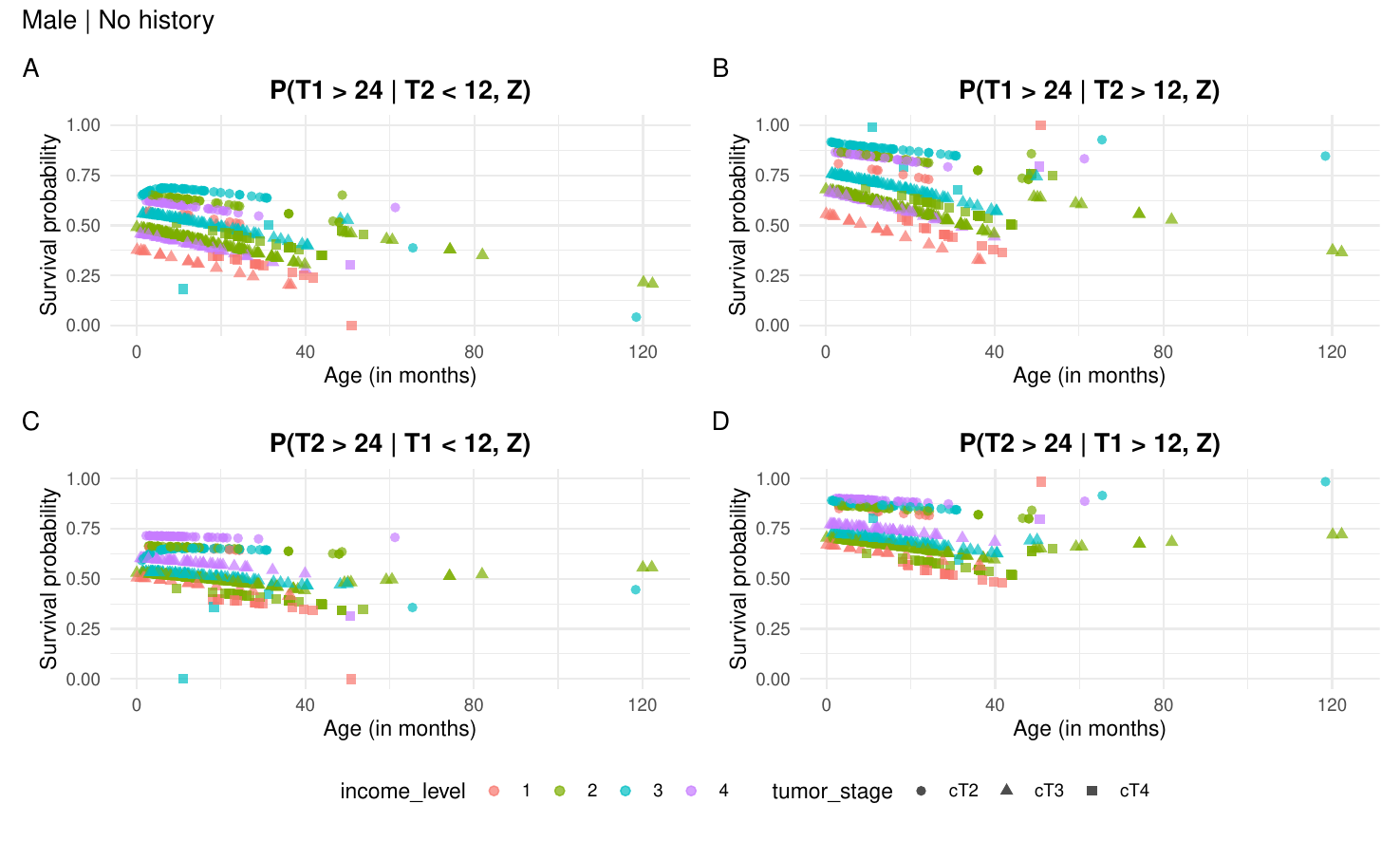}
		\caption{Estimated conditional survival probabilities under the copula-based approach of \cite{marra_copula_2020}. Panels A and C correspond to $\Pr(T_i>24 \mid T_j \leq 12, Z)$, while panels B and D correspond to $\Pr(T_i>24 \mid T_j > 12, Z)$. The results reflect the positive quadrant dependence implied by the selected Clayton copula.}
		\label{fig:cond_prob_cop}
	\end{figure}

	Figure~\ref{fig:cond_prob_cop} presents the estimated conditional survival probabilities obtained from the copula-based approach described in Section~\ref{sec:data}. These results complement the comparison in the main text and illustrate the implied dependence structure under the selected copula model.
	
	\subsection{Sensitivity analysis without the exchangeability constraint}\label{app:nonexchangeable_retinoblastoma}
	
	As a sensitivity analysis, we repeated the retinoblastoma data analysis without imposing exchangeability between the right and left eyes. This unconstrained analysis allows the two marginal survival functions, as well as the joint survival surface, to differ by eye. It therefore provides a useful comparison to the main analysis in Section~\ref{sec:data}, where exchangeability was adopted as the primary working assumption.

	We fitted both the simple and generalized Lehmann models without imposing symmetry constraints. As before, we used the time points
	$$\{(0,12), (24,0), (24,12), (12,0), (0,24), (12,24), (6,6), (12,12), (18,18), (24,24)\}.$$
	Tables~\ref{tab:simple_Lehmann_results}--\ref{tab:lehmann_gamma} report the corresponding regression estimates.
	
	\begin{table}[h]
		\centering
		\caption{Regression estimates for the simple Lehmann model at 10 bivariate time points. Positive coefficients indicate lower bivariate survival probability.}
		\begin{tabular}{lccc}
			\hline
			\textbf{Time point / Covariate} & \textbf{Estimate} & \textbf{SE} & \textbf{p-value} \\
			\hline
			$(0,12)$ (Intercept) & -2.437 & 0.249 & $<0.001$ \\
			$(24,0)$              & 0.334  & 0.106 & 0.002 \\
			$(24,12)$             & 1.026  & 0.065 & $<0.001$ \\
			$(12,0)$              & 0.221  & 0.107 & 0.041 \\
			$(0,24)$              & 0.212  & 0.035 & $<0.001$ \\
			$(12,24)$             & 1.077  & 0.066 & $<0.001$ \\
			$(6,6)$               & 0.721  & 0.072 & $<0.001$ \\
			$(12,12)$             & 0.949  & 0.064 & $<0.001$ \\
			$(18,18)$             & 1.064  & 0.066 & $<0.001$ \\
			$(24,24)$             & 1.137  & 0.067 & $<0.001$ \\
			\midrule
			Age at diagnosis           & 0.020  & 0.004 & $<0.001$ \\
			Income level 2             & -0.399 & 0.144 & 0.005 \\
			Income level 3             & -0.619 & 0.165 & $<0.001$ \\
			Income level 4             & -0.225 & 0.175 & 0.196 \\
			Tumor stage cT3            & 1.297  & 0.128 & $<0.001$ \\
			Tumor stage cT4            & 0.942  & 0.194 & $<0.001$ \\
			Over 4      & 0.515  & 0.562 & 0.294 \\
			Sex (female)               & 0.043  & 0.090 & 0.630 \\
			Family history (no)    & 0.221  & 0.190 & 0.237 \\
			Age $\times$ Over 4        & -0.018 & 0.009 & 0.027 \\
			\hline
		\end{tabular}
		\label{tab:simple_Lehmann_results}
	\end{table}
	
	\begin{table}[h]
		\centering
		\caption{Regression estimates for the marginal Cox models (Margin 1 and Margin 2) at 4 time points estimated using the univariate pseudo-observations approach. Positive coefficients indicate lower marginal survival probability.}
		\begin{tabular}{lccc}
			\hline
			\textbf{Time point / Covariate}  & \textbf{Estimate} & \textbf{SE} & \textbf{p-value} \\
			\hline
			Margin 1 (t=6, intercept)    & -2.642 & 0.396 & $<0.001$ \\
			Margin 2 (t=6, intercept)    & -2.346 & 0.338 & $<0.001$ \\
			Margin 1 (t=12)   & 0.179  & 0.034 & $<0.001$ \\
			Margin 2 (t=12)   & 0.281  & 0.043 & $<0.001$ \\
			Margin 1 (t=18)   & 0.256  & 0.038 & $<0.001$ \\
			Margin 2 (t=18)   & 0.405  & 0.051 & $<0.001$ \\
			Margin 1 (t=24)   & 0.291  & 0.041 & $<0.001$ \\
			Margin 2 (t=24)   & 0.492  & 0.054 & $<0.001$ \\
			\midrule
			Margin 1: Age at diagnosis              & 0.020  & 0.006 & $<0.001$ \\
			Margin 2: Age at diagnosis             & 0.017  & 0.006 & $0.006$ \\
			Margin 1: Income level 2 & -0.323 & 0.201 & $0.108$ \\
			Margin 2: Income level 2  & -0.237 & 0.205 & $0.249$ \\
			Margin 1: Income level 3  & -0.678 & 0.237 & $0.004$ \\
			Margin 2: Income level 3  & -0.295 & 0.235 & $0.210$ \\
			Margin 1: Income level 4  & -0.198 & 0.252 & $0.431$ \\
			Margin 2: Income level 4  & -0.230 & 0.273 & $0.400$ \\
			Margin 1: Tumor stage cT3 & 1.277  & 0.182 & $<0.001$ \\
			Margin 2: Tumor stage cT3 & 1.043  & 0.181 & $<0.001$ \\
			Margin 1: Tumor stage cT4 & 0.866  & 0.272 & $0.001$ \\
			Margin 2: Tumor stage cT4 & 0.983  & 0.262 & $<0.001$ \\
			Margin 1: Over 4            & -0.385 & 0.844 & $0.648$ \\
			Margin 2: Over 4            & 0.688  & 0.742 & $0.354$ \\
			Margin 1: Sex (female)             & -0.057 & 0.128 & $0.658$ \\
			Margin 2: Sex (female)            & 0.159  & 0.131 & $0.226$ \\
			Margin 1: Family history (no) & 0.549 & 0.327 & $0.093$ \\
			Margin 2: Family history (no) & -0.124 & 0.259 & $0.633$ \\
			Margin 1: Age $\times$ Over 4        & -0.010 & 0.012 & $0.441$ \\
			Margin 2: Age $\times$ Over 4        & -0.017 & 0.012 & $0.152$ \\
			\hline
		\end{tabular}
		\label{tab:gen_Lehmann_results}
	\end{table}
	
	\begin{table}[h]
		\centering
		\caption{Regression estimates for the joint parameter $\gamma$ in the generalized Lehmann model at 6 bivariate time points.}
		\begin{tabular}{lccc}
			\hline
			\textbf{Time point / Covariate}  & \textbf{Estimate} & \textbf{SE} & \textbf{p-value} \\
			\hline
			(24,12) (Intercept)  & -5.240 & 2.521 & 0.038 \\
			(12,24)              & 0.147  & 0.144 & 0.308 \\
			(6,6)                & -0.506 & 0.266 & 0.057 \\
			(12,12)              & -0.015 & 0.071 & 0.832 \\
			(18,18)              & 0.029  & 0.085 & 0.734 \\
			(24,24)              & 0.153  & 0.133 & 0.251 \\
			\midrule
			Age at diagnosis     & 0.046  & 0.027 & 0.082 \\
			Income level 2       & -0.819 & 1.013 & 0.419 \\
			Income level 3       & -1.909 & 1.505 & 0.205 \\
			Income level 4       & 0.485  & 1.011 & 0.632 \\
			Tumor stage cT3      & 3.116  & 1.971 & 0.114 \\
			Tumor stage cT4      & 0.618  & 2.575 & 0.810 \\
			Over 4               & 4.282  & 1.884 & 0.023 \\
			Sex (female)         & -0.394 & 0.584 & 0.500 \\
			Family history (no)  & 0.290  & 1.289 & 0.822 \\
			Age $\times$ Over 4  & -0.067 & 0.037 & 0.073 \\
			\hline
		\end{tabular}
		\label{tab:lehmann_gamma}
	\end{table}
	
	%\begin{table}[h]
	%	\centering
	%	\caption{Regression estimates for the joint parameter $\gamma$ in the generalized Lehmann model at 6 bivariate time points.}
	%	\begin{tabular}{lccc}
		%		\hline
		%		\textbf{Time point / Covariate}  & \textbf{Estimate} & \textbf{SE} & \textbf{p-value} \\
		%		\hline
		%		(24,12) (Intercept)  & -5.240 & 2.463 & 0.033 \\
		%		(12,24)              & 0.147  & 0.131 & 0.262 \\
		%		(6,6)                & -0.506 & 0.228 & 0.026 \\
		%		(12,12)              & -0.015 & 0.068 & 0.825 \\
		%		(18,18)              & 0.029  & 0.083 & 0.728 \\
		%		(24,24)              & 0.153  & 0.122 & 0.210 \\
		%		\midrule
		%		Age at diagnosis     & 0.046  & 0.026 & 0.071 \\
		%		Income level 2       & -0.819 & 0.960 & 0.393 \\
		%		Income level 3       & -1.909 & 1.459 & 0.191 \\
		%		Income level 4       & 0.485  & 0.932 & 0.603 \\
		%		Tumor stage cT3      & 3.116  & 1.946 & 0.109 \\
		%		Tumor stage cT4      & 0.618  & 2.537 & 0.808 \\
		%		Over 4               & 4.282  & 1.693 & 0.012 \\
		%		Sex (female)         & -0.394 & 0.559 & 0.482 \\
		%		Family history (no)  & 0.290  & 1.255 & 0.817 \\
		%		Age $\times$ Over 4  & -0.067 & 0.034 & 0.051 \\
		%		\hline
		%	\end{tabular}
	%	\label{tab:lehmann_gamma}
	%\end{table}
	
	Table~\ref{tab:simple_Lehmann_results} presents the results for the simple Lehmann model. The first ten rows correspond to the bivariate time points, with the intercept at $(0,12)$ taken as the baseline; coefficients at other time points represent deviations from this reference. These intercepts are linked to the baseline bivariate survival function through $b_0(t_1^0,t_2^0)=\log\{-\log(S^0(t_1^0,t_2^0))\}$. As expected, the coefficients increase with time, reflecting decreasing survival probabilities along the natural partial order on the bivariate time plane. The estimated covariate effects are very similar to those obtained under the exchangeable analysis (Table~\ref{tab:sim_Lehmann_exchangeable}).
	
	For the generalized Lehmann model, comparison of the marginal estimates in Table~\ref{tab:gen_Lehmann_results} with those from the exchangeable analysis (Table~\ref{tab:gen_Lehmann_exchangeable_marginals}) shows that the unconstrained model includes twice as many parameters, reflecting that $S_1(t\mid Z)$ and $S_2(t\mid Z)$ are allowed to differ. As expected, each coefficient in Table~\ref{tab:gen_Lehmann_exchangeable_marginals} lies between the corresponding estimates for the right and left eyes in Table~\ref{tab:gen_Lehmann_results}, since imposing exchangeability effectively averages the two marginal effects. The estimates for the joint parameter $\gamma$ in the unconstrained analysis (Table~\ref{tab:lehmann_gamma}) are also very similar to those obtained under the exchangeable constraint (Table~\ref{tab:lehmann_gamma_exchangeable}).
	
\begin{figure}[h]
		\centering
		\begin{subfigure}[t]{0.99\textwidth}
			\centering
			\includegraphics[width=0.99\linewidth]{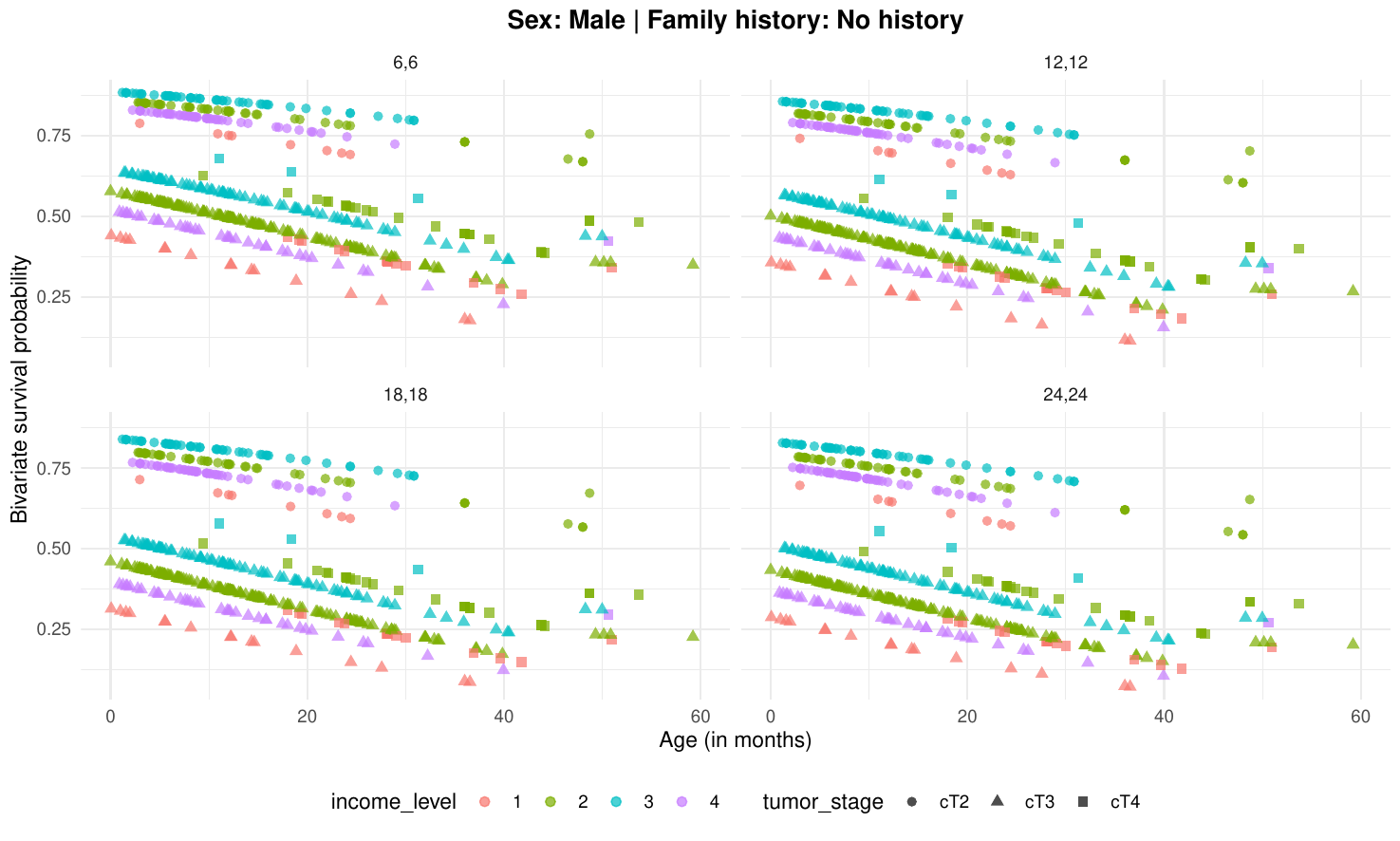}
			\caption{Simple Lehmann model}
			\label{fig:fig-a}
		\end{subfigure}
		\hfill
		\begin{subfigure}[t]{0.99\textwidth}
			\centering
			\includegraphics[width=0.99\linewidth]{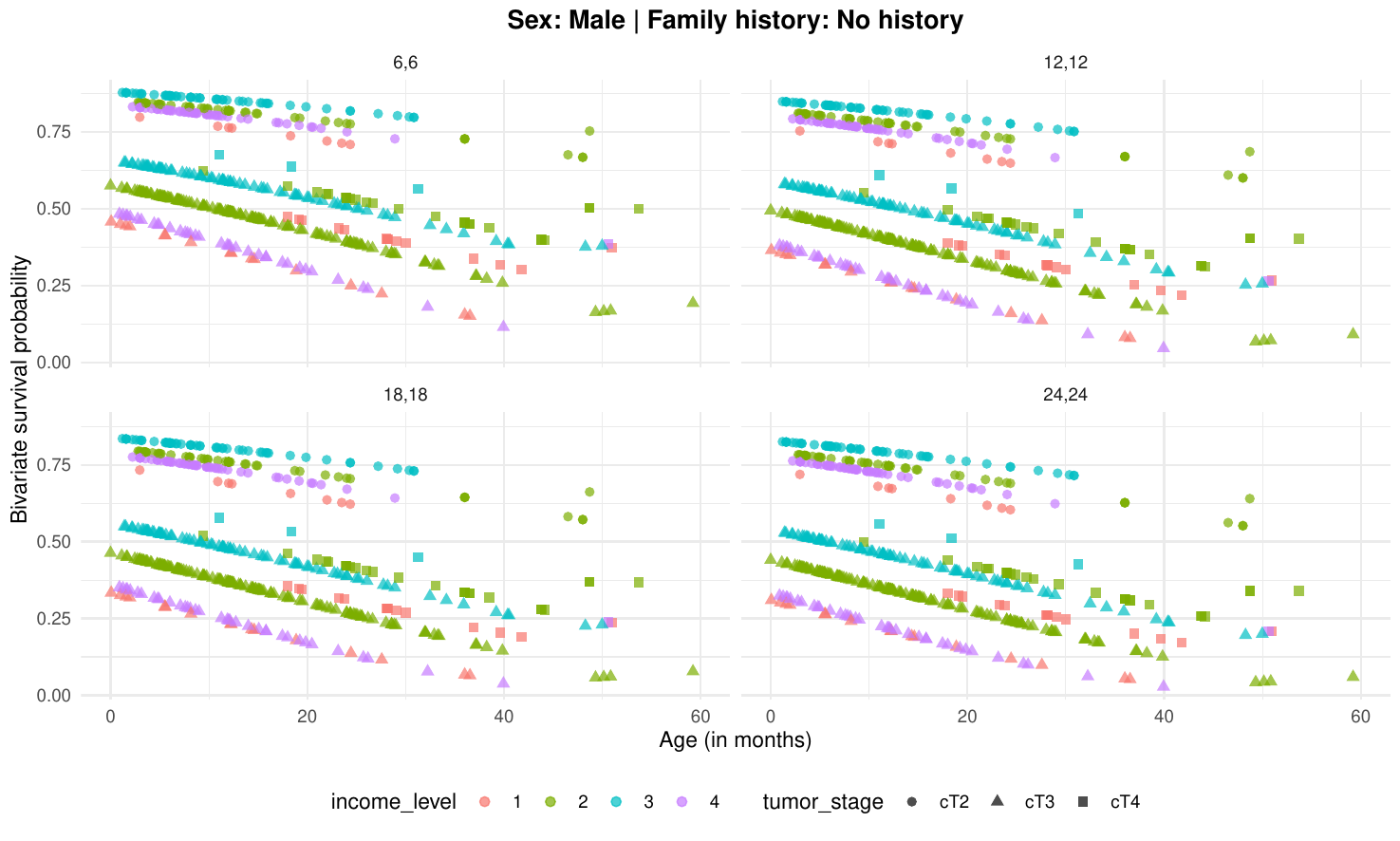}
			\caption{Generalized Lehmann model}
			\label{fig:fig-b}
		\end{subfigure}
		\caption{Predicted bivariate survival probabilities at 6, 12, 18, and 24 months as functions of age at presentation, income level, and tumor stage for males with no family history of retinoblastoma. Top: simple Lehmann model; bottom: generalized Lehmann model.}
		\label{fig:predicted_probs_males_no_hist}
	\end{figure}

	Figure~\ref{fig:predicted_probs_males_no_hist} displays the predicted bivariate survival probabilities for both models (top: simple; bottom: generalized) at the four bivariate time points $(6,6)$, $(12,12)$, $(18,18)$, and $(24,24)$, shown as functions of age (x-axis), income level (color), and tumor stage (shape). These predictions correspond to males without a family history of retinoblastoma; similar patterns were observed across other combinations of sex and family history.
	
	Both unconstrained Lehmann models yielded highly similar predicted probabilities. As in the exchangeable analysis, bivariate survival decreased monotonically with time as we moved farther from the origin. Among patients diagnosed before age four, predicted survival probabilities declined with increasing age at presentation for each time point. Survival was highest for tumor stage $cT2$, with lower probabilities for $cT3$ and $cT4$; again, $cT4$ showed slightly higher predicted survival than $cT3$, potentially reflecting deaths prior to enucleation or treatment with palliative chemotherapy alone.
	
	Predicted survival also varied by income level: upper-middle income (level 3) exceeded lower-middle (level 2), which in turn exceeded high income (level 4). As noted in the main analysis, this pattern may reflect lower overall mortality among higher-income patients \citep{fabian_global_2022}, possibly due to broader access to treatment options, including enucleation. As before, death prior to enucleation represents a competing risk that is not modeled explicitly here.
	
	Next, we estimated the survival probability of the right eye ($T_1$) conditional on either failure or survival of the left eye ($T_2$), and vice versa. Specifically, we computed
	$P(T_1>24 \mid T_2\leq 12, Z)$,
	$P(T_1>24 \mid T_2>12, Z)$,
	$P(T_2>24 \mid T_1\leq 12, Z)$, and
	$P(T_2>24 \mid T_1>12, Z)$.
	These quantities were obtained from
	\begin{align*}
		P(T_1>24 \mid T_2\leq 12, Z)
		&=\frac{S(24,0 \mid Z)-S(24,12 \mid Z)}{1-S(0,12 \mid Z)}
		=\frac{S_1(24 \mid Z)-S(24,12 \mid Z)}{1-S_2(12 \mid Z)}, \\
		P(T_1>24 \mid T_2>12, Z)
		&=\frac{S(24,12 \mid Z)}{S(0,12 \mid Z)}
		=\frac{S(24,12 \mid Z)}{S_2(12 \mid Z)},
	\end{align*}
	and similarly,
	\begin{align*}
		P(T_2>24 \mid T_1\leq 12, Z)
		&=\frac{S(0,24 \mid Z)-S(12,24 \mid Z)}{1-S(12,0 \mid Z)}
		=\frac{S_2(24 \mid Z)-S(12,24 \mid Z)}{1-S_1(12 \mid Z)}, \\
		P(T_2>24 \mid T_1>12, Z)
		&=\frac{S(12,24 \mid Z)}{S(12,0 \mid Z)}
		=\frac{S(12,24 \mid Z)}{S_1(12 \mid Z)}.
	\end{align*}
	
	\begin{figure}[h]
		\centering
		\begin{subfigure}[t]{0.99\textwidth}
			\centering
			\includegraphics[width=0.99\linewidth]{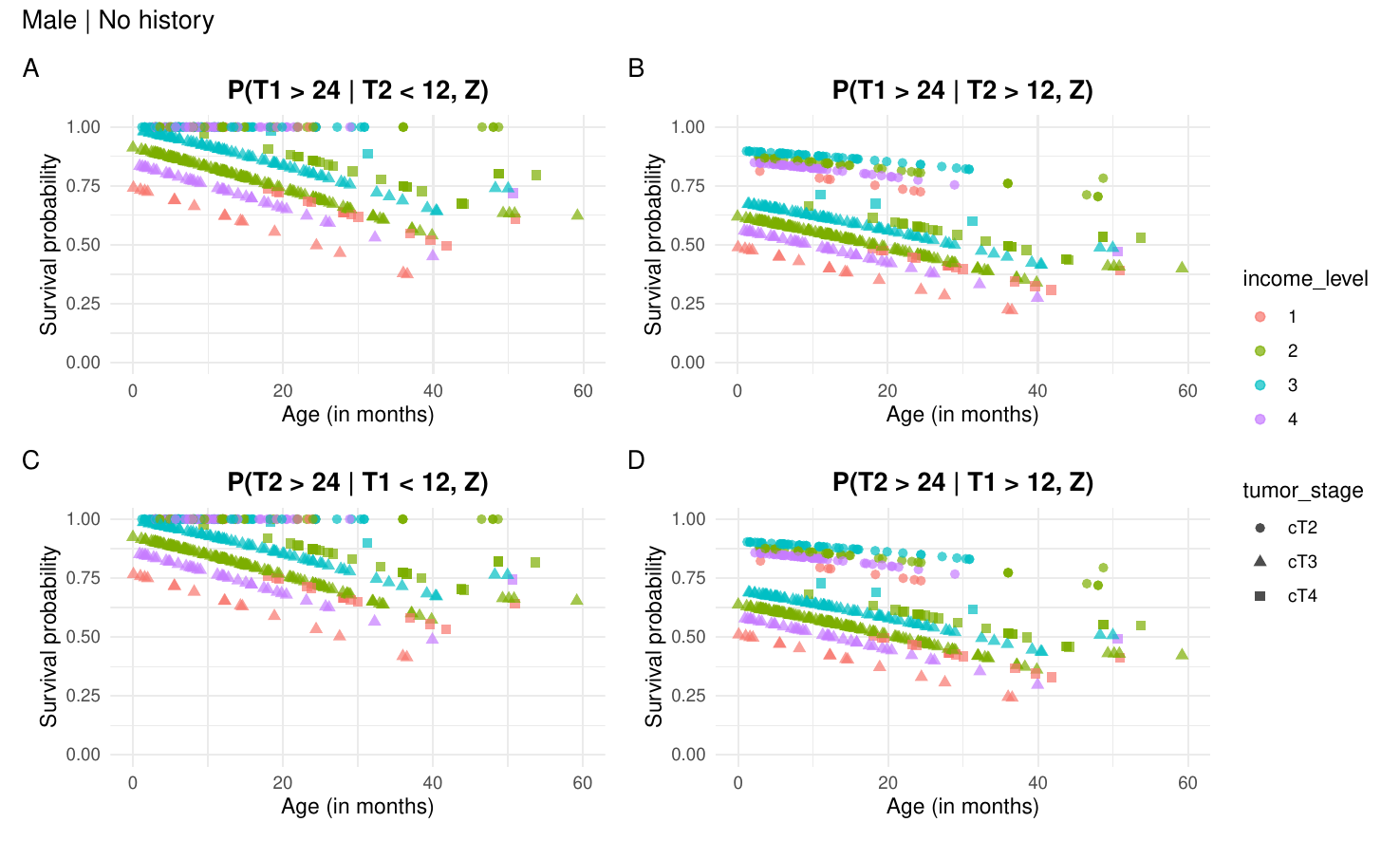}
			\caption{Simple Lehmann model}
		\end{subfigure}
		\hfill
		\begin{subfigure}[t]{0.99\textwidth}
			\centering
			\includegraphics[width=0.99\linewidth]{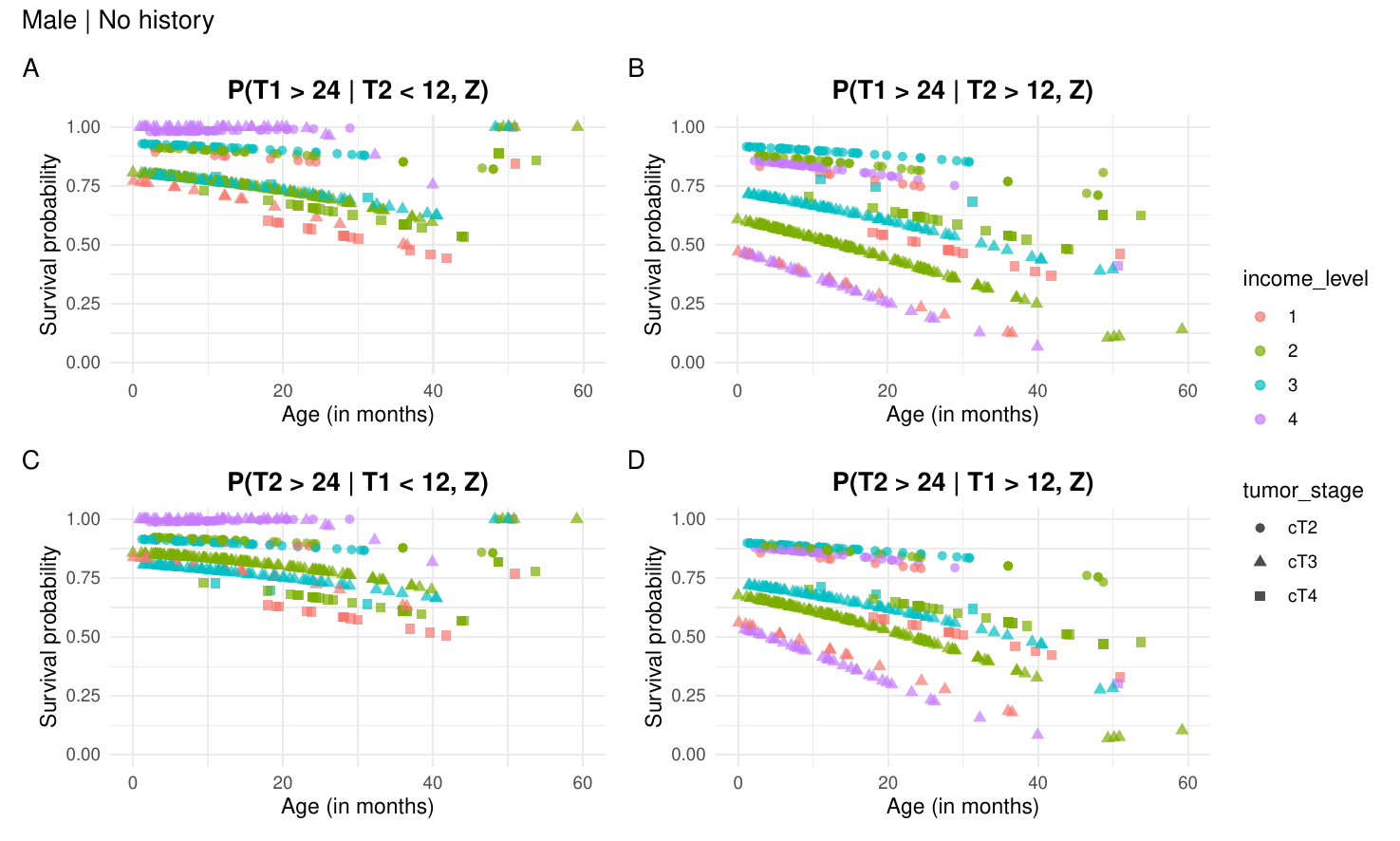}
			\caption{Generalized Lehmann model}
		\end{subfigure}
		\caption{Predicted conditional survival probabilities for males with no family history of retinoblastoma. Top: simple Lehmann model; bottom: generalized Lehmann model.}
		\label{fig:cond_probs_males_no_hist}
	\end{figure}
	
	Figure~\ref{fig:cond_probs_males_no_hist} presents the estimated conditional survival probabilities from the simple (top) and generalized (bottom) Lehmann models, adjusted for all covariates. Results are shown for males with no family history of retinoblastoma, with similar patterns observed across other combinations of sex and family history. Conditional survival probabilities generally decreased with increasing age at presentation, and the separation between tumor stages and income levels largely persisted.
	
	When conditioning on enucleation of one eye (panels A and C in Figure~\ref{fig:cond_probs_males_no_hist}), the simple Lehmann model estimates that patients with early-stage disease ($cT2$) have an effectively zero probability of requiring enucleation in the second eye, consistent with the low likelihood of bilateral surgery in mild cases. Under the generalized Lehmann model, high-income patients diagnosed before age two with tumor stage $cT3$ show survival probabilities similar to those with stage $cT2$ and higher than patients from other income levels. Notably, patients with stage $cT3$ diagnosed around age four exhibit unexpectedly high survival, likely reflecting instability in the estimated level change for diagnoses after age four.
	
	Importantly, the unconstrained analysis reveals an approximate symmetry:
	\begin{align*}
		P(T_2>24\mid T_1>12,Z)&\approx P(T_1>24\mid T_2>12,Z), \\
		P(T_2>24\mid T_1\leq 12,Z)&\approx P(T_1>24\mid T_2\leq 12,Z),
	\end{align*}
	suggesting that the two eyes are close to exchangeable even without imposing this restriction. This empirical finding supports the choice of the exchangeable specification as the primary analysis in Section~\ref{sec:data}.
	
	As in the main analysis, the conditional survival probability of one eye beyond two years, given that the other eye was enucleated within the first year, exceeds the corresponding probability conditional on the other eye surviving the first year:
	\[
	P(T_i>24\mid T_j\leq 12,Z)>P(T_i>24\mid T_j>12,Z), \qquad i\neq j.
	\]
	Equivalently,
	\[
	S_i(24\mid Z)S_j(12\mid Z)>S_{ij}(24,12\mid Z),
	\]
	which indicates NQD between the times to enucleation of the two eyes.
	
\begin{figure}[h]
		\centering
		\begin{subfigure}[t]{0.99\textwidth}
			\centering
			\includegraphics[width=0.99\linewidth]{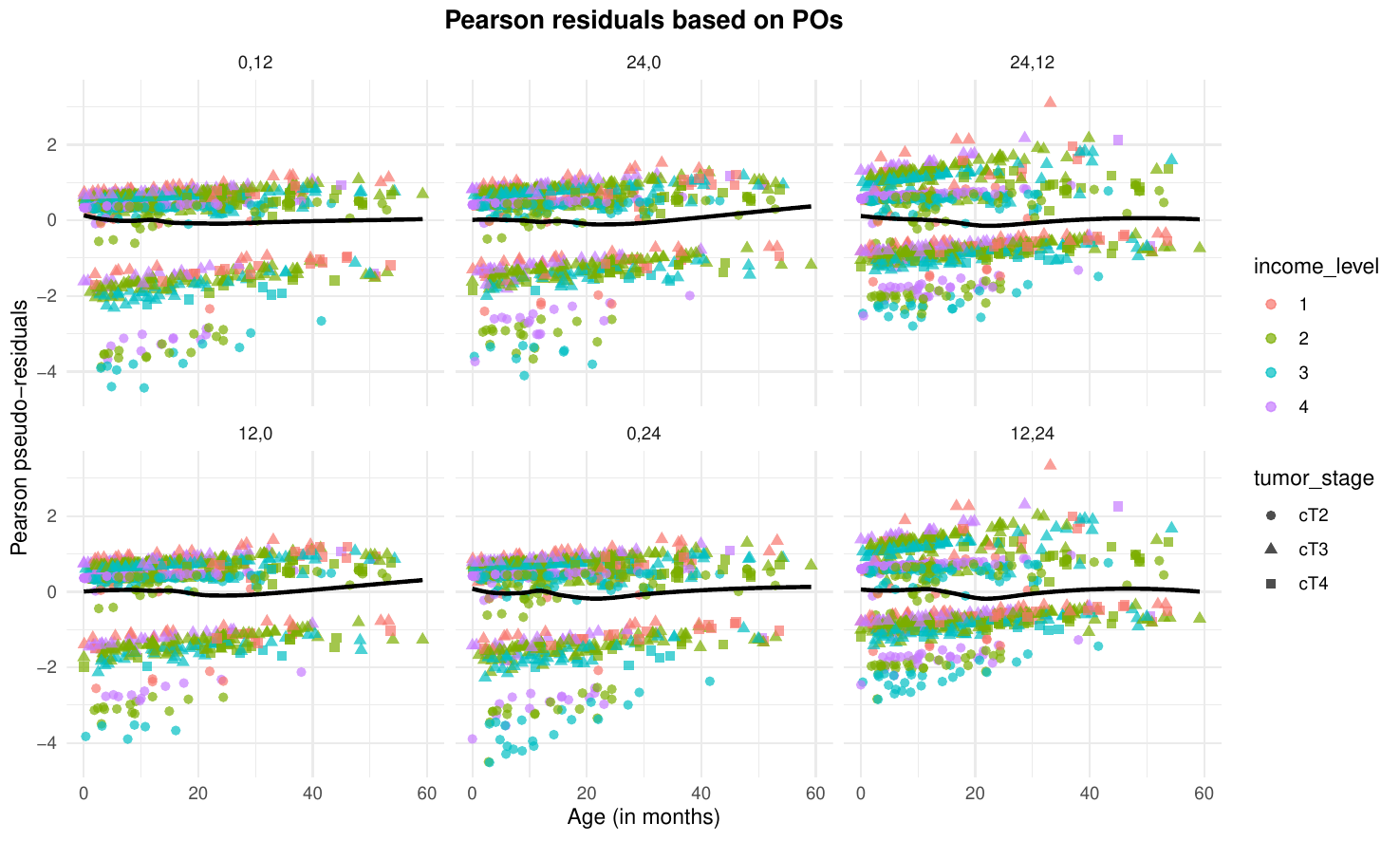}
			\caption{Simple Lehmann model}
		\end{subfigure}
		\hfill
		\begin{subfigure}[t]{0.99\textwidth}
			\centering
			\includegraphics[width=0.99\linewidth]{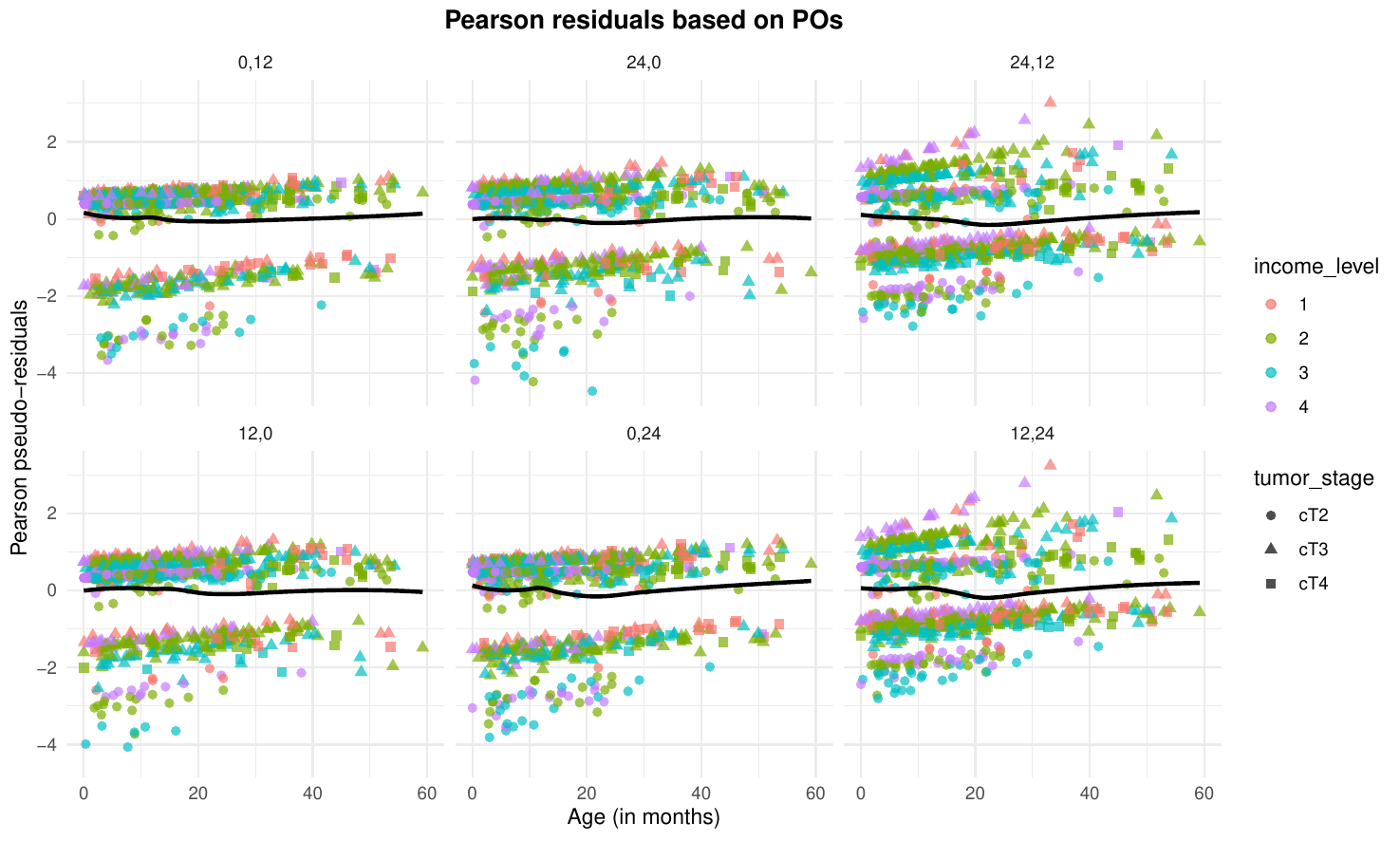}
			\caption{Generalized Lehmann model}
		\end{subfigure}
		\caption{Graphical goodness-of-fit based on pseudo-residuals at six time points, plotted against age at diagnosis (in months). Top: simple Lehmann model; bottom: generalized Lehmann model.}
		\label{fig:GOF}
	\end{figure}

	Finally, Figure~\ref{fig:GOF} shows the graphical diagnostics from Section~\ref{app:GOF} for the unconstrained simple and generalized Lehmann models. In both cases, the pseudo-residuals are centered around zero with no apparent trend, indicating an adequate fit.
	
\end{appendices}
\bibliographystyle{plainnat}
\bibliography{Cox2D}

\end{document}